\documentclass[12pt]{article} 

\pdfoutput=1

\usepackage[top=80pt,bottom=85pt,left=85pt,right=85pt]{geometry}
\usepackage{amssymb}
\usepackage{amsmath}
\usepackage{mathtools}
\usepackage{setspace}
\usepackage{accents}
\usepackage{comment}
\usepackage[english]{babel}

\usepackage[utf8]{inputenc}
\usepackage{uniinput}
\usepackage{changepage,cite}
\usepackage[plain]{fancyref}

\usepackage{makecell} 
\usepackage{longtable} 
\usepackage[normalem]{ulem} 
\usepackage{transparent} 
\usepackage{bm} 

\usepackage[usenames]{xcolor}
\usepackage{graphicx,subcaption}
\usepackage{setspace}
\usepackage{float}
\usepackage{booktabs}
\usepackage{rotating}
\usepackage[vcentermath]{youngtab}

\usepackage[debug,pageanchor=false]{hyperref}
\definecolor{link}{rgb}{.8,.15,.1}
\definecolor{pigment}{rgb}{0.36, 0.54, 0.66}
\definecolor{pigment2}{rgb}{0.19, 0.55, 0.91}
\definecolor{pigment3}{rgb}{0.2, 0.2, 0.6}
\definecolor{light-gray}{gray}{0.75}
\hypersetup{colorlinks=true,linkcolor=link,citecolor=link,urlcolor=link,linktocpage}
\usepackage{cleveref}

\usepackage{array,multirow,booktabs,longtable}
\usepackage{mathrsfs}
\usepackage{tikz-cd} 
\usetikzlibrary{backgrounds, arrows,calc,shapes,decorations.pathreplacing, decorations.markings, automata,positioning,matrix}

\newcommand\vertarrowbox[3][6ex]{%
  \begin{array}[t]{@{} c c @{}} #2 \\
  \left\downarrow\vcenter{\hrule height #1}\right.\kern-\nulldelimiterspace & #3
  \end{array}%
}

\tikzset{
        cvertex/.style={circle,draw=black,inner sep=1pt,outer sep=3pt},
        vertex/.style={circle,fill=black,inner sep=1pt,outer sep=3pt},
        star/.style={circle,fill=yellow,inner sep=0.75pt,outer sep=0.75pt},
        tvertex/.style={inner sep=1pt,font=\scriptsize},
        gap/.style={inner sep=0.5pt,fill=white}}

\tikzstyle{mybox} = [draw=black, fill=blue!10, very thick,
    rectangle, rounded corners, inner sep=10pt, inner ysep=20pt]
\tikzstyle{boxtitle} =[fill=blue!50, text=white,rectangle,rounded corners]

\def\node#1#2{\overset{#1}{\underset{#2}{\circ}}}
\def\ver#1#2{\overset{{\llap{$\scriptstyle#1$}\displaystyle\circ{\rlap{$\scriptstyle#2$}}}}{\scriptstyle\vert}}

\def\verT#1{\overset{\overset{\displaystyle\curvearrowright}{\displaystyle #1}}{\scriptstyle \vert}}
\def\vernocirc#1{\overset{\displaystyle #1}{\scriptstyle \vert}}

\usetikzlibrary{arrows}
\tikzstyle{every picture}+=[remember picture]
\tikzstyle{na} = [baseline=-.5ex]
\tikzstyle{mine}= [arrows={angle 90}-{angle 90},thick]

\def\Llleftarrow{%
\lower2pt\hbox{\begingroup
\tikz
\draw[shorten >=0pt,shorten <=0pt] (0,3pt) -- ++(-1em,0) (0,1pt) -- ++(-1em-1pt,0) (0,-1pt) -- ++(-1em-1pt,0) (0,-3pt) -- ++(-1em,0) (-1em+1pt,5pt) to[out=-105,in=45] (-1em-2pt,0) to[out=-45,in=105] (-1em+1pt,-5pt);
\endgroup}
}

\newcommand{\rr}{\mathbb{R}}
\newcommand{\cc}{\mathbb{C}}
\newcommand{\zz}{\mathbb{Z}}
\newcommand{\pp}{\mathbb{P}}

\renewcommand{\Re}{\mathrm{Re}}
\renewcommand{\Im}{\mathrm{Im}}

\DeclareMathOperator{\rk}{rk}

\DeclareMathOperator{\Tr}{Tr}
\DeclareMathOperator{\SU}{SU}
\DeclareMathOperator{\USp}{USp}
\DeclareMathOperator{\U}{U}
\DeclareMathOperator{\SO}{SO}

\DeclareMathOperator{\depth}{depth}

\newcommand{\todo}[1]{}
\renewcommand{\todo}[1]{{\color{red} TODO: {#1}}}
\newcommand{\red}[1]{}
\renewcommand{\red}[1]{{\color{red} {#1}}}
\newcommand{\blue}[1]{}
\renewcommand{\blue}[1]{{\color{blue} {#1}}}

\newcommand{\su}[1]{}
\renewcommand{\su}[1]{{\mathfrak{su}({#1})}}
\newcommand{\uu}[1]{}
\renewcommand{\uu}[1]{{\mathfrak{u}({#1})}}
\newcommand{\so}[1]{}
\renewcommand{\so}[1]{{\mathfrak{so}({#1})}}
\newcommand{\usp}[1]{}
\renewcommand{\usp}[1]{{\mathfrak{usp}({#1})}}
\def\f{\mathfrak{f}}

\renewenvironment{rcases}
  {\left.\begin{aligned}}
  {\end{aligned}\right\rbrace}
  
\makeatletter
\renewcommand\xleftrightarrow[2][]{%
  \ext@arrow 9999{\longleftrightarrowfill@}{#1}{#2}}
\newcommand\longleftrightarrowfill@{%
  \arrowfill@\leftarrow\relbar\rightarrow}
\makeatother

\makeatletter
\@addtoreset{equation}{section}
\makeatother

\graphicspath{{figures/}}

\begin{document}


\begin{titlepage}

\begin{center}

\vskip .3in \noindent

{\Large \textbf{Hierarchy of RG flows in 6d $(1,0)$ orbi-instantons}}

\bigskip

{\large \textsc{part i}}

\bigskip

Marco Fazzi and Suvendu Giri

\bigskip
{\small 
Dipartimento di Fisica, Universit\`a di Milano--Bicocca, \\ Piazza della Scienza 3, I-20126 Milano, Italy \\
\vspace{.25cm}
INFN, sezione di Milano--Bicocca,  Piazza della Scienza 3, I-20126 Milano, Italy\\ 
}

\vskip .3cm
{\small \tt \href{mailto:marco.fazzi@mib.infn.it}{marco.fazzi@mib.infn.it}} \hspace{.5cm} {\small \tt \href{mailto:suvendu.giri@unimib.it}{suvendu.giri@unimib.it}}

\vskip .6cm
     	{\bf Abstract }
\vskip .1in
\end{center}

\noindent
$N$ M5-branes probing the intersection between the orbifold $\cc^2/\Gamma_\text{ADE}$ and an $E_8$ wall give rise to 6d $(1,0)$ SCFTs known as ADE-type orbi-instantons.  At fixed $N$ and order of the orbifold, each element of $\text{Hom}(\Gamma_\text{ADE},E_8)$ defines a different SCFT.  The SCFTs are connected by Higgs branch RG flows, which generically reduce the flavor symmetry of the UV fixed point.  We determine the full hierarchy of these RG flows for type A, i.e. $\cc^2/\zz_k$, for any value of $N$ and $k$.  The hierarchy takes the form of an intricate Hasse diagram: each node represents an IR orbi-instanton (homomorphism),  and each edge an allowed flow, compatibly with the 6d $a$-theorem.  The partial order is defined via quiver subtraction of the 3d magnetic quivers associated with the 6d SCFTs,  which is equivalent to performing a so-called Kraft--Procesi transition between homomorphisms.

\vfill
\eject

\end{titlepage}


\tableofcontents
\newpage


\section{Introduction}

The construction and classification of six-dimensional superconformal field theories (6d SCFTs) via string and M/F-theory is a momentous result of the recent past. This endeavor has sent shock waves through all lower dimensional classification efforts, providing new tools and perspectives to construct theories,  geometrize (and thus understand the origin of) the dualities they enjoy,  and map out their interdependencies across dimensions.

The simplest geometric engineering of a 6d SCFT requires putting Type IIB string theory on the orbifold $\cc^2/\Gamma_\text{ADE}$, with $\Gamma$ a finite subgroup of $\SU(2)$.  This setup preserves $(2,0)$ supersymmetry, and all $(2,0)$ SCFTs can be classified according to their ADE-type \cite{Witten:1995zh,Strominger:1995ac,Witten:1995em}.  For $(1,0)$ supersymmetry the landscape of models is much richer (the first string and brane constructions date back to \cite{Seiberg:1996vs,Seiberg:1996qx,Ganor:1996mu,Intriligator:1997kq,Blum:1997mm,Intriligator:1997dh,Brunner:1997gk,Brunner:1997gf,Hanany:1997gh}), and a full classification has been achieved only very recently in the context of F-theory \cite{Heckman:2015bfa} (see \cite{Heckman:2018jxk} for a review circa 2018, whose notations we will adopt throughout this paper).

One of the most pressing dynamical questions dealing with (S)CFTs is the structure of the renormalization group (RG) flow of which they are fixed points.\footnote{As M.~Strassler once put it \cite{Strassler:2003qg}, `Go with the flow'.}  For 6d SCFTs we distinguish two types of supersymmetry-preserving deformations which trigger an RG flow out of the ultraviolet (UV) SCFT onto its moduli space \cite{Cordova:2015fha,Louis:2015mka,Cordova:2016emh,Cordova:2016xhm}:  tensor branch flows obtained by giving vacuum expectation values (vevs) to tensor multiplet scalars,\footnote{These parameterize the separations between M5s or NS5s in brane constructions -- see table \ref{tab:M5}.} where the deep infrared (IR) theory is a quiver gauge theory (which breaks conformal invariance); and Higgs branch flows obtained by giving vevs to hypermultiplet scalars which are charged under the flavor symmetry,  where the endpoint of the flow is a new IR SCFT with generically smaller flavor symmetry.\footnote{There also exist mixed branches. We will comment on them where appropriate.} A quantity which captures the intuition that along RG flows high-energy degrees of freedom are integrated out is the $a$ anomaly of the SCFT, which can be understood as a measure of the degrees of freedom of the theory.  Along any RG flow the ``$a$-theorem'' should be satisfied, namely $\Delta a = a_\text{UV} - a_\text{IR}> 0$.  (In fact $a\geq 0$, and the equality holds if and only if there are no local degrees of freedom \cite{Hofman:2008ar,Elvang:2012st}.) This is a direct generalization of Zamolodchikov's result in two dimensions \cite{Zamolodchikov:1986gt} and its four-dimensional analog for the $\mathcal{N}=1$ $a$ anomaly \cite{Komargodski:2011vj,Komargodski:2011xv}.

Tensor branch flows are well-understood for all 6d SCFTs (starting from \cite[Sec. 7]{Heckman:2013pva}), and the corresponding $a$-theorem was proven in \cite{Cordova:2015fha}. On the contrary, the Higgs branch case has remained elusive, and a proof of the $a$-theorem is known \cite{Mekareeya:2016yal} only for a certain class of models known as T-brane theories \cite{DelZotto:2014hpa}.\footnote{See \cite[Sec. 8]{Heckman:2018jxk} for a summary of known results on tensor branch and Higgs branch flows.} (Prior evidence was provided in field theory in \cite{Gaiotto:2014lca,Heckman:2015axa}, and then confirmed holographically in \cite{Cremonesi:2015bld,Apruzzi:2017nck}.)

Most interesting for our purposes in the $(1,0)$ landscape is the theory of $N$ M5-branes probing the intersection between a Ho\v{r}ava--Witten $E_8$ wall and the orbifold $\cc^2/\Gamma_\text{ADE}$,  a theory we will refer to as ``ADE-type orbi-instanton'' following \cite{Heckman:2018pqx}.  This is the theory of $N$ small $E_8$ instantons probing the orbifold \cite{Witten:1995gx,Seiberg:1996vs,Ganor:1996mu}.  Its M-theory engineering is specified by the following data: the number $N$ of M5-branes (i.e. the number of tensor multiplets of the SCFT), the order $k$ of the orbifold (for type A or D), and a choice of ``boundary condition'' at infinity, namely an element of $\text{Hom}(\Gamma_\text{ADE},E_8)$ specifying how the orbifold group embeds into $E_8$ (whose details will be spelled out in section \ref{sec:kac}). 

In the F-theory language of \cite{Heckman:2015bfa} we can think of an orbi-instanton as a decoration of a rank-$N$ E-string theory
\begin{equation}\label{eq:Estring}
[E_8]\ \underbrace{1\ 2 \cdots\ 2}_{N}\ 
\end{equation}
to
\begin{equation}\label{eq:Estrnonres}
[E_8]\ \underbrace{\overset{\mathfrak{g}}{1}\ \overset{\mathfrak{g}}{2} \cdots \overset{\mathfrak{g}}{2}}_{N}\ [G]\ ,
\end{equation}
where $G$ is a group with ADE Lie algebra $\mathfrak{g}$ associated to the finite orbifold group $\Gamma_\text{ADE}$ via the usual $\SU(2)$ McKay correspondence.\footnote{For an introduction on the ``geometric'' $\SU(2)$ McKay correspondence see e.g. the old but pedagogical reference \url{https://math.ucr.edu/home/baez/joris_van_hoboken_platonic.pdf}, or \cite[Chap. 6]{leuschke}.  The F-theory engineering of \eqref{eq:Estrnonres} was first constructed in \cite{Aspinwall:1997ye}, and their matter content derived in \cite{Intriligator:1997dh} by requiring gauge anomaly cancellation on the tensor branch. This assumed a ``trivial'' boundary condition which preserves the full $E_8$; see \eqref{eq:1k} and \eqref{eq:fullymasslessE8} below.} The decorated $\overset{\mathfrak{g}}{1}$ curve adjacent to $E_8$ ``gauges'' the rank-1 E-string $[E_8]1$ on the left of \eqref{eq:Estring} (the $E_8$ algebra being supported on a $-12$ noncompact curve), i.e.  gauges a subalgebra $\mathfrak{g}$ of $E_8$, leaving behind a commutant.  Exactly which commutant is preserved as a flavor symmetry factor of the SCFT is specified by the M-theory boundary condition.  The rightmost noncompact $-2$ curve supporting $G$ provides another flavor symmetry factor of the SCFT, which for us will just be a spectator.\footnote{That is, we will not consider RG flows triggered by vevs for hypermultiplets charged under it.} Importantly, \cite[Sec. 7]{Heckman:2015bfa} argued that there exists one F-theory configuration of decorated curves per boundary condition on the M-theory side (i.e. the information carried by the boundary condition is fully geometrized in F-theory), and \cite{Frey:2018vpw} used this observation to construct a dictionary between M-theory and F-theory for orbi-instantons of any ADE type.  For type A, \cite{Mekareeya:2017jgc} gave an algorithm which allows to construct the full tensor branch and matter content of the 6d SCFT,\footnote{See also \cite{Zafrir:2015rga,Ohmori:2015tka,Hayashi:2015zka} for previous results.} and to identify the SCFT Higgs branch with a certain $E_8$ moduli space (including its quaternionic dimension).\footnote{See also \cite{Hanany:2018uhm}.  For the undecorated rank-$N$ E-string the Higgs branch was already identified in \cite[Sec. 5]{Cordova:2015fha}. The rank-1 E-string can also be seen as arising at the end of a Higgsing tree of 6d SCFTs with larger tensor branch \cite[Fig. 5]{DelZotto:2018tcj}.}

In spite of much effort to construct these orbi-instantons,  the structure of the Higgs branch RG flows that connect them is still terra incognita.  As is clear from the following short list of references, the known results are few and far between.  The existence of Higgs branch RG flows between orbi-instantons of fixed type and at fixed order of $\Gamma_\text{ADE}$ but with different boundary conditions was pointed out already in \cite[Sec. 6.1.2]{DelZotto:2014hpa} and \cite[Sec. 7.5]{Heckman:2015bfa}, and the first explicit examples (analyzed via the 6d anomaly polynomial) were presented in \cite[Sec. 4.2 \& 4.3]{Heckman:2015ola}.  Later \cite[Sec. 4]{Frey:2018vpw} proposed an explicit hierarchy of RG flows in the  $k=4$ case of A-type orbi-instantons where, as we will see, there are ten inequivalent choices of boundary condition (corresponding to ten different F-theory configurations).  These are ``mixed'' flows however, which sometimes can also involve a small instanton transition reducing the number $N$ of tensor multiplets by one unit (turning a tensor into twenty-nine hypermultiplets \cite{Ganor:1996mu}).  We will come back to this point in appendix \ref{app:RG}. Much more recently \cite[Sec. 4]{Giacomelli:2022drw} has analyzed the $k=2,3$ and partially the $k=4,6$ cases in type A while also changing $N$, proposing a hierarchy of flows for these values of $k$ (with the expectation that a similar logic applies to any $k$).\footnote{In \cite{Giacomelli:2020gee} it was noted that for the torus compactifications of A-type orbi-instantons engineering the 4d $\mathcal{N}=3$ S-folds known as $\mathcal{T}^N_{G,k}$ and $\mathcal{S}^N_{G,k}$ (where $G$ corresponds to the strong-coupling summand $[G]$ of $\f$, in the notation introduced in section \ref{subsub:IIAk6}), different choices of boundary condition are related by RG flow.}

Abstracting from specific examples, the question as to whether there exists a hierarchy of Higgs branch RG flows for orbi-instantons of any type, and whether this translates into a Hasse diagram of elements of $\text{Hom}(\Gamma_\text{ADE},E_8)$ was asked in \cite[p. 31]{Heckman:2018pqx},\footnote{Much like what happens for T-brane theories with nilpotent orbits of the flavor symmetry algebra $\mathfrak{g}$ \cite{Heckman:2016ssk}, where the relevant homomorphisms are elements (embeddings) of $\text{Hom}(\su{2},\mathfrak{g})$.}  but no answer was provided.  In this paper we will provide an answer to these questions for type A (and propose a way forward to analyze type D in the conclusions).  We fix the number $N$ of M5's (i.e. our flows do \emph{not} involve small instanton transitions) and the order $k$ of the orbifold  and derive the full hierarchy of Higgs branch RG flows between orbi-instantons with different boundary conditions, at the origin of their tensor branch. Namely, we only deal with Higgs branches ``at infinite (gauge) coupling'', in the parlance launched by \cite{Cremonesi:2015lsa}.

On the mathematics side this translates into a natural proposal for a Hasse diagram of (injective) homomorphisms (i.e. embeddings) $\rho \in \text{Hom}(\zz_k,E_8)$ for type A ($\text{Hom}(\mathbb{D}_{k},E_8)$ for type D),  where the partial ordering $\rho_1 \succ \rho_2$ is physically obtained via an operation performed on the tensor branch of the SCFTs known as ``quiver subtraction'' \cite{Cabrera:2018ann}.  In the case of nilpotent orbits $\mathcal{O}_\mu$ of a Lie algebra $\mathfrak{g}$ (i.e. homomorphisms $\mu: \mathfrak{su}(2) \to \mathfrak{g}$ by Jacobson--Morozov \cite{collingwood1993nilpotent}), quiver subtraction between the tensor branch descriptions of two SCFTs (two 6d ``electric quivers'') is equivalent \cite{Cabrera:2016vvv,Cabrera:2017njm,Hanany:2018uhm} to performing a so-called ``Kraft--Procesi transition'' between two orbits \cite{KP0,KP1,KP2},\footnote{The transition modifies the singularity type of a nilpotent orbit closure (see \cite[Thm. 3.2]{KP1} for $\mathfrak{su}$,  \cite[Thm.  2]{KP2} for $\mathfrak{so}$ and $\mathfrak{usp}$, and \cite[Thm. 1.2]{fu-juteau-levy-sommers} for the exceptional cases) to that of the next along the partially ordered set -- see e.g. the tables in \cite[Sec. 7]{Cabrera:2016vvv} for a few $\mathfrak{su}$ examples.} and the partial ordering $\mu_1 \succ \mu_2$ (given by inclusion of their closures $\overline{\mathcal{O}}_{\mu_2} \subset \overline{\mathcal{O}}_{\mu_1}$ as varieties) is exactly reproduced in field theory by the hierarchy of allowed Higgs branch RG flows of T-brane theories.  That is, the Hasse diagram of nilpotent orbits mimics closely the hierarchy of RG flows of \cite{Heckman:2016ssk}.  In the case of $\text{Hom}(\zz_k,E_8)$ there is \emph{no} known notion of Kraft--Procesi transition, even though the physics of 6d SCFTs suggests a similar concept should exist.  We will come back to this point in the conclusions.

In practice,  to perform the subtraction we first construct the 3d ``magnetic quiver'' \cite{Cabrera:2019izd} associated with the 6d electric quiver of an SCFT (i.e. its full tensor branch description); we propose that a flow $\mathcal{T}_1 \to \mathcal{T}_2$ exists between two SCFTs $\mathcal{T}_i$ if we can subtract the corresponding 3d magnetic quivers (in a technical sense introduced in section \ref{sec:mag}), as first noted in \cite{Giacomelli:2022drw} for A-type orbi-instantons and in the general spirit of \cite{Bourget:2019aer}. When this happens, we have an ordering $\rho_1 \succ \rho_2$,  the orbi-instanton $\mathcal{T}_i$ being defined by the homomorphism $\rho_i \in \text{Hom}(\zz_k,E_8)$ (for fixed $N$), i.e. by the boundary condition in M-theory.  To add strength to this proposal we check that $\Delta a>0$ (i.e. compatibility with the $a$-theorem) for each allowed flow by computing the exact $a$ anomaly of the SCFTs.  At each step of the RG flow hierarchy we write down the flavor symmetry that is preserved by the boundary condition. In this way, we are able to produce very intricate Hasse diagrams, an example of which is given in figure \ref{fig:flowk6}.  The remainder of this paper is dedicated to explaining how to obtain such partially ordered diagrams for any $N,k$ as well as the meaning of the labels and decorations that appear therein.  We  unearth many subtleties for high values of $k$, which were overlooked by previous references (since they focused on specific cases for which they do not arise). We will dedicate due space to explain their physical meaning.

This paper is organized as follows. We begin in section \ref{sec:kac} by constructing orbi-instantons of type A in M-theory: we introduce Kac labels, their Type IIA realization, and their associated 3d magnetic quivers. In section \ref{sec:k=6} we showcase the technology just introduced in the $k=6$ case, produce the Hasse diagram of homomorphisms, and point out a number of physically interesting facts. In section \ref{sec:highk} we move to generic $k$, and discuss some subtleties which only arise for high-enough $k$. Finally we present our conclusions in section \ref{sec:conc}. In appendix \ref{app:RG} we collect the Hasse diagrams for $k=2,\ldots,20$, and draw a comparison with the result of \cite[Fig. 1]{Frey:2018vpw} for $k=4$. The figures appearing in the appendix are also included as ancillary files with the \texttt{arXiv} submission for the convenience of the reader.

\section{Orbi-instantons of type A and Kac labels}
\label{sec:kac}

\subsection{M-theory engineering}
\label{sub:Mth}

Consider $N$ M5's probing the intersection between an $E_8$ wall (sometimes denoted M9) and a singularity $\cc^2/\Gamma_\text{ADE}$ (with $\Gamma \subset \SU(2)$ a finite ADE group). Let us focus on type A, i.e. the orbifold is $\cc^2/\zz_k$.  The M-theory brane scan, as well as its reduction to Type IIA (which will be important in the next section), is presented in table \ref{tab:M5}.  This setup provides the M-theory engineering of $N$ small $E_8$ instantons in the heterotic string probing the singularity $\cc^2/\zz_k$, which requires a choice of boundary condition \cite[Sec. 6.1.2]{DelZotto:2014hpa}: the latter corresponds to turning on a flat $E_8$ connection (holonomy) at infinity on $\cc^2/\zz_k$. These flat connections are one-to-one with homomorphisms $\pi_1(S^3/\zz_k) \to E_8$ which are in turn one-to-one with homomorphisms $\rho: \zz_k \to E_8$, which are classified \cite{kac1990infinite}.  
\begin{table}[ht!]
\centering
{\renewcommand{\arraystretch}{1.2}%
\begin{tabular}{@{}l c c c c c c  @{}}
& & $\rr$ & \multicolumn{4}{c}{$\cc^2/\Gamma_\text{ADE}$} \\
& $x^{0\ldots 5}$ & $x^6$ & \multicolumn{1}{|c}{$x^7$} & $x^8$ & $x^9$ & $S^1_\text{M}$\\
\toprule\toprule 
M5-branes & $\times$ & $\cdot$ & $\cdot$ & $\cdot$ & $\cdot$ & $\cdot$ \\
M9-wall & $\times$ & $\cdot$ & $\times$ & $\times$ & $\times$ & $\times$ \\
\toprule
NS5-branes & $\times$ & $\cdot$ & $\cdot$ & $\cdot$ & $\cdot$ & \\
D6-branes & $\times$ & $\times$ & $\cdot$ & $\cdot$ & $\cdot$ &  \\
O8-plane & $\times$ & $\cdot$ & $\times$ & $\times$ & $\times$  \\
D8-branes & $\times$ & $\cdot$ & $\times$ & $\times$ & $\times$ & \\
\toprule
\end{tabular}}
\caption{A $\cdot$ means the brane is sitting at a point along that direction; $\times$ means it is infinitely extended along that (noncompact) direction.  The tensor branch of the SCFT is parameterized by the separations along $x^6$ of the M5's (NS5's).  We choose to place the M9 or O8 at the origin of $x^6$: when all M5's (NS5's) are coincident at $x^6=0$ we reach the CFT point of the orbi-instanton since their separations are proportional to the gauge couplings (see e.g. \eqref{eq:FIcoup}),  i.e. we are at the origin of the tensor branch and the CFT possesses a Higgs branch at infinite coupling. }
\label{tab:M5}
\end{table}
In fact a holonomy at spatial infinity $S^3/\zz_k$ (i.e. the sphere surrounding the singularity of $\cc^2/\zz_k$) around the $\zz_k$ one-cycle is a representation $\rho: \zz_k \to E_8$ which can be nicely encoded in a choice of ``Kac label'' \cite{Mekareeya:2017jgc}, i.e. a choice of positive integers
\begin{equation}\label{eq:kaccoord}
\node{}{n_1}-\node{}{n_2}-\node{}{n_3}-\node{}{n_4}-\node{}{n_5}-\node{\ver{}{n_{3'}}}{n_6}-\node{}{n_{4'}}-\node{}{n_{2'}}
\end{equation}
such that
\begin{equation}\label{eq:E8partition}
k = \sum_{j=0}^5 a_j n_{j+1} + 4 n_{4'} + 3 n_{3'} + 2 n_{2'} \ .
\end{equation}
That is, the integers $n_i$ are the multiplicities of the Coxeter labels $a_i$ of the affine $E_8$ Dynkin diagram
\begin{equation}\label{eq:E81}
E_8^{(1)}: \quad \node{1}{\alpha_0}-\node{2}{\alpha_1}-\node{3}{\alpha_2}-\node{4}{\alpha_3}-\node{5}{\alpha_4}-\node{6\ver{3'}{\alpha_8}}{\alpha_5}-\node{4'}{\alpha_6}-\node{2'}{\alpha_7}\ .
\end{equation}
In turn, the Coxeter labels are integers such that $\sum_{j=0}^8 a_j A^{ji}=0$ for the Cartan matrix $A^{ji}$ of $E_8^{(1)}$.\footnote{For an introduction to Kac--Moody algebras see e.g. \cite[Sec. 7]{fuchs2003symmetries} and the original reference \cite{kac1990infinite}. For simple calculations in Sage (e.g. calculating the explicit form of $A^{ji}$ for $E_8^{(1)}$) see \url{https://doc.sagemath.org/html/en/thematic_tutorials/lie/affine.html}.}

In other words, the requirement \eqref{eq:E8partition} defines a partition of $k$ in terms of the integers $\{1,\ldots,6,4',2',3'\}$ only,\footnote{We will use a prime on $2,3,4$ as a bookkeeping device to distinguish them from their ``unprimed'' version.}  with multiplicities $\{n_1,\ldots,n_6,n_{4'},n_{2'},n_{3'}\}$.  We will denote the most general such partition (Kac label) by
\begin{equation}\label{eq:longk}
k = [1^{n_1},2^{n_2},3^{n_3},4^{n_4},5^{n_5},6^{n_6}, 4'^{n_{4'}},2'^{n_{2'}},3'^{n_{3'}}]
\end{equation}
where of course some of the parts may be absent.\footnote{As is standard,  exponentiation by $n_i$ means multiplication by $n_i$ (as in \eqref{eq:E8partition}), so that $a_i^0=0$.} The subalgebra $\f$ of $E_8$ that is unbroken by the Kac label is the commutant of the image $\rho(\zz_k)\subset E_8$, and its Dynkin diagram is easily obtained by deleting the nodes appearing in the partition \eqref{eq:longk} (i.e. the nodes for which $n_i$ is nonzero), together with an Abelian subalgebra making the total rank 8 \cite[Sec. 8.6]{kac1990infinite}, i.e. a summand of the form $\bigoplus_i \mathfrak{u}(1)_i$.\footnote{The maximal subalgebras that do not contain $\mathfrak{u}(1)$ summands are the semisimple regular ones; those that do are non-semisimple regular. The former are preserved by Kac labels with a single part, the latter by those with more than one part.} (We will comment on its physical interpretation in section \ref{sub:u1}.) These are the so-called pseudo-Levi sub\-algebras of $E_8$, which can be obtained via the Borel--de Siebenthal algorithm.  E.g.  Kac label
\begin{equation}\label{eq:1k}
k=[1^k] = [\underbrace{1,\ldots,1}_{\text{$k$ times}}]
\end{equation}
exists for any $k$, and preserves the full $E_8$ flavor symmetry coming from the wall (since it ``kills'' only the extending node $\alpha_0$ of $E_8^{(1)}$ with Coxeter label $1$ and multiplicity $k$, leaving behind the full $E_8$ Dynkin).

Let us now go back to the F-theory configuration \eqref{eq:Estrnonres}. For type A, $\Gamma_\text{ADE} = \zz_k$ (i.e. $\mathfrak{g}=\su{k}$ in that formula); then for the above choice \eqref{eq:1k} of Kac label the full $E_8$ flavor symmetry is preserved and the fully blown-up configuration of curves (i.e. performing all necessary base blow-ups, as explained in \cite{Aspinwall:1997ye,Heckman:2013pva}) becomes
\begin{equation}\label{eq:fullymasslessE8}
[E_8]\ {1}\ \overset{\mathfrak{su}(1)}{2}\ \overset{\mathfrak{su}(2)}{2} \cdots \overset{\mathfrak{su}(k-1)}{2}\ \overset{\su{k}}{\underset{[N_\text{f}=1]}{2}}\ \overset{\su{k}}{2}\cdots \overset{\su{k}}{2}\ [\SU(k)]\ .
\end{equation}
Notice that this requires $N>k$, and we will assume this is the case throughout.  We label the compact curves starting from 0, i.e. $-1$ is the zeroth curve and does not support any gauge algebra.  The first $-2$ curve sits in position 1, and supports a trivial $\su{1}$ algebra.  The leftmost $\su{k}$ compact curve (i.e. the one in position $k$) has $N_\text{f}=1$ fundamental hypermultiplets, as required by gauge anomaly cancellation on the tensor branch.  The curves $[E_8] 1 2$ make up a rank-2 E-string as in \eqref{eq:Estring}.

On the other hand, the choice of Kac label $[2,1^{k-2}]$, which exists for all $k> 2$,  preserves $E_7 \oplus \mathfrak{u}(1)$ (since it kills both $\alpha_0$ and $\alpha_1$),\footnote{For $k=2$, $[2,1^{k-2}] =[2]$, which instead preserves $E_7 \oplus \su{2}$.} and corresponds to the F-theory configuration
\begin{equation}\label{eq:fullymasslessE7}
[E_7]\ {1}\ \overset{\mathfrak{su}(2)}{\underset{[N_\text{f}=1]}{2}}\ \overset{\mathfrak{su}(3)}{2} \cdots \overset{\mathfrak{su}(k-1)}{2}\ \overset{\su{k}}{\underset{[N_\text{f}=1]}{2}}\ \overset{\mathfrak{su}(k)}{2}\cdots \overset{\su{k}}{2}\ [\SU(k)]\ ,
\end{equation}
with a partially gauged rank-1 E-string on the left (since $E_8$ is broken to $E_7$, which is now supported on a $-8$ noncompact curve).

We close this section with a comment on notation. In the mathematics literature the decorated affine Dynkin diagram \eqref{eq:kaccoord} is known as (a choice of) ``Kac coordinates'' or ``Kac diagram'', and plays a central role in $\zz_k$-gradings (more generally, gradings by an Abelian group)\footnote{If $A$ is any group, then an $A$-grading of $\mathfrak{g}$ is a decomposition of $\mathfrak{g}$ as a direct sum of subspaces $\mathfrak{g}_a$ such that $[\mathfrak{g}_a, \mathfrak{g}_b]$ is contained in $\mathfrak{g}_{ab}$. Since $[\mathfrak{g}_a,\mathfrak{g}_b]=[\mathfrak{g}_b,\mathfrak{g}_a]$,  we must have $\mathfrak{g}_{ab}=\mathfrak{g}_{ba}$ so $A$ is Abelian. Therefore any grading of $\mathfrak{g}$ is a grading by an abelian group $A$, and this is the same as a homomorphism from $A$ to $\text{Aut}(\mathfrak{g})$ \cite{kac1990infinite}. \label{foot:grade}} of complex Lie algebras $\mathfrak{g}$, which is the same as homomorphisms from $\zz_k$ to the (inner and outer) automorphisms group $\text{Aut}(\mathfrak{g})$ of $\mathfrak{g}$ \cite{kac1990infinite} (see also \cite{levy1,levy2,reeder-levy-yu-gross}).  (The pseudo-Levi subalgebras of $\mathfrak{g}$ are just the fixed point subalgebras of a $\zz_k$-grading for some $k$,  and more specifically are fixed points for an inner automorphism. The fixed point subalgebras for the outer automorphisms do not have a name in the literature.) In fact Kac has solved the problem of classifying $\zz_k$-gradings of $E_8(\cc)$, i.e. the complex $E_8$ Lie algebra. In the rest of this paper we will instead use the \emph{compact real form} of $E_8$, as well as real forms of the other Lie algebras, i.e. $\mathfrak{su}$, $\mathfrak{so}$, $\mathfrak{usp}$, $E_{6}$, $E_7$. Even when we write $G$, we mean the Lie algebra $\mathfrak{g}$ associated with the compact Lie group. (We decided to keep the notation $[G]$ to make contact with the existing extensive literature on the subject.) Notice however that the global structure of the (zero-form) flavor symmetry \emph{group} plays an important role in the finer classification of 6d SCFTs. For orbi-instantons, this group (together with its consequences for four-dimensional compactifications) was recently determined in \cite[Sec. 4]{Heckman:2022suy}.

\subsection{Type IIA configurations}
\label{sec:IIA}

It is interesting to perform a reduction from the M-theory configuration described above to Type IIA string theory. This will make transparent the different origins of various flavor symmetry factors, and will allow us to identify certain universal features of the Hasse diagram of homomorphisms in  $\text{Hom}(\zz_k,E_8)$.

Consider M-theory on $S^1_\text{M} \times S^1/\zz_2$ with gauge symmetry $E_8\times E_8$ \cite{Horava:1995qa,Horava:1996ma}.  The gauge degrees of freedom are confined to the two ten-dimensional boundaries of $S^1/\zz_2$, and the theory in the ``bulk'' (i.e. between them) is just Type IIA. In fact in perturbative Type IIA (reducing along $S^1_\text{M}$), each of the two $E_8$ walls becomes an O8$^-$-plane plus 8 D8-branes, which are needed to cancel the total D8 charge (i.e. Romans mass $F_0$) in the compact space provided by the finite interval.\footnote{We work in conventions whereby the charge of an O$p^\pm$ is $\pm 2^{p-5}$ that of a D$p$-brane, so 8 D8's are needed to cancel the charge of an O8$^-$, which carries negative tension and charge.}  This is now a Type I' background, with string coupling proportional to the radius of $S^1_\text{M}$.  To see why this is the case,  let us focus on one of the two ends (O8-planes) of the interval (i.e. the interval $S^1/\zz_2$ effectively becomes the semi-infinite line $\rr_+$ parameterized by $x^6$ in table \ref{tab:M5}). Because D8-branes are a source for the dilaton (i.e. Type I' string coupling), the latter is piecewise-linear with discontinuous derivative at the D8 locations along the interval.  If all of the 8 D8's sit on the orientifold we have constant dilaton and a perturbative $\mathfrak{so}(16)$ gauge symmetry in 9d. We can break $\mathfrak{so}(16) \to \mathfrak{so}(14) \oplus \mathfrak{u}(1)$ by pulling one D8 out (leaving 7 on top of the O8$^-$); if we further tune its location to a critical value it is possible to make the dilaton (i.e. string coupling) blow up exactly at the location of the O8$^-$ \cite{Gorbatov:2001pw}.  When this happens, there is a non-perturbative enhancement $\mathfrak{so}(14) \oplus \mathfrak{u}(1) \to E_8$ of the gauge symmetry, and the whole O8-D8 system lifts to an M9 wall (since the radius of $S^1_\text{M}$ which the former wraps is also blowing up). To see the enhancement one has to consider a number of D0-branes becoming tensionless at the O-plane (since $T_\text{D0} \sim {1}/{g_\text{s}^{I'}}$),  furnishing the remaining states needed to complete the $E_8$ BPS algebra \cite{Bergman:1997py}.  The setup generalizes straightforwardly. For example, to see the enhancement $\mathfrak{so}(12) \oplus \mathfrak{su}(2) \to E_7$ we pull out 2 D8's and tune both their positions to the critical value \cite[Sec. 3]{Gorbatov:2001pw}.

\begin{figure}[ht!]
\centering
\begin{tikzpicture}[scale=1,baseline]
\draw[fill=black] (0,0) circle (0.5cm) node[black,midway, xshift =0.65cm, yshift=.5cm] {\footnotesize $N$};

\draw[dashed,black,very thick] (-.25,-1)--(-.25,1) node[black,midway, xshift =0cm, yshift=1.5cm] {\footnotesize O8$^-$} node[black,midway, xshift =0cm, yshift=1.5cm] {};
\draw[solid,black,very thick] (.25,-1)--(.25,1) node[black,midway, xshift =0cm, yshift=+1.5cm] {\footnotesize $7$};
\draw[solid,black,very thick] (1.5,-1)--(1.5,1) node[black,midway, xshift =0cm, yshift=+1.5cm] {\footnotesize $1$};

\draw[solid,black,thick] (0,0)--(2.5,0) node[black,midway,xshift=1cm,yshift=0.25cm] {\footnotesize $k$};

\draw[solid,black,thick,->] (-3,0)--(-2,0) node[black,midway,xshift=0.5cm,yshift=.325cm] {\footnotesize $x^6$};
\draw[solid,black,thick,->] (-3,0)--(-3,1) node[black,midway,xshift=0.5cm,yshift=.5cm] {\footnotesize $x^{789}$};

\end{tikzpicture}

\begin{tikzpicture}[scale=1,baseline]
\node at (0,0) {};
\draw[fill=black] (0.5,0) circle (0.075cm);
\draw[fill=black] (1.5,0) circle (0.075cm);
\draw[fill=black] (2.5,0) circle (0.075cm);
\draw[fill=black] (3.5,0) circle (0.075cm);
\draw[fill=black] (4.5,0) circle (0.075cm);
\draw[fill=black] (5.5,0) circle (0.075cm);
\draw[fill=black] (6.5,0) circle (0.075cm);
\draw[fill=black] (7.5,0) circle (0.075cm);
\draw[fill=black] (8.5,0) circle (0.075cm);


\draw[solid,black,thick] (0.5,0)--(1.5,0) node[black,midway,yshift=0.2cm] {\footnotesize $1$};
\draw[solid,black,thick] (1.5,0)--(2.5,0) node[black,midway,yshift=0.2cm] {\footnotesize $2$};
\draw[loosely dotted,thick,black] (2.5,0)--(3.5,0) node[black,midway] {};
\draw[solid,black,thick] (3.5,0)--(4.5,0) node[black,midway,yshift=0.2cm] {\footnotesize $k-1$};
\draw[solid,black,thick] (4.5,0)--(5.5,0) node[black,midway,xshift=0.2cm,yshift=0.2cm] {\footnotesize $k$};
\draw[solid,thick,black] (5.5,0)--(6.5,0) node[black,midway,yshift=0.2cm] {\footnotesize $k$};
\draw[loosely dotted,thick,black] (6.5,0)--(7.5,0) node[black,midway] {};
\draw[solid,black,thick] (7.5,0)--(8.5,0) node[black,midway,yshift=0.2cm] {\footnotesize $k$};
\draw[solid,black,thick] (8.5,0)--(9.5,0) node[black,midway,yshift=0.2cm] {\footnotesize $k$};

\draw[dashed,black,very thick] (0,-.5)--(0,.5) node[black,midway, xshift =0cm, yshift=-1.5cm] {} node[black,midway, xshift =0cm, yshift=1.5cm] {};
\draw[solid,black,very thick] (0.25,-.5)--(0.25,.5) node[black,midway, xshift =0cm, yshift=+.75cm] {\footnotesize $7$};
\draw[solid,black,very thick] (4.5+0.375,-.5)--(4.5+0.375,.5) node[black,midway, xshift =0cm, yshift=+.75cm] {\footnotesize $1$};
\end{tikzpicture}

\begin{equation}
[E_8]\ {1}\ \overset{\mathfrak{su}(1)}{2}\ \overset{\mathfrak{su}(2)}{2} \cdots \overset{\mathfrak{su}(k-1)}{2}\ \overset{\su{k}}{\underset{[N_\text{f}=1]}{2}}\ \overset{\su{k}}{2}\cdots \overset{\su{k}}{2}\ [\SU(k)]\nonumber
\end{equation}
\caption{Type IIA reduction of $N$ M5's probing the M9-$\cc^2/\zz_k$ intersection with  Kac label $[1^k]$. Solid vertical lines represent D8-branes (with their number written on top); the dashed vertical line is the O8$^-$; solid horizontal lines represent D6-branes (with their number written on top); a circle represents an NS5 (all $N$ of them being stacked together in the top frame).  The ``zeroth'' gauge algebra is empty, i.e. there are no D6-branes crossing (or ending on) the O8.  \textbf{Top:} origin of the tensor branch (SCFT). \textbf{Middle \& Bottom:} low-energy quiver gauge theory with $\mathfrak{su}$ gauge algebras (i.e. generic point on the tensor branch of the SCFT).}
\label{fig:IIAE8}
\end{figure}
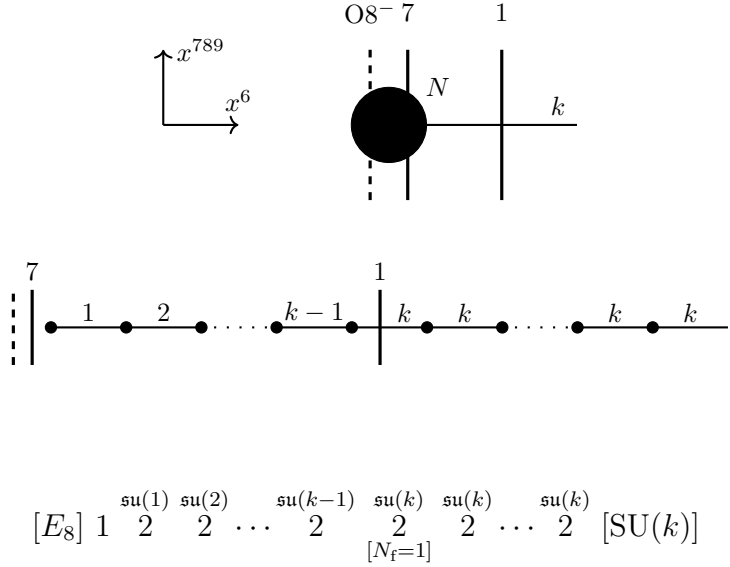

\cite[Sec. 5]{Hanany:1997gh} used the above results on string dualities to claim that $N$ M5-branes probing the intersection between an M9-wall (i.e. only one side of the $E_8\times E_8$ Ho\v{r}ava--Witten setup) and a $\cc^2/\zz_k$ orbifold (with $S^1_\text{M}$ now a fiber of the orbifold,  $\cc \cong \cc^* \times S^1$) reduce to $N$ NS5's intersecting $k$ D6-branes, both probing an O8-D8 system, with the eighth D8 sitting away from the orientifold at a critical value along $x^6$ such that the Type IIA dilaton diverges at the O-plane.  In the language of the previous section, the M-theory setup has trivial boundary condition at infinity, i.e. Kac label $k=[1^k]$.  Table \ref{tab:M5} contains the IIA ``brane scan'', whereas the brane configuration is drawn in figure \ref{fig:IIAE8}.  

Notice that, upon T-dualizing to Type IIB and lifting to F-theory, we obtain precisely the configuration in \eqref{eq:fullymasslessE8} with the single hypermultiplet representing a perturbative D7-brane (T-dual to the eighth D8), i.e. an $I_1$ locus of the F-theory discriminant, and the $E_8$ flavor symmetry being carried by an $E_8$ seven-brane wrapping a noncompact $-12$ curve. The origin of the tensor branch of the SCFT is the point where all NS5's are coincident; the generic point on the tensor branch corresponds instead to separating all NS5's (i.e. blowing up all compact $-2$ curves in the F-theory configuration).

With these general rules in mind, it is now possible to construct Type IIA configurations corresponding to orbi-instantons with any allowed boundary condition. That is, every Kac label corresponds to one Type IIA O8-D8-D6-NS5 brane setup. Rather conveniently to us, these setups have already been classified in \cite{Cabrera:2019izd} by exploiting simple combinatorics which takes as input the Kac label and produces the setup as an output. (On the other hand,  the F-theory quiver can be obtained from the algorithm of \cite{Mekareeya:2017jgc}.) The classification requires the introduction of one last ingredient, known as the O8$^*$-plane \cite{Gorbatov:2001pw},\footnote{This is necessary to describe the non-perturbative extension of Type I' \cite{Morrison:1996xf,Douglas:1996xp} which allows to construct the 5d rank-1 SCFTs known as $\tilde{E}_1$ and $E_0$, thereby completing the original list of \cite{Seiberg:1996bd}.} which carries $-9$ D8 charge and can be thought of as the product of pulling out an extra D8-brane from the O8$^-$.

\subsubsection{\texorpdfstring{Some O8-D8-D6-NS5 brane configurations for $k=6$}{Some O8-D8-D6-NS5 brane configurations for k=6}}
\label{subsub:IIAk6}

Let us now showcase a few entries in the M-theory/Type IIA/F-theory dictionary which will be instructive for our purposes. Pick $k=6$. A selection of Kac labels, i.e. of SCFTs (having fixed $N$ and $k$), and their IIA engineerings is tabulated in table \ref{tab:k6Kacquiv}.%
\setlength{\tabcolsep}{4pt}

\LTpre=\smallskipamount
\LTpost=\smallskipamount
\begin{longtable}[c]{@{}lclllc@{}}
			\footnotesize  $[6]$ & {\footnotesize $[\SU(3) × \SU(2)]$} & 
			\begin{tikzpicture}[scale=.75,baseline]
				\node at (0,0) {};
				\draw[fill=black] (0.5,0) circle (0.05cm);
				\draw[fill=black] (1.25,0) circle (0.05cm);
				\draw[fill=black] (2,0) circle (0.05cm);
				\draw[fill=black] (2.75,0) circle (0.05cm);
				
				\draw[solid,black,thick] (0.5,0)--(1.25,0) node[black,midway,xshift=-0.15cm,yshift=0.2cm] {\footnotesize $6$};
				\draw[solid,black,thick] (1.25,0)--(2,0) node[black,midway,yshift=0.2cm] {\footnotesize $6$};
				\draw[loosely dotted,black,thick] (2,0)--(2.75,0) node[black,midway, ] {};
 			\draw[solid,black,thick] (2.75,0)--(3.5,0) node[black,midway,yshift=0.2cm] {\footnotesize $6$};
				\draw[dashed,black,very thick] (0,-.5)--(0,.5) node[black,midway] {};
				\draw[solid,black,very thick] (0.2,-.5)--(0.2,.5) node[black,midway, xshift =0cm, yshift=+.75cm] {\footnotesize $2$};
				\draw[solid,black,very thick] (1,-.5)--(1,.5) node[black,midway, xshift =0cm, yshift=+.75cm] {\footnotesize $6$};
			\end{tikzpicture}
            &
			\begin{tikzpicture}[scale=.75,baseline]
				\draw[solid,black,thick] (0.4,.6)--(2,.6) node[] {};
				\draw[solid,black,thick] (0.6,.4)--(2,.4) node[] {};
				\draw[solid,black,thick] (0.8,0.2)--(2,0.2) node[] {};
				\draw[solid,black,thick] (1,0)--(2,0) node[] {};
				\draw[solid,black,thick] (1.2,-0.2)--(2,-0.2) node[] {};
				\draw[solid,black,thick] (1.4,-0.4)--(2,-0.4) node[] {};
				
				\draw[dashed,black,very thick] (0,-.75)--(0,.75) node[black,midway, xshift =0cm, yshift=-1.5cm] {} node[black,midway, xshift =0cm, yshift=1.5cm] {};
				\draw[solid,black,very thick] (0.2,-.75)--(0.2,.75) node[black,midway, yshift=.75cm]{\footnotesize $2$};
				\draw[solid,black,very thick] (0.4,-.75)--(0.4,.75) node[black,midway, xshift =0cm, yshift=+.75cm]{};
				\draw[solid,black,very thick] (0.6,-.75)--(0.6,.75) node[black,midway, xshift =0cm, yshift=+.75cm]{};
				\draw[solid,black,very thick] (0.8,-.75)--(0.8,.75) node[black,midway, xshift =0cm, yshift=+.75cm]{};
				\draw[solid,black,very thick] (1,-.75)--(1,.75) node[black,midway, xshift =0cm, yshift=+.75cm]{};
				\draw[solid,black,very thick] (1.2,-.75)--(1.2,.75) node[black,midway, xshift =0cm, yshift=+.75cm]{};
				\draw[solid,black,very thick] (1.4,-.75)--(1.4,.75) node[black,midway, xshift =0cm, yshift=+.75cm]{};
			\end{tikzpicture}			
			& ${\fontsize{0.25pt}{0.5pt}\selectfont \yng(1,1,1,1,1,1)}$ & \footnotesize $\su{6}$
			\vspace*{-30pt}
			\\ 
			\footnotesize  $[5,1]$ & {\footnotesize $[\SU(5)]$} & 
			\begin{tikzpicture}[scale=.75,baseline]
				\node at (0,0) {};
				\draw[fill=black] (0.5,0) circle (0.05cm);
				\draw[fill=black] (1.25,0) circle (0.05cm);
				\draw[fill=black] (2,0) circle (0.05cm);
				\draw[fill=black] (2.75,0) circle (0.05cm);
				\draw[fill=black] (3.5,0) circle (0.05cm);
				
				\draw[solid,black,thick] (0.5,0)--(1.25,0) node[black,midway,xshift=-0.15cm,yshift=0.2cm] {\footnotesize $5$};
				\draw[solid,black,thick] (1.25,0)--(2,0) node[black,midway,xshift=-0.15cm,yshift=0.2cm] {\footnotesize $6$};
				\draw[solid,black,thick] (2,0)--(2.75,0) node[black,midway,yshift=0.2cm] {\footnotesize $6$};
				\draw[loosely dotted,black,thick] (2.75,0)--(3.5,0) node[black,midway] {};
				\draw[solid,black,thick] (3.5,0)--(4.25,0) node[black,midway,yshift=0.2cm] {\footnotesize $6$};
				
				\draw[dashed,black,very thick] (0,-.5)--(0,.5) node[black,midway] {};
				\draw[solid,black,very thick] (0.2,-.5)--(0.2,.5) node[black,midway, xshift =0cm, yshift=+.75cm] {\footnotesize $3$};
				\draw[solid,black,very thick] (1,-.5)--(1,.5) node[black,midway, xshift =0cm, yshift=+.75cm] {\footnotesize $4$};
				\draw[solid,black,very thick] (1.75,-.5)--(1.75,.5) node[black,midway, xshift =0cm, yshift=+.75cm] {\footnotesize $1$};
			\end{tikzpicture}  
&
\begin{tikzpicture}[scale=.75,baseline]
				\draw[dashed,black,very thick] (0.2,-.75)--(0.2,.75) node[black,midway, xshift =0cm, yshift=+.75cm]{};
				\draw[solid,black,very thick] (0.4,-.75)--(0.4,.75) node[black,midway, xshift =0cm, yshift=.75cm]{\footnotesize $3$};
				\draw[solid,black,very thick] (0.6,-.75)--(0.6,.75) node[black,midway, xshift =0cm, yshift=+.75cm]{};
				\draw[solid,black,very thick] (0.8,-.75)--(0.8,.75) node[black,midway, xshift =0cm, yshift=+.75cm]{};
				\draw[solid,black,very thick] (1,-.75)--(1,.75) node[black,midway, xshift =0cm, yshift=+.75cm]{};
				\draw[solid,black,very thick] (1.2,-.75)--(1.2,.75) node[black,midway, xshift =0cm, yshift=+.75cm]{};
				\draw[solid,black,very thick] (1.4,-.75)--(1.4,.75) node[black,midway, xshift =0cm, yshift=+.75cm]{};
				
				\draw[solid,black,thick] (0.6,.6)--(2,.6) node[] {};
				\draw[solid,black,thick] (0.6,.4)--(2,.4) node[] {};
				\draw[solid,black,thick] (0.8,0.2)--(2,0.2) node[] {};
				\draw[solid,black,thick] (1,0)--(2,0) node[] {};
				\draw[solid,black,thick] (1.2,-0.2)--(2,-0.2) node[] {};
				\draw[solid,black,thick] (1.4,-0.4)--(2,-0.4) node[] {};
				
			\end{tikzpicture}   					
			& ${\fontsize{0.25pt}{0.5pt}\selectfont \yng(2,1,1,1,1)}$ & \footnotesize \makecell{$\mathfrak{s} (\uu{4}\oplus \uu{1})$ \\ $\cong \su{4}\oplus \uu{1}$} 			
			\\ 
			
			\footnotesize  $[4,2]$ & {\footnotesize $[\SO(10)]$} & 
			\begin{tikzpicture}[scale=.75,baseline]
				\node at (0,0) {};
				\draw[fill=black] (0.5,0) circle (0.05cm);
				\draw[fill=black] (1.25,0) circle (0.05cm);
				\draw[fill=black] (2,0) circle (0.05cm);
				\draw[fill=black] (2.75,0) circle (0.05cm);
				\draw[fill=black] (3.5,0) circle (0.05cm);
				
				\draw[solid,black,thick] (0.5,0)--(1.25,0) node[black,midway,xshift=-0.15cm,yshift=0.2cm] {\footnotesize $4$};
				\draw[solid,black,thick] (1.25,0)--(2,0) node[black,midway,xshift=-0.15cm,yshift=0.2cm] {\footnotesize $6$};
				\draw[solid,black,thick] (2,0)--(2.75,0) node[black,midway,yshift=0.2cm] {\footnotesize $6$};
				\draw[loosely dotted,black,thick] (2.75,0)--(3.5,0) node[black,midway] {};
				\draw[solid,black,thick] (3.5,0)--(4.25,0) node[black,midway,yshift=0.2cm] {\footnotesize $6$};
				
				\draw[dashed,black,very thick] (0,-.5)--(0,.5) node[black,midway] {};
				\draw[solid,black,very thick] (0.2,-.5)--(0.2,.5) node[black,midway, xshift =0cm, yshift=+.75cm] {\footnotesize $4$};
				\draw[solid,black,very thick] (1,-.5)--(1,.5) node[black,midway, xshift =0cm, yshift=+.75cm] {\footnotesize $2$};
				\draw[solid,black,very thick] (1.75,-.5)--(1.75,.5) node[black,midway, xshift =0cm, yshift=+.75cm] {\footnotesize $2$};
			\end{tikzpicture}  
			&
			\begin{tikzpicture}[scale=.75,baseline]
				\draw[dashed,black,very thick] (0.4,-.75)--(0.4,.75) node[black,midway, xshift =0cm, yshift=-1cm]{};
				\draw[solid,black,very thick] (0.6,-.75)--(0.6,.75) node[black,midway, xshift =0cm, yshift=.75cm]{\footnotesize $4$};
				\draw[solid,black,very thick] (0.8,-.75)--(0.8,.75) node[black,midway, xshift =0cm, yshift=+.75cm]{};
				\draw[solid,black,very thick] (1,-.75)--(1,.75) node[black,midway, xshift =0cm, yshift=+.75cm]{};
				\draw[solid,black,very thick] (1.2,-.75)--(1.2,.75) node[black,midway, xshift =0cm, yshift=+.75cm]{};
				\draw[solid,black,very thick] (1.4,-.75)--(1.4,.75) node[black,midway, xshift =0cm, yshift=+.75cm]{};
				
				\draw[solid,black,thick] (0.8,.6)--(2,.6) node[] {};
				\draw[solid,black,thick] (0.8,.4)--(2,.4) node[] {};
				\draw[solid,black,thick] (1,0.2)--(2,0.2) node[] {};
				\draw[solid,black,thick] (1,0)--(2,0) node[] {};
				\draw[solid,black,thick] (1.2,-0.2)--(2,-0.2) node[] {};
				\draw[solid,black,thick] (1.4,-0.4)--(2,-0.4) node[] {};
				
			\end{tikzpicture}
			& ${\fontsize{0.25pt}{0.5pt}\selectfont \yng(2,2,1,1)}$ & \footnotesize \makecell{$\mathfrak{s}(\uu{2}\oplus\uu{2} \cong $ \\ $\su{2}\oplus\su{2} \oplus \uu{1}$}
			\vspace*{-15pt}
		   \\
			
			\footnotesize  $[4,1²]$ & {\footnotesize $[\SO(10)]$} & 
			\begin{tikzpicture}[scale=.75,baseline]
				\node at (0,0) {};
				\draw[fill=black] (0.5,0) circle (0.05cm);
				\draw[fill=black] (1.25,0) circle (0.05cm);
				\draw[fill=black] (2,0) circle (0.05cm);
				\draw[fill=black] (2.75,0) circle (0.05cm);
				\draw[fill=black] (3.5,0) circle (0.05cm);
				\draw[fill=black] (4.25,0) circle (0.05cm);
				
				\draw[solid,black,thick] (0.5,0)--(1.25,0) node[black,midway,xshift=-0.15cm,yshift=0.2cm] {\footnotesize $4$};
				\draw[solid,black,thick] (1.25,0)--(2,0) node[black,midway,yshift=0.2cm] {\footnotesize $5$};
				\draw[solid,black,thick] (2,0)--(2.75,0) node[black,midway,xshift=-0.15cm,yshift=0.2cm] {\footnotesize $6$};
				\draw[solid,black,thick] (2.75,0)--(3.5,0) node[black,midway,yshift=0.2cm] {\footnotesize $6$};
				\draw[loosely dotted,black,thick] (3.5,0)--(4.25,0) node[black,midway]{};
				\draw[solid,black,thick] (4.25,0)--(5,0) node[black,midway,yshift=0.2cm] {\footnotesize $6$};
				
				\draw[dashed,black,very thick] (0,-.5)--(0,.5) node[black,midway] {};
				\draw[solid,black,very thick] (0.2,-.5)--(0.2,.5) node[black,midway, xshift =0cm, yshift=+.75cm] {\footnotesize $4$};
				\draw[solid,black,very thick] (1,-.5)--(1,.5) node[black,midway, xshift =0cm, yshift=+.75cm] {\footnotesize $3$};
				\draw[solid,black,very thick] (2.5,-.5)--(2.5,.5) node[black,midway, xshift =0cm, yshift=+.75cm] {\footnotesize $1$};
			\end{tikzpicture}  
			&
			\begin{tikzpicture}[scale=.75,baseline]
				\draw[dashed,black,very thick] (0.4,-.75)--(0.4,.75) node[black,midway, xshift =0cm, yshift=+.75cm]{};
				\draw[solid,black,very thick] (0.6,-.75)--(0.6,.75) node[black,midway, xshift =0cm, yshift=.75cm]{\footnotesize $4$};
				\draw[solid,black,very thick] (0.8,-.75)--(0.8,.75) node[black,midway, xshift =0cm, yshift=+.75cm]{};
				\draw[solid,black,very thick] (1,-.75)--(1,.75) node[black,midway, xshift =0cm, yshift=+.75cm]{};
				\draw[solid,black,very thick] (1.2,-.75)--(1.2,.75) node[black,midway, xshift =0cm, yshift=+.75cm]{};
				\draw[solid,black,very thick] (1.4,-.75)--(1.4,.75) node[black,midway, xshift =0cm, yshift=+.75cm]{};
				
				\draw[solid,black,thick] (0.8,.6)--(2,.6) node[] {};
				\draw[solid,black,thick] (0.8,.4)--(2,.4) node[] {};
				\draw[solid,black,thick] (0.8,0.2)--(2,0.2) node[] {};
				\draw[solid,black,thick] (1,0)--(2,0) node[] {};
				\draw[solid,black,thick] (1.2,-0.2)--(2,-0.2) node[] {};
				\draw[solid,black,thick] (1.4,-0.4)--(2,-0.4) node[] {};
				
			\end{tikzpicture}
			
			& ${\fontsize{0.25pt}{0.5pt}\selectfont  \yng(3,1,1,1)}$ & \footnotesize \makecell{$\mathfrak{s} (\uu{3}\oplus \uu{1})$ \\ $\cong \su{3}\oplus \uu{1}$}  
			 \\

			\footnotesize  $[3²]$ & {\footnotesize $[E_6]$} & 
			\begin{tikzpicture}[scale=.75,baseline]
				\node at (0,0) {};
				\draw[fill=black] (0.5,0) circle (0.05cm);
				\draw[fill=black] (1.25,0) circle (0.05cm);
				\draw[fill=black] (2,0) circle (0.05cm);
				\draw[fill=black] (2.75,0) circle (0.05cm);
				\draw[fill=black] (3.5,0) circle (0.05cm);
				
				\draw[solid,black,thick] (0.5,0)--(1.25,0) node[black,midway,yshift=0.2cm] {\footnotesize $3$};
				\draw[solid,black,thick] (1.25,0)--(2,0) node[black,midway,xshift=-0.15cm,yshift=0.2cm] {\footnotesize $6$};
				\draw[solid,black,thick] (2,0)--(2.75,0) node[black,midway,yshift=0.2cm] {\footnotesize $6$};
				\draw[loosely dotted,black,thick] (2.75,0)--(3.5,0) node[black,midway] {};
				\draw[solid,black,thick] (3.5,0)--(4.25,0) node[black,midway,yshift=0.2cm] {\footnotesize $6$};
				
				\draw[dashed,black,very thick] (0,-.5)--(0,.5) node[black,midway] {};
				\draw[solid,black,very thick] (0.2,-.5)--(0.2,.5) node[black,midway, xshift =0cm, yshift=+.75cm] {\footnotesize $5$};
				\draw[solid,black,very thick] (1.75,-.5)--(1.75,.5) node[black,midway, xshift =0cm, yshift=+.75cm] {\footnotesize $3$};
			\end{tikzpicture}  
			&
			\begin{tikzpicture}[scale=.75,baseline]
				\draw[dashed,black,very thick] (0.6,-.75)--(0.6,.75) node[black,midway, xshift =0cm, yshift=+.75cm]{};
				\draw[solid,black,very thick] (0.8,-.75)--(0.8,.75) node[black,midway, xshift =0cm, yshift=.75cm]{\footnotesize $5$};
				\draw[solid,black,very thick] (1,-.75)--(1,.75) node[black,midway, xshift =0cm, yshift=+.75cm]{};
				\draw[solid,black,very thick] (1.2,-.75)--(1.2,.75) node[black,midway, xshift =0cm, yshift=+.75cm]{};
				\draw[solid,black,very thick] (1.4,-.75)--(1.4,.75) node[black,midway, xshift =0cm, yshift=+.75cm]{};
				
				\draw[solid,black,thick] (1,.6)--(2,.6) node[] {};
				\draw[solid,black,thick] (1,.4)--(2,.4) node[] {};
				\draw[solid,black,thick] (1.2,0.2)--(2,0.2) node[] {};
				\draw[solid,black,thick] (1.2,0)--(2,0) node[] {};
				\draw[solid,black,thick] (1.4,-0.2)--(2,-0.2) node[] {};
				\draw[solid,black,thick] (1.4,-0.4)--(2,-0.4) node[] {};
				
			\end{tikzpicture}	
			
			& ${\fontsize{0.25pt}{0.5pt}\selectfont \yng(2,2,2)}$ & \footnotesize $\mathfrak{su}(3)$ 
			 \\

			\footnotesize  $[3,2,1]$ & {\footnotesize $[E_6]$} & 
			\begin{tikzpicture}[scale=.75,baseline]
				\node at (0,0) {};
				\draw[fill=black] (0.5,0) circle (0.05cm);
				\draw[fill=black] (1.25,0) circle (0.05cm);
				\draw[fill=black] (2,0) circle (0.05cm);
				\draw[fill=black] (2.75,0) circle (0.05cm);
				\draw[fill=black] (3.5,0) circle (0.05cm);
				\draw[fill=black] (4.25,0) circle (0.05cm);
				
				\draw[solid,black,thick] (0.5,0)--(1.25,0) node[black,midway,xshift=-0.15cm,yshift=0.2cm] {\footnotesize $3$};
				\draw[solid,black,thick] (1.25,0)--(2,0) node[black,midway,xshift=-0.15cm,yshift=0.2cm] {\footnotesize $5$};
				\draw[solid,black,thick] (2,0)--(2.75,0) node[black,midway,xshift=-0.15cm,yshift=0.2cm] {\footnotesize $6$};
				\draw[solid,black,thick] (2.75,0)--(3.5,0) node[black,midway,yshift=0.2cm] {\footnotesize $6$};
				\draw[loosely dotted,black,thick] (3.5,0)--(4.25,0) node[black,midway]{};
				\draw[solid,black,thick] (4.25,0)--(5,0) node[black,midway,yshift=0.2cm] {\footnotesize $6$};
				
				\draw[dashed,black,very thick] (0,-.5)--(0,.5) node[black,midway] {};
				\draw[solid,black,very thick] (0.2,-.5)--(0.2,.5) node[black,midway, xshift =0cm, yshift=+.75cm] {\footnotesize $5$};
				\draw[solid,black,very thick] (1,-.5)--(1,.5) node[black,midway, xshift =0cm, yshift=+.75cm] {\footnotesize $1$};
				\draw[solid,black,very thick] (1.75,-.5)--(1.75,.5) node[black,midway, xshift =0cm, yshift=+.75cm] {\footnotesize $1$};
				\draw[solid,black,very thick] (2.5,-.5)--(2.5,.5) node[black,midway, xshift =0cm, yshift=+.75cm] {\footnotesize $1$};
			\end{tikzpicture} 
			&
			\begin{tikzpicture}[scale=.75,baseline]
				\draw[dashed,black,very thick] (0.6,-.75)--(0.6,.75) node[black,midway, xshift =0cm, yshift=+.75cm]{};
				\draw[solid,black,very thick] (0.8,-.75)--(0.8,.75) node[black,midway, xshift =0cm, yshift=.75cm]{\footnotesize $5$};
				\draw[solid,black,very thick] (1,-.75)--(1,.75) node[black,midway, xshift =0cm, yshift=+.75cm]{};
				\draw[solid,black,very thick] (1.2,-.75)--(1.2,.75) node[black,midway, xshift =0cm, yshift=+.75cm]{};
				\draw[solid,black,very thick] (1.4,-.75)--(1.4,.75) node[black,midway, xshift =0cm, yshift=+.75cm]{};
				
				\draw[solid,black,thick] (1,.6)--(2,.6) node[] {};
				\draw[solid,black,thick] (1.2,.4)--(2,.4) node[] {};
				\draw[solid,black,thick] (1.2,0.2)--(2,0.2) node[] {};
				\draw[solid,black,thick] (1.4,0)--(2,0) node[] {};
				\draw[solid,black,thick] (1.4,-0.2)--(2,-0.2) node[] {};
				\draw[solid,black,thick] (1.4,-0.4)--(2,-0.4) node[] {};
				
			\end{tikzpicture} 
			& ${\fontsize{0.25pt}{0.5pt}\selectfont \yng(3,2,1)}$ & \footnotesize \makecell{$\mathfrak{s}(\uu{1}\oplus\uu{1}\oplus\uu{1})$ \\ $\cong \uu{1} \oplus \uu{1}$}			
			\\

		\footnotesize  $[3,1³]$ & {\footnotesize $[E_6]$} & 
		\begin{tikzpicture}[scale=.75,baseline]
			\node at (0,0) {};
			\draw[fill=black] (0.5,0) circle (0.05cm);
			\draw[fill=black] (1.25,0) circle (0.05cm);
			\draw[fill=black] (2,0) circle (0.05cm);
			\draw[fill=black] (2.75,0) circle (0.05cm);
			\draw[fill=black] (3.5,0) circle (0.05cm);
			\draw[fill=black] (4.25,0) circle (0.05cm);
			\draw[fill=black] (5,0) circle (0.05cm);
			
			\draw[solid,black,thick] (0.5,0)--(1.25,0) node[black,midway,xshift=-0.15cm,yshift=0.2cm] {\footnotesize $3$};
			\draw[solid,black,thick] (1.25,0)--(2,0) node[black,midway,yshift=0.2cm] {\footnotesize $4$};
			\draw[solid,black,thick] (2,0)--(2.75,0) node[black,midway,yshift=0.2cm] {\footnotesize $5$};
			\draw[solid,black,thick] (2.75,0)--(3.5,0) node[black,midway,xshift=-0.15cm,yshift=0.2cm] {\footnotesize $6$};
			\draw[solid,black,thick] (3.5,0)--(4.25,0) node[black,midway,yshift=0.2cm] {\footnotesize $6$};
			\draw[loosely dotted,black,thick] (4.25,0)--(5,0) node[black,midway]{};
		\draw[solid,black,thick] (5,0)--(5.75,0) node[black,midway,yshift=0.2cm] {\footnotesize $6$};
			
			\draw[dashed,black,very thick] (0,-.5)--(0,.5) node[black,midway] {};
			\draw[solid,black,very thick] (0.2,-.5)--(0.2,.5) node[black,midway, xshift =0cm, yshift=+.75cm] {\footnotesize $5$};
			\draw[solid,black,very thick] (1,-.5)--(1,.5) node[black,midway, xshift =0cm, yshift=+.75cm] {\footnotesize $2$};
			\draw[solid,black,very thick] (3.25,-.5)--(3.25,.5) node[black,midway, xshift =0cm, yshift=+.75cm] {\footnotesize $1$};
		\end{tikzpicture} 
		& 
		\begin{tikzpicture}[scale=.75,baseline]
			\draw[dashed,black,very thick] (0.6,-.75)--(0.6,.75) node[black,midway, xshift =0cm, yshift=+.75cm]{};
			\draw[solid,black,very thick] (0.8,-.75)--(0.8,.75) node[black,midway, xshift =0cm, yshift=.75cm]{\footnotesize $5$};
			\draw[solid,black,very thick] (1,-.75)--(1,.75) node[black,midway, xshift =0cm, yshift=+.75cm]{};
			\draw[solid,black,very thick] (1.2,-.75)--(1.2,.75) node[black,midway, xshift =0cm, yshift=+.75cm]{};
			\draw[solid,black,very thick] (1.4,-.75)--(1.4,.75) node[black,midway, xshift =0cm, yshift=+.75cm]{};
			
			\draw[solid,black,thick] (1,.6)--(2,.6) node[] {};
			\draw[solid,black,thick] (1,.4)--(2,.4) node[] {};
			\draw[solid,black,thick] (1,0.2)--(2,0.2) node[] {};
			\draw[solid,black,thick] (1,0)--(2,0) node[] {};
			\draw[solid,black,thick] (1.2,-0.2)--(2,-0.2) node[] {};
			\draw[solid,black,thick] (1.4,-0.4)--(2,-0.4) node[] {};
			
		\end{tikzpicture}
		& ${\fontsize{0.25pt}{0.5pt}\selectfont \yng(4,1,1)}$ & \footnotesize \makecell{$\mathfrak{s} (\uu{2}\oplus \uu{1})$ \\ $\cong \su{2}\oplus \uu{1}$} 
		 \\
		\footnotesize $[2³]$ & {\footnotesize $[E_7]$} & 
		\begin{tikzpicture}[scale=.75,baseline]
			\node at (0,0) {};
			\draw[fill=black] (0.5,0) circle (0.05cm);
			\draw[fill=black] (1.25,0) circle (0.05cm);
			\draw[fill=black] (2,0) circle (0.05cm);
			\draw[fill=black] (2.75,0) circle (0.05cm);
			\draw[fill=black] (3.5,0) circle (0.05cm);
			\draw[fill=black] (4.25,0) circle (0.05cm);
			
			\draw[solid,black,thick] (0.5,0)--(1.25,0) node[black,midway,yshift=0.2cm] {\footnotesize $2$};
			\draw[solid,black,thick] (1.25,0)--(2,0) node[black,midway,yshift=0.2cm] {\footnotesize $4$};
			\draw[solid,black,thick] (2,0)--(2.75,0) node[black,midway,xshift=-0.15cm,yshift=0.2cm] {\footnotesize $6$};
			\draw[solid,black,thick] (2.75,0)--(3.5,0) node[black,midway,yshift=0.2cm] {\footnotesize $6$};
			\draw[loosely dotted,black,thick] (3.5,0)--(4.25,0) node[black,midway]{};
			\draw[solid,black,thick] (4.25,0)--(5,0) node[black,midway,yshift=0.2cm] {\footnotesize $6$};
			
			\draw[dashed,black,very thick] (0,-.5)--(0,.5) node[black,midway] {};
			\draw[solid,black,very thick] (0.2,-.5)--(0.2,.5) node[black,midway, xshift =0cm, yshift=+.75cm] {\footnotesize $6$};
			\draw[solid,black,very thick] (2.5,-.5)--(2.5,.5) node[black,midway, xshift =0cm, yshift=+.75cm] {\footnotesize $2$};
		\end{tikzpicture} 
		& 
		\begin{tikzpicture}[scale=.75,baseline]
			\draw[dashed,black,very thick] (0.8,-.75)--(0.8,.75) node[black,midway, xshift =0cm, yshift=+.75cm]{};
			\draw[solid,black,very thick] (1,-.75)--(1,.75) node[black,midway, xshift =0cm, yshift=.75cm]{\footnotesize $6$};
			\draw[solid,black,very thick] (1.2,-.75)--(1.2,.75) node[black,midway, xshift =0cm, yshift=+.75cm]{};
			\draw[solid,black,very thick] (1.4,-.75)--(1.4,.75) node[black,midway, xshift =0cm, yshift=+.75cm]{};
			
			\draw[solid,black,thick] (1.2,.6)--(2,.6) node[] {};
			\draw[solid,black,thick] (1.2,.4)--(2,.4) node[] {};
			\draw[solid,black,thick] (1.2,0.2)--(2,0.2) node[] {};
			\draw[solid,black,thick] (1.4,0)--(2,0) node[] {};
			\draw[solid,black,thick] (1.4,-0.2)--(2,-0.2) node[] {};
			\draw[solid,black,thick] (1.4,-0.4)--(2,-0.4) node[] {};
			
		\end{tikzpicture}
		&${\fontsize{0.25pt}{0.5pt}\selectfont \yng(3,3)}$ & \footnotesize $\mathfrak{su}(2)$ 
		 \\
			
			\footnotesize $[2²,1²]$ & {\footnotesize $[E_7]$} & 
			\begin{tikzpicture}[scale=.75,baseline]
				\node at (0,0) {};
				\draw[fill=black] (0.5,0) circle (0.05cm);
				\draw[fill=black] (1.25,0) circle (0.05cm);
				\draw[fill=black] (2,0) circle (0.05cm);
				\draw[fill=black] (2.75,0) circle (0.05cm);
				\draw[fill=black] (3.5,0) circle (0.05cm);
				\draw[fill=black] (4.25,0) circle (0.05cm);
				\draw[fill=black] (5,0) circle (0.05cm);
				
				\draw[solid,black,thick] (0.5,0)--(1.25,0) node[black,midway,yshift=0.2cm] {\footnotesize $2$};
				\draw[solid,black,thick] (1.25,0)--(2,0) node[black,midway,xshift=-0.15cm,yshift=0.2cm] {\footnotesize $4$};
				\draw[solid,black,thick] (2,0)--(2.75,0) node[black,midway,yshift=0.2cm] {\footnotesize $5$};
				\draw[solid,black,thick] (2.75,0)--(3.5,0) node[black,midway,xshift=-0.15cm,yshift=0.2cm] {\footnotesize $6$};
				\draw[solid,black,thick] (3.5,0)--(4.25,0) node[black,midway,yshift=0.2cm]{\footnotesize $6$};
				\draw[loosely dotted,black,thick] (4.25,0)--(5,0) node[black,midway,yshift=0.2cm] {};
                 \draw[solid,black,thick] (5,0)--(5.75,0) node[black,midway,yshift=0.2cm] {\footnotesize $6$};
				
				\draw[dashed,black,very thick] (0,-.5)--(0,.5) node[black,midway] {};
				\draw[solid,black,very thick] (0.2,-.5)--(0.2,.5) node[black,midway, xshift =0cm, yshift=+.75cm] {\footnotesize $6$};
				\draw[solid,black,very thick] (1.75,-.5)--(1.75,.5) node[black,midway, xshift =0cm, yshift=+.75cm] {\footnotesize $1$};
				\draw[solid,black,very thick] (3.25,-.5)--(3.25,.5) node[black,midway, xshift =0cm, yshift=+.75cm] {\footnotesize $1$};
			\end{tikzpicture} 
			& 
			\begin{tikzpicture}[scale=.75,baseline]
				\draw[dashed,black,very thick] (0.8,-.75)--(0.8,.75) node[black,midway, xshift =0cm, yshift=+.75cm]{};
				\draw[solid,black,very thick] (1,-.75)--(1,.75) node[black,midway, xshift =0cm, yshift=.75cm]{\footnotesize $6$};
				\draw[solid,black,very thick] (1.2,-.75)--(1.2,.75) node[black,midway, xshift =0cm, yshift=+.75cm]{};
				\draw[solid,black,very thick] (1.4,-.75)--(1.4,.75) node[black,midway, xshift =0cm, yshift=+.75cm]{};
				
				\draw[solid,black,thick] (1.2,.6)--(2,.6) node[] {};
				\draw[solid,black,thick] (1.2,.4)--(2,.4) node[] {};
				\draw[solid,black,thick] (1.4,0.2)--(2,0.2) node[] {};
				\draw[solid,black,thick] (1.4,0)--(2,0) node[] {};
				\draw[solid,black,thick] (1.4,-0.2)--(2,-0.2) node[] {};
				\draw[solid,black,thick] (1.4,-0.4)--(2,-0.4) node[] {};
				
			\end{tikzpicture}
			& ${\fontsize{0.25pt}{0.5pt}\selectfont \yng(4,2)}$ & \footnotesize \makecell{$\mathfrak{s}(\uu{1}\oplus\uu{1})$ \\ $\cong \uu{1}$}
			 \\
			
			\footnotesize $[2,1^4]$ & {\footnotesize $[E_7]$} & 
			\begin{tikzpicture}[scale=.75,baseline]
				\node at (0,0) {};
				\draw[fill=black] (0.5,0) circle (0.05cm);
				\draw[fill=black] (1.25,0) circle (0.05cm);
				\draw[fill=black] (2,0) circle (0.05cm);
				\draw[fill=black] (2.75,0) circle (0.05cm);
				\draw[fill=black] (3.5,0) circle (0.05cm);
				\draw[fill=black] (4.25,0) circle (0.05cm);
				\draw[fill=black] (5,0) circle (0.05cm);
				
				\draw[solid,black,thick] (0.5,0)--(1.25,0) node[black,midway,xshift=-0.15cm,yshift=0.2cm] {\footnotesize $2$};
				\draw[solid,black,thick] (1.25,0)--(2,0) node[black,midway,yshift=0.2cm] {\footnotesize $3$};
				\draw[solid,black,thick] (2,0)--(2.75,0) node[black,midway,yshift=0.2cm] {\footnotesize $4$};
				\draw[solid,black,thick] (2.75,0)--(3.5,0) node[black,midway,yshift=0.2cm] {\footnotesize $5$};
				\draw[solid,black,thick] (3.5,0)--(4.25,0) node[black,midway,xshift=-0.15cm,yshift=0.2cm] {\footnotesize $6$};
				\draw[loosely dotted,black,thick] (4.25,0)--(5,0) node[black,midway]{};
				\draw[solid,black,thick] (5,0)--(5.75,0) node[black,midway,yshift=0.2cm] {\footnotesize $6$};
				
				\draw[dashed,black,very thick] (0,-.5)--(0,.5) node[black,midway] {};
				\draw[solid,black,very thick] (0.2,-.5)--(0.2,.5) node[black,midway, xshift =0cm, yshift=+.75cm] {\footnotesize $6$};
				\draw[solid,black,very thick] (1,-.5)--(1,.5) node[black,midway, xshift =0cm, yshift=+.75cm] {\footnotesize $1$};
				\draw[solid,black,very thick] (4,-.5)--(4,.5) node[black,midway, xshift =0cm, yshift=+.75cm] {\footnotesize $1$};
			\end{tikzpicture} 
			&
			\begin{tikzpicture}[scale=.75,baseline]
				\draw[dashed,black,very thick] (0.8,-.75)--(0.8,.75) node[black,midway, xshift =0cm, yshift=+.75cm]{};
				\draw[solid,black,very thick] (1,-.75)--(1,.75) node[black,midway, xshift =0cm, yshift=.75cm]{\footnotesize $6$};
				\draw[solid,black,very thick] (1.2,-.75)--(1.2,.75) node[black,midway, xshift =0cm, yshift=+.75cm]{};
				\draw[solid,black,very thick] (1.4,-.75)--(1.4,.75) node[black,midway, xshift =0cm, yshift=+.75cm]{};
				
				\draw[solid,black,thick] (1.2,.6)--(2,.6) node[] {};
				\draw[solid,black,thick] (1.4,.4)--(2,.4) node[] {};
				\draw[solid,black,thick] (1.4,0.2)--(2,0.2) node[] {};
				\draw[solid,black,thick] (1.4,0)--(2,0) node[] {};
				\draw[solid,black,thick] (1.4,-0.2)--(2,-0.2) node[] {};
				\draw[solid,black,thick] (1.4,-0.4)--(2,-0.4) node[] {};
				
			\end{tikzpicture}
			& ${\fontsize{0.25pt}{0.5pt}\selectfont \yng(5,1)}$ & \footnotesize \makecell{$\mathfrak{s}(\uu{1}\oplus\uu{1})$ \\ $\cong \uu{1}$}
			 \\

			\footnotesize $[1^6]$ & {\footnotesize $[E_8]$} & 
			\begin{tikzpicture}[scale=.75,baseline]
				\node at (0,0) {};
				\draw[fill=black] (0.5,0) circle (0.05cm);
				\draw[fill=black] (1.25,0) circle (0.05cm);
				\draw[fill=black] (2,0) circle (0.05cm);
				\draw[fill=black] (2.75,0) circle (0.05cm);
				\draw[fill=black] (3.55,0) circle (0.05cm);
				\draw[fill=black] (4.25,0) circle (0.05cm);
				\draw[fill=black] (5,0) circle (0.05cm);
				
				\draw[solid,black,thick] (0.5,0)--(1.25,0) node[black,midway,yshift=0.2cm] {\footnotesize $1$};
				\draw[solid,black,thick] (1.25,0)--(2,0) node[black,midway,yshift=0.2cm] {\footnotesize $2$};
				\draw[loosely dotted,black,thick] (2,0)--(2.75,0) node[black,midway] {};
				\draw[solid,black,thick] (2.75,0)--(3.5,0) node[black,midway,xshift=-.15cm,yshift=0.2cm] {\footnotesize $6$};
				\draw[solid,black,thick] (3.5,0)--(4.25,0) node[black,midway,yshift=0.2cm] {\footnotesize $6$};
				\draw[loosely dotted,black,thick] (4.25,0)--(5,0) node[black,midway] {};
				\draw[solid,black,thick] (5,0)--(5.75,0) node[black,midway,yshift=0.2cm] {\footnotesize $6$};
				
				\draw[dashed,black,very thick] (0,-.5)--(0,.5) node[black,midway] {};
				\draw[solid,black,very thick] (0.2,-.5)--(0.2,.5) node[black,midway, xshift =0cm, yshift=+.75cm] {\footnotesize $7$};
				\draw[solid,black,very thick] (3.25,-.5)--(3.25,.5) node[black,midway, xshift =0cm, yshift=+.75cm] {\footnotesize $1$};
			\end{tikzpicture}  
			&
			\begin{tikzpicture}[scale=.75,baseline]
				\draw[dashed,black,very thick] (1,-.75)--(1,.75) node[black,midway, xshift =0cm, yshift=+.75cm]{};
				\draw[solid,black,very thick] (1.2,-.75)--(1.2,.75) node[black,midway, xshift =0cm, yshift=.75cm]{\footnotesize $7$};
				\draw[solid,black,very thick] (1.4,-.75)--(1.4,.75) node[black,midway, xshift =0cm, yshift=+.75cm]{};
				
				\draw[solid,black,thick] (1.4,.6)--(2,.6) node[] {};
				\draw[solid,black,thick] (1.4,.4)--(2,.4) node[] {};
				\draw[solid,black,thick] (1.4,0.2)--(2,0.2) node[] {};
				\draw[solid,black,thick] (1.4,0)--(2,0) node[] {};
				\draw[solid,black,thick] (1.4,-0.2)--(2,-0.2) node[] {};
				\draw[solid,black,thick] (1.4,-0.4)--(2,-0.4) node[] {};
			\end{tikzpicture}
			 & ${\fontsize{0.25pt}{0.5pt}\selectfont \yng(6)}$ & \footnotesize $\su{1} = \emptyset$\\ 
\caption{Hasse diagram of nilpotent orbits of $\su{6}$ identified within the Higgs branch RG flow hierarchy of A-type orbi-instantons for $k=6$. A D8 on which $n$ D6's end sits in position $n$ (starting from $n=0$) in the IIA setup; it corresponds to a row of $n$ boxes in the Young tableau. A substack of $m$ D8's has 9d gauge symmetry $\su{m}$; multiple substacks have gauge symmetry given by \eqref{eq:suflavor}.}
	\label{tab:k6Kacquiv}
\end{longtable}%
These examples hopefully make clear one feature which will be universal for all $k$'s: the commutant of Kac label inside $E_8$ generically contains a ``strong-coupling'' summand $[G]$, which can be thought of as being associated with the D8's that are on top of the O8 but do not cross any D6 (once the dilaton diverges and we lift everything to M-theory), plus a ``weak-coupling'' summand $\mathfrak{g}$ which is immediately visible in perturbative IIA and as such is identified with the worldvolume gauge symmetry of a substack of D8's, those that have left the O8 and are crossing a stack of D6's further inside the quiver.  (Open strings stretched from the D8's to the stack of D6's the former cross provide fundamental hypermultiplets.) The former is tabulated in the second column of the table, whereas the latter in the last. Looking at the IIA configurations (in the third column) engineering the tensor branch of the given SCFT, we see that the strong-coupling factor is ``associated'' with the zeroth D6 segment,  which is empty (i.e. there are no D6's there and the gauge algebra is of course trivial). In the F-theory language, the $-1$ curve is undecorated and the strong-coupling factor comes from $[G]1$ as in \eqref{eq:fullymasslessE7}. On the other hand, if the $-1$ curve were decorated (in other words, if the zeroth segment were populated by D6's, engineering a nontrivial gauge algebra) --as we will see for other Kac labels not included in table \ref{tab:k6Kacquiv}-- then the D8's which cross that segment would engineer a weak-coupling $\mathfrak{su}$ or $\mathfrak{so}$ flavor symmetry summand (depending on whether there is or there is not a half-NS5 stuck on the O8).

Another important feature we wish to highlight from the above selection of Kac labels is the following. For $6=[1^6]$ the substack that has left the O8 is composed of a single D8, realizing the $N_\text{f}=1$ hypermultiplet in \eqref{eq:fullymasslessE8}. One can reach this configuration by subsequent Higgsings starting from the $[6]$ label in the top row.  Indeed looking a the fourth column we see that by subsequently sliding off to infinity (along $x^{789}$) trapped segments of D6's we can reach all configurations which sit in the rows below. (One then goes back to the brane configuration in the third column via simple Hanany--Witten moves.) The weak-coupling summand $\mathfrak{g}$ of the flavor symmetry algebra can then be read off as the 9d gauge symmetry on the D8 worldvolumes:
\begin{equation}\label{eq:suflavor}
\mathfrak{g} = \mathfrak{s} \left( \bigoplus_i \uu{f_i} \right)\ ,
\end{equation}
where the summation is over D8 substacks that cross some nonempty D6 segment and $f_i$ is the number of branes in the $i$-th substack. (The $\mathfrak{s}(\cdot)$ simply removes the center-of-mass $\uu{1}$.\footnote{A physical way to see this is to consider the Wess--Zumino term in the action of the D8 stack. Expanding it, one notices that the center-of-mass $\uu{1}$ on the stack participates in a St\"uckleberg term with the RR potential $\tilde{C}_7 \equiv C_7 \wedge e^{B_2}$, which renders it massive and therefore decouples it from the low-energy spectrum of the 6d theory \cite[Sec. 3.2]{Bergman:2020bvi}.})  Further,  to each IIA configuration (i.e. ending pattern of D6's on D8's) one can associate (the transpose of) a Young tableau \cite{DelZotto:2014hpa}: each row represents a D8 on which a number of D6's end, the number being equal to the number of boxes in that row.  Via the well-known correspondence between Young tableaux, integer partitions of an integer, and nilpotent orbits of an $\mathfrak{su}$ algebra \cite{collingwood1993nilpotent} (see also \cite[Sec. 3.1]{Heckman:2016ssk}, whose conventions about partitions and tableaux we adopt here), one can also understand \eqref{eq:suflavor} as the centralizer of the $\su{6}$ nilpotent orbit associated with the integer partition of $6$ given by the Kac label in the first column. It is then clear that the Higgsings tabulated in table \ref{tab:k6Kacquiv} realize the Hasse diagram of nilpotent orbits of $\su{6}$, a point we will spell out in greater detail in section \ref{sub:su6}.

One may also wonder what happens if we Higgs a strong-coupling summand of the flavor symmetry algebra and whether this generates the full Hasse diagram of nilpotent orbits of this factor (which can be classical or exceptional). However there are two problems with this expectation. The first is that very few nilpotent orbits of such a summand would have dimension smaller or equal to the Higgs branch dimension of a rank-1 or rank-2 E-string (quaternionic dimension 30 and 59 respectively),  i.e. $[E_8]1$ and $[E_8]12$ respectively.  Second,  doing so would necessarily generate a small instanton transition whereby the E-string is turned into a bunch of hypermultiplets, thereby reducing the number $N$ of tensor multiplets.  We find it most transparent to describe the full hierarchy of RG flows for \emph{fixed} $N$, and generate one Hasse diagram of homomorphisms in $\text{Hom}(\zz_k,E_8)$ per value of $N$. This is a somewhat different perspective with respect to the one used in \cite{Frey:2018vpw,Giacomelli:2022drw} where flows at fixed $k$ but varying $N$ are described for a few cases of small $k$.  In section \ref{sub:6N} we will show how one can produce such ``mixed'' flows by only considering flows at fixed $N$ followed by a ``jump'' between the hierarchy of RG flows at $N$ and $N-1$, so that we never need to discuss the case of mixed flows separately.

Let us close with a simple observation which will play a role in section \ref{sub:theta}. For $k=6$, Kac labels contain all unprimed integers $1,\ldots,6$, meaning we can embed the entire list of integer partitions of $6$ inside the former. Looking at table \ref{tab:k6Kacquiv} we see that we can read off such partitions directly from the Type IIA configurations engineering the tensor branch of the SCFTs, by subtracting adjacent gauge algebra ranks. E.g.  $[5,1]$ is obtained as $5-0=5$, $6-5=1$, $6-6=0$, and so on.

\subsection{Magnetic quivers}
\label{sec:mag}

We will now introduce the 3d magnetic quiver associated with the 6d electric one,  where by the latter we mean the low-energy quiver gauge theory describing the  SCFT on its tensor branch, the prototypical example of which is \eqref{eq:fullymasslessE8} (in the F-theory language we have adopted).

The electric quiver can be obtained by reading off the spectrum of fundamental strings stretched between D6-branes in an O8-D8-D6-NS5 engineering of the SCFT. In analogy to 3d $\mathcal{N}=4$ D3-D5-NS5 setups in Type IIB 
\cite{Hanany:1996ie}, where the electric quiver is obtained by stretching F1's between D3-D5's and the magnetic one by stretching D1's between D3-NS5's (i.e. applying mirror symmetry \cite{Kapustin:1999ha}), one can also define a magnetic quiver for D6-D8-NS5 setups which are T-dual to the D3-D5-NS5 ones (along $x^{345}$ in table \ref{tab:M5}).  Moving away all NS5's from the D6's and suspending the latter between the D8's, we can see that the magnetic object is now a D4-brane (T-dual along $x^{345}$ to a D1). One can then read off the theory on the worldvolume of the stretched D4's following the simple rules introduced in \cite{Cabrera:2019izd}, thereby producing the 3d magnetic quiver of the 6d SCFT. For orbi-instantons these are $\mathcal{N}=4$ unitary quivers \cite{Mekareeya:2017jgc}. Importantly for us, as shown in \cite{Cabrera:2019izd} the Coulomb branch of the 3d magnetic quiver ``at infinite coupling'' captures the Higgs branch of the 6d electric quiver at the origin of the tensor branch, i.e.  the point where all gauge couplings are infinite and new massless degrees of freedom arise from light D2-branes.  (A 6d observer sees the D2's stretched along $x^{016}$ as instantons for the D6 gauge theory \cite{Douglas:1995bn}, propagating as strings inside $x^{0\ldots5}$.) In turn, the geometry and quaternionic dimension of the Coulomb branch can easily be computed in terms of closures of nilpotent orbits of classical or exceptional Lie algebras, for which the technology is well-established \cite{collingwood1993nilpotent}.

For orbi-instantons of type A the magnetic quivers have been worked out in full generality (i.e. for any $k$ and Kac label) in \cite[Sec. 3.6]{Cabrera:2019izd}, and we will heavily exploit their results. In particular, once we compute the magnetic quivers we can proceed to ``subtract'' them, which simply means subtracting the ranks of the $\U(n)$ gauge nodes of the two quivers in an ordered way. (We will see a few concrete examples below.) 

Quiver subtraction was then shown \cite{Cabrera:2019dob} to be equivalent to a ``Kraft--Procesi (KP) transition'' between nilpotent orbits,\footnote{This quiver ``arithmetic'' has been further extended by introducing the notions of quiver addition \cite{Rogers:2019pqe,Gledhill:2021cbe,Bourget:2021siw}, implosion \cite{Dancer:2020wll,Bourget:2021zyc,Bourget:2021qpx},  inversion \cite{Grimminger:2020dmg}, and folding \cite{Hanany:2020jzl,Bourget:2020bxh,Bourget:2021xex}.} which for T-brane theories (defined by a pair of nilpotent orbits of a Lie algebra) is the same as a Higgs branch RG flow \cite{Heckman:2016ssk}.  Motivated by this and by the fact that Higgsings in Lagrangian quivers with 8 supercharges can always be organized into a Hasse diagram (via decomposition of the Higgs branch into symplectic leaves) \cite{Bourget:2019aer},\footnote{See also \cite[Sec. 3]{Cabrera:2019dob} for a quick summary of this idea.} we expect the subtraction (when allowed) to provide an ordering $\rho_1 \succ \rho_2$ between homomorphisms $\rho_i:\zz_k \to E_8$ at fixed $k$ and $N$ (i.e. varying only the Kac label giving the boundary condition in M-theory).

\subsubsection{Quiver subtraction and Kraft--Procesi transitions}

Let us show how quiver subtraction works in practice by means of a couple of examples.  Consider Kac labels $[6]$ and $[5,1]$ for $k=6$ in table \ref{tab:k6Kacquiv}. The electric quivers can be read off of the IIA brane setups; in F-theory notation they read
\begin{subequations}\label{eq:651elec}
\begin{align}
&[6] \leftrightarrow \mathfrak{f}=\mathfrak{su}(6)\oplus \mathfrak{su}(3)\oplus\mathfrak{su}(2): & &[\SU(3)\times \SU(2)] \ {1}\ \overset{\mathfrak{su}(6)}{\underset{[N_\text{f}=6]}{2}}\ \overset{\mathfrak{su}(6)}{2} \cdots  \overset{\mathfrak{su}(6)}{2}\ [\SU(6)]\ , 
\end{align} 
\begin{align}
&[5,1]\leftrightarrow \mathfrak{f}=\mathfrak{su}(5)\oplus \mathfrak{su}(4)\oplus\mathfrak{u}(1): & &[\SU(5)] \ {1}\ \overset{\mathfrak{su}(5)}{\underset{[N_\text{f}=4]}{2}}\ \overset{\mathfrak{su}(6)}{\underset{[N_\text{f}=1]}{2}}\ \overset{\mathfrak{su}(6)}{2} \cdots  \overset{\mathfrak{su}(6)}{2}\ [\SU(6)]\ .
\end{align}
\end{subequations}
The magnetic quivers at infinite coupling (i.e. at the origin of the tensor branch) read instead
\begin{subequations}\label{eq:651magquiv}
\begin{align}
&[6]: & {\scriptstyle 1 - 2 - 3 - 4 -5 -6 - (N+5) -(2N+4) -(3N+3) - (4N+2) -(5N+1)-\overset{\overset{\scriptstyle 3N}{\vert}}{6N}-4N-2N} \\
&[5,1]: & {\scriptstyle 1 - 2 - 3 - 4 -5 -6 - (N+4) - (2N+3) - (3N+2) - (4N+1)-5N-\overset{\overset{\scriptstyle 3N}{\vert}}{6N}-4N-2N} \label{eq:51magquiv}
\end{align}
\end{subequations}
where $-$ or $|$ denotes a hypermultiplet, and $n$ a $\U(n)$ gauge group.
The quivers in \eqref{eq:651magquiv} are unbalanced \cite{Gaiotto:2008ak}, namely the condition that the number of flavors equals twice the number of colors, $N_\text{f} = 2N_\text{c}$,  is not satisfied by some of the gauge nodes; e.g. $6+(2N+3) \neq 2 \cdot (N+4)$.  

Subtracting the two quivers node-by-node in the obvious way (i.e. subtracting the ranks of $\U$ groups in the same position) we get
\begin{equation}\label{eq:subtr1}
1 - 1 - 1 - 1 - 1\ .
\end{equation}
Since the ranks of all gauge groups in \eqref{eq:subtr1} are non-negative, this is an allowed subtraction. However \eqref{eq:subtr1} is also unbalanced; ``rebalancing'' it with the rules in  \cite[App. A.2]{Bourget:2019aer} and \cite[Sec. 3.2]{Bourget:2020mez},\footnote{See also \cite[Sec. 3]{vanBeest:2021xyt} and \cite{Gledhill:2021cbe}.} i.e. adding flavors as necessary such that $N_\text{f} = 2N_\text{c}$ at each gauge node,\footnote{With the understanding that gauge nodes neighboring a given $\U(n)$ node act as flavors for the latter. The 6d RG flow $[6] \to [5,1]$ can be understood as a complex Fayet--Iliopoulos deformation of the 3d magnetic quiver associated with $[6]$, which in turn is equivalent to rebalancing the product of the quiver subtraction \cite[Sec. 3.2]{Bourget:2020mez}.} we obtain the affine Dynkin diagram of $\mathfrak{su}(6)$,
\begin{equation}\label{eq:su6subtr}
\begin{array}{c} \raisebox{-10.5pt}{\rotatebox{30}{$\rule{40pt}{0.4pt}$}} \raisebox{6pt}{1} \raisebox{9pt} {\rotatebox{-30}{$\rule{40pt}{0.4pt}$}} \\[-6pt]  1 - 1 -1 -1 -1  \end{array}\ ,
\end{equation}
which means to go from the $[6]$ magnetic quiver to the $[5,1]$ a matter field charged under $\su{6}$ is taking a vev.  We denote this flow by an edge labeled by $\mathfrak{a}_5$ in the Hasse diagram of RG flows for $k=6$.  Remembering that in our conventions the $[6]$ is associated with the partition $(\,{\fontsize{0.25pt}{0.5pt}\selectfont \yng(6)}\,)^{\raisebox{1pt}{\scriptsize \text{t}}}$ and $[5,1]$ with $\left(\, {\fontsize{0.05pt}{0.5pt}\selectfont \yng(5,1)}\,\right)^{\raisebox{2pt}{\scriptsize \text{t}}}$, taking transposes once more we see that this is the well-known $\mathfrak{a}_5$ KP transition from the integer partition $[2,1^4]$ of $6$ to $[1^6]$ \cite[Sec. 6]{KP1}: namely $\overline{\mathcal{O}}_{[1^6]} \subset \overline{\mathcal{O}}_{[2,1^4]}$ and the transverse Slodowy slice $S \subseteq \overline{\mathcal{O}}_{[2,1^4]}$ to $\mathcal{O}_{[1^6]}$ is a so-called minimal singularity (i.e. the closure of the minimal nilpotent orbit) given by collapsing the cotangent bundle of $\pp^5$ \cite{kempf,KP2}.  Because we initially took transposes, we are actually flowing from $[6]$ to $[5,1]$, which is just the transpose of the statement $\overline{\mathcal{O}}_{[1^6]} \subset \overline{\mathcal{O}}_{[2,1^4]}$ (of quaternionic dimension $0$ and $5$ respectively).  This is a general feature of the Hasse diagrams we will produce. 

In general we have an $\mathfrak{a}_i$ KP transition whenever the product of the quiver subtraction (after rebalancing) is of the form
\begin{equation}\label{eq:ai}
\mathfrak{a}_i: \begin{array}{c} \raisebox{-10.5pt}{\rotatebox{30}{$\rule{40pt}{0.4pt}$}} \raisebox{6pt}{1} \raisebox{9pt} {\rotatebox{-30}{$\rule{40pt}{0.4pt}$}} \\[-6pt]  \underbrace{1 - 1 - \cdots  -1 -1}_{i}  \end{array}\ .
\end{equation}
In the next section we will also see examples of $\mathfrak{d}_i$ KP transitions,  where as the notation implies the product of the subtraction is an affine Dynkin diagram of D type:
\begin{equation}\label{eq:di}
\mathfrak{d}_i:  \underbrace{1 - \overset{\vernocirc{1}}{2} - 2 - \cdots  -2 - \overset{\vernocirc{1}}{2} -1}_{i-1}\ .
\end{equation}
Finally, in figure \ref{fig:frey-rudelius} we will also see examples of $\mathfrak{e}_i$ KP transitions.\footnote{See \cite[Tab. 1]{Bourget:2021siw} for a complete list of possible types of KP transitions.}

Consider now the quiver subtraction between two 3d magnetic quivers corresponding to electric ones given by Kac labels which involve primed parts.  For instance, take $[4',2']$ and $[2'^3]$. The electric quivers are
\begin{subequations}
\begin{align}
&[4',2'] \leftrightarrow \mathfrak{f}=\mathfrak{su}(8)\oplus \mathfrak{u}(1): & & \overset{\mathfrak{su}(6)}{\underset{[N_{\fontsize{0.25pt}{0.5pt}\selectfont \yng(1,1)}=1,N_\text{f}=8]}{1}}\ \overset{\mathfrak{su}(6)}{2}\ \overset{\mathfrak{su}(6)}{2} \cdots  \overset{\mathfrak{su}(6)}{2}\ [\SU(6)]\ ,  \\
&[2'^3] \leftrightarrow \mathfrak{f}=\mathfrak{so}(16): & & \overset{\mathfrak{usp}(6)}{\underset{[N_\text{f}=8]}{1}}\ \overset{\mathfrak{su}(6)}{2}\ \overset{\mathfrak{su}(6)}{2}\ \overset{\mathfrak{su}(6)}{2} \cdots  \overset{\mathfrak{su}(6)}{2}\ [\SU(6)] \ .\label{eq:k62'3}
\end{align}
\end{subequations}%
\begin{figure}[ht!]
\centering
\begin{tikzpicture}[scale=1,baseline]
				\draw[fill=black] (0,0) circle (0.075cm) node[xshift=-0.3cm] {${\fontsize{0.25pt}{0.5pt}\selectfont \yng(1,1)}$};
				\draw[fill=black] (0.75,0) circle (0.075cm);
				\draw[fill=black] (1.5,0) circle (0.075cm);
				\draw[fill=black] (2.25,0) circle (0.075cm);
				
				\draw[solid,black,thick] (0,0)--(0.75,0) node[black,midway,xshift=0.1cm,yshift=0.2cm] {\footnotesize $6$};
				\draw[solid,black,thick] (0.75,0)--(1.5,0) node[black,midway,yshift=0.2cm] {\footnotesize $6$};
				\draw[loosely dotted,black,thick] (1.5,0)--(2.25,0) node[black,midway, ] {};
				\draw[solid,black,thick] (2.25,0)--(3,0) node[black,midway,yshift=0.2cm] {\footnotesize $6$};
				
				\draw[dashed,black,very thick] (0,-.5)--(0,.5) node[black,midway] {};
				\draw[solid,black,very thick] (0.2,-.5)--(0.2,.5) node[black,midway, xshift =0cm, yshift=+.75cm] {\footnotesize $8$};
\end{tikzpicture}
			\hspace{3cm}
\begin{tikzpicture}[scale=1,baseline]
				\draw[fill=black] (0.75,0) circle (0.075cm);
				\draw[fill=black] (1.5,0) circle (0.075cm);
				\draw[fill=black] (2.25,0) circle (0.075cm);
				
				\draw[solid,black,thick] (0,0)--(0.75,0) node[black,midway,xshift=0.1cm,yshift=0.2cm] {\footnotesize $6$};
				\draw[solid,black,thick] (0.75,0)--(1.5,0) node[black,midway,yshift=0.2cm] {\footnotesize $6$};
				\draw[loosely dotted,black,thick] (1.5,0)--(2.25,0) node[black,midway, ] {};
				\draw[solid,black,thick] (2.25,0)--(3,0) node[black,midway,yshift=0.2cm] {\footnotesize $6$};
				
				\draw[dashed,black,very thick] (0,-.5)--(0,.5) node[black,midway] {};
				\draw[solid,black,very thick] (0.2,-.5)--(0.2,.5) node[black,midway, xshift =0cm, yshift=+.75cm] {\footnotesize $8$};
\end{tikzpicture}
\caption{Type IIA brane configurations engineering the $k=6$ orbi-instantons with Kac label $[4',2']$ and $[2'^3]$ respectively.  The setup for $[4',2']$ engineers a two-index antisymmetric of $\su{6}$ (coming from D6-image D6 strings),  and there is a half-NS5 stuck on the orientifold. We can Higgs the $\su{6}$ down to $\usp{6}$ by sliding off to infinity the stuck NS5 along the O8: this corresponds to an A-type KP transition (see e.g. $[4',2'] \xrightarrow{A_2} [2'^3]$ in figure \ref{fig:flowk6}).}
\label{fig:IIAk64'2'2'3}
\end{figure}
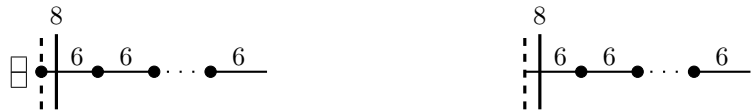%
Here ${\fontsize{0.25pt}{0.5pt}\selectfont \yng(1,1)}$ denotes one (full) hypermultiplet in the two-index antisymmetric representation of $\mathfrak{su}(6)$, which is complex and as such carries a $\U(1)$ flavor symmetry with algebra $\mathfrak{u}(1)$.  (The antisymmetric hypermultiplet comes from open strings stretched between the physical D6's in the zeroth segment and their images ``beyond'' the O8 \cite{Hanany:1997gh}.\footnote{See \cite{Landsteiner:1997ei} for a similar observation for 4d $\mathcal{N}=2$ O6-D6-D4-NS5 brane setups.}) In fact we know that $[4',2']$ preserves $\mathfrak{su}(8)\oplus \mathfrak{u}(1)$, and this explains the physical origin of the second summand.  (For a more precise statement see the discussion in section \ref{sub:u1}.) On the other hand the $\mathfrak{so}(16)$ symmetry preserved by $[2'^3]$ (since the fundamental of $\mathfrak{usp}$ is pseudo-real) is already rank-8, and in fact there are no hypermultiplets charged under any other gauge algebra but the $N_\text{f}=8$ of $\usp{6}$.  Both flavor symmetries have a weak-coupling origin, as is clear from their IIA setups drawn in figure \ref{fig:IIAk64'2'2'3}: both have a nonempty zeroth gauge algebra (i.e. decorated $-1$ curve) engineered by the D6 segment ending on the O8 (respectively, half-NS5 stuck on the O8) with flavor hypermultiplets charged under it. 

The magnetic quivers at infinite coupling read
\begin{subequations}
\begin{align}
&[4',2']: & & {\scriptstyle 1 - 2 - 3 - 4 -5 -6 - (N+6) -(2N+6) -(3N+6) - (4N+6) -(5N+6)-\overset{\overset{\scriptstyle 3N+3}{\vert}}{(6N+6)}-(4N+3)-(2N+1)} \label{eq:4'2'magquiv}\ ,\\
&[2'^3]: & & {\scriptstyle 1 - 2 - 3 - 4 -5 -6 - (N+6) -(2N+6) -(3N+6) - (4N+6) -(5N+6)-\overset{\overset{\scriptstyle 3N+3}{\vert}}{(6N+6)}-(4N+3)-(2N)}\ .
\end{align}
\end{subequations}
Subtracting the two we get a single unbalanced node, $1$.  Since its origin in \eqref{eq:4'2'magquiv} is an underbalanced node by 1 flavor, to rebalance it with the rules in \cite[Sec. 3.2]{Bourget:2020mez} we need to add an extra flavor on top of the 2 which would saturate the $N_\text{f}=2N_\text{c}$ equality, so that we end up with $1 - \fbox{$3$}$ (the box representing flavor hypermultiplets). The Coulomb branch of this 3d $\mathcal{N}=4$ quiver is the $A_2$ Kleinian singularity,  i.e. $\cc^2/\zz_3$ \cite{Cabrera:2017njm}.  More generally, whenever the product of the subtraction is of the form
\begin{equation}\label{eq:Ai}
A_{i-1}: 1 - \fbox{$i$}\ ,
\end{equation}
the KP transition is of type $A_{i-1}: \cc^2/\Gamma_\text{A} = \cc^2/\zz_{i}$, i.e. the transverse Slodowy slice to the orbit is an A-type Kleinian singularity.\footnote{Due to mirror symmetry, a 3d magnetic quiver whose Coulomb branch is $\mathfrak{a}_i$ has Higgs branch given by $A_i$. Likewise for $\mathfrak{d}_i, \mathfrak{e}_i$ and $D_i,E_i$ respectively. See e.g. \cite[Tab. 2]{Grimminger:2020dmg}.} 

From \cite[Sec. 3]{KP1} we expect another $A_2$ KP transition between e.g. the integer partitions $[4,1^2]$ and $[3,2,1]$ of $6$; taking transposes and matching with the entries in our table \ref{tab:k6Kacquiv} we see that the Kac labels involved in such a transition should be $[3,1^3]$ and $[3,2,1]$ respectively. Computing the magnetic quivers we obtain
\begin{subequations}
\begin{align}
&[3,2,1]: & &
{\scriptstyle 1 - 2 - 3 - 4 -5 -6 - (N+3) -(2N+1) -3N - 4N - 5N-\overset{\overset{\scriptstyle 3N}{\vert}}{6N}-4N-2N}\ ,\\
&[3,1^3]: & & {\scriptstyle 1 - 2 - 3 - 4 -5 -6 - (N+2) -(2N+1) -3N - 4N - 5N-\overset{\overset{\scriptstyle 3N}{\vert}}{6N}-4N-2N}\ .
\end{align}
\end{subequations}
Subtracting and rebalancing we get $1 - \fbox{$3$}$, hence an $A_2$ KP transition as expected from $[4,1^2] \xrightarrow{A_2} [3,2,1]$, though with $[3,2,1]$ flowing into $[3,1^3]$.

\section{\texorpdfstring{An interesting case study: $k=6$}{An interesting case study: k=6}}
\label{sec:k=6}

\begin{figure}[!htbp]
\centering
\def\svgwidth{0.5\columnwidth}
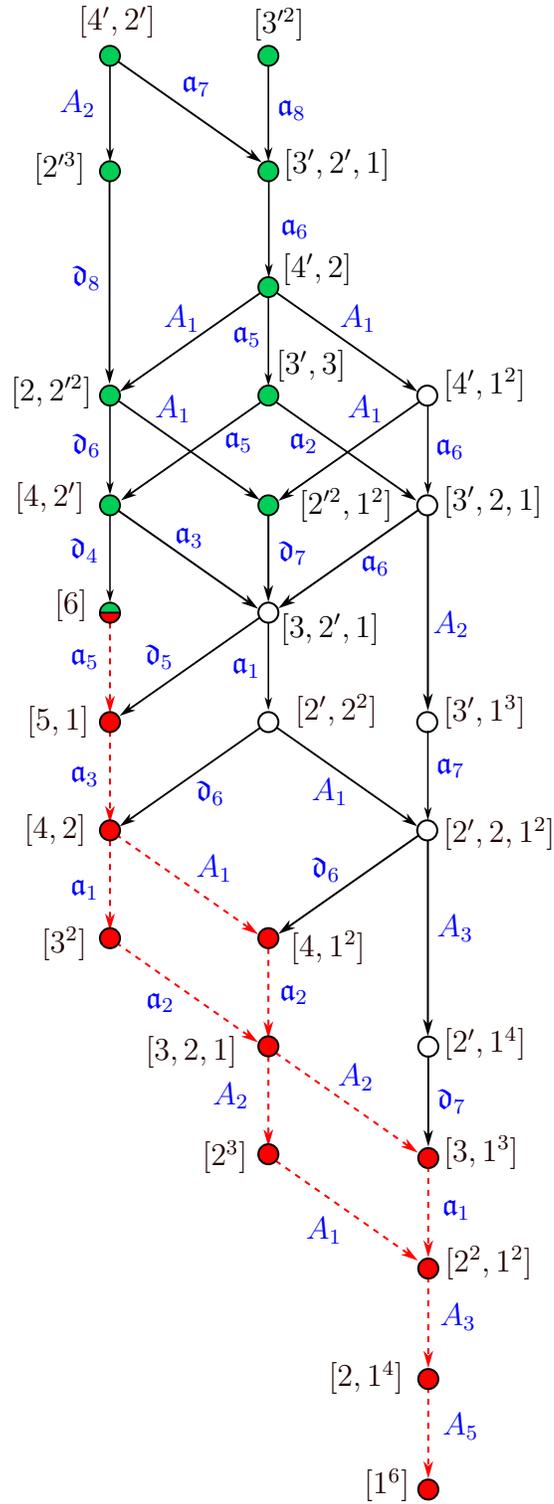
\caption{The Hasse diagram of homomorphisms in $\text{Hom}(\zz_6,E_8)$, i.e.  hierarchy of Higgs branch RG flows of A-type orbi-instantons for $N$ M5-branes and $k=6$.  The legend of colors and decorations is included in section \ref{sub:6legend}.}
\label{fig:flowk6}
\end{figure}

We are finally in a position to showcase the full hierarchy of infinite-coupling Higgs branch RG flows for A-type $k=6$ orbi-instantons at fixed $N$.  The hierarchy of flows (i.e. Hasse diagram) is drawn in figure \ref{fig:flowk6}.
The algorithm to obtain such a diagram for any $N,k$ is as follows. (We will see in the next section a few subtleties and novelties that only arise for $k\geq 7$.)
\begin{itemize}
\item[\emph{i)}] Write down all possible Kac labels for the chosen $k$ as in \eqref{eq:longk}.
\item[\emph{ii)}] Write down the 6d electric quiver for each Kac label, i.e. orbi-instanton tensor branch,  following the algorithm in \cite{Mekareeya:2017jgc}.
\item[\emph{iii)}] Compute the 3d infinite-coupling magnetic quivers associated with the 6d electric quivers according to the rules in \cite{Cabrera:2019izd}.
\item[\emph{iv)}] Compute all possible subtractions between 3d magnetic quivers according to the rules in \cite{Cabrera:2018ann} and rebalance if necessary.  For a chosen 3d magnetic quiver (i.e. Kac label) only a subset of all other quivers can be consistently subtracted from it. Associate an edge to an allowed subtraction, where the edge is labeled by the (re)balanced quiver subtraction, i.e. KP transition of type \eqref{eq:ai}, \eqref{eq:di}, or \eqref{eq:Ai}.
\end{itemize}
Following these simple steps we arrive at an \emph{undecorated} Hasse diagram, where each node is labeled by its Kac label. 

\subsection{\texorpdfstring{The $a$ anomaly}{The a anomaly}}

To add strength to the proposed algorithm we check consistency with the $a$-theorem,  i.e. we compute the 6d $a$ anomaly of each SCFT and check that $\Delta a >0$ for any allowed subtraction. The $a$ anomaly of orbi-instantons can be extracted from the anomaly inflow calculation in \cite[Sec. 3.4]{Mekareeya:2017jgc}, and reads:
\begin{equation}\label{eq:a6danalytic}
\footnotesize
\begin{split}
  a= &-\frac{3}{2} \langle \bm{w},\bm{ρ} \rangle_M - \frac{16}{7} \langle \bm{w},\bm{w} \rangle_M \left[ 15 \left(N-σ\right) + k \left( 6\left(N-σ\right)² + 6k \left(N-σ\right) + 2 k² + 13 \right) \right] + \\
  	&+ \frac{48}{7} \langle \bm{w}, \bm{w} \rangle_M ² \left[ k + \left(N-σ\right)\right]
  	+ \frac{1}{420} \big[ -502 + 18900 \left(N-σ\right) + 14415 k\, +\\
	& + k² \left(251 + 3680 + 576 k³\right) + 1440k \left(N-σ\right) \left(9 + 2k²\right) 
    + 2880 \left(N-σ\right)² \left( 5 + 2k² \right) + \\
	& + 3840 \left(N-σ\right)³ \big] + 400 \sum_{\bm{α}\in Δ^+} \langle \bm{w} \cdot \bm{\omega},\bm{α} \rangle^3 - 96 \sum_{\bm{α}\in Δ^+} \langle \bm{w} \cdot \bm{\omega},\bm{α} \rangle^5\ ,
\end{split}
\end{equation}
where $\langle \bm{w},\bm{w} \rangle_M \equiv \bm{w} \cdot M \cdot \bm{w}$ and $\langle \bm{w},\bm{ρ} \rangle_M \equiv \bm{w} \cdot M \cdot \bm{ρ}$ are inner products of 
\begin{subequations}
\begin{align}
\bm{w} & \equiv  (n_2, n_3, n_4, n_5, n_6, n_{4'}, n_{2'},n_{3'})\ , \\
\bm{ρ} & \equiv   (1,1,1,1,1,1,1,1)\ 
\end{align}
\end{subequations}
with respect to the inverse Cartan matrix $M=A(E_8)^{-1}$ of $E_8$.  (A $\cdot$ denotes matrix multiplication.) We choose the following basis for the simple roots:
\begin{equation}
	E = \begin{pmatrix}
		  0 & 1 & -1 & 0 & 0 & 0 & 0 & 0\\
		  0 & 0 & 1 & -1 & 0 & 0 & 0 & 0\\
		  0 & 0 & 0 & 1 & -1 & 0 & 0 & 0\\
		  0 & 0 & 0 & 0 & 1 & -1 & 0 & 0\\
		  0 & 0 & 0 & 0 & 0 & 1 & -1 & 0\\
		  0 & 0 & 0 & 0 & 0 & 0 & 1 & -1\\
		  \frac{1}{2} & -\frac{1}{2} & -\frac{1}{2} & -\frac{1}{2} & -\frac{1}{2} & -\frac{1}{2} & -\frac{1}{2} &  \frac{1}{2}\\
		  0 & 0 & 0 & 0 & 0 & 0 & 1 & 1\\
	\end{pmatrix}\ ,
\end{equation}
so that the Cartan matrix reads $A(E_8) = E \cdot E^\text{t}$.  The fundamental weights $\bm{\omega}$ of $E_8$ are the rows of $(E^\text{t}){}^{-1}$,  $Δ^+$ is the set of 120 positive roots $\bm{α}$ of $E_8$, and $\langle \, ,  \rangle$ denotes the $\rr^8$ inner product.  The quantity $σ \equiv \sum_{i=1}^{6} n_i + p$ depends on the specific Kac label, where $p$ has to be determined case by case. In the ``case'' notation of \cite[Sec. 3.2]{Mekareeya:2017jgc}, we have:
\begin{equation}
\small
	p=
    \begin{cases}
      \left(n_{3'}+n_{4'}\right)/2, & \text{for cases 1 and 5}\  ,\\
      \left(n_{3'}+n_{4'}-1\right)/2, & \text{for cases 2 and 3}\ ,\\
	  \left(n_{3'}+2n_{4'}+n_{2'}-l\right)/3 & \text{for case 4, with $l (\equiv n_{3'}+2n_{4'}+n_{2'} \mod 3)=\{ 0,1,2\}$}\ .
    \end{cases}
\end{equation}
As is clear,  $a$ can only depend on $N,k$ and the choice of Kac label (via $σ$ and $\bm{w}$), and its large-$N$ leading term is easily shown to scale like $N^3$ and to be universal, as it should \cite{Cremonesi:2015bld,Apruzzi:2017nck}.  The values of \eqref{eq:a6danalytic} for $k=6$ are tabulated in table \ref{tab:a6d}.%
\setlength{\tabcolsep}{4pt}
\begin{longtable}[c]{@{}ll ll@{}}
 	Kac label & \hspace*{25pt}   exact 6d $a$ anomaly &  Kac label & \hspace*{25pt}  exact 6d $a$ anomaly \\ \toprule
	\footnotesize$[6]$ & \footnotesize $\frac{2304}{7}N^3-288 N^2-\frac{261}{7} N+\frac{251}{6}$ & \footnotesize$[3,2',1]$ & \footnotesize $\frac{2304}{7} N^3-288 N^2-\frac{261}{7} N+\frac{30887}{210}$ \\
	\footnotesize$[3'^2]$ & \footnotesize$\frac{2304}{7} N^3+\frac{3744}{7} N^2+\frac{1179}{7} N+\frac{352}{105}$ & \footnotesize$[3',2,1]$ &\footnotesize $\frac{2304}{7} N^3-\frac{864}{7} N^2-\frac{741}{7} N+\frac{41273}{210}$ \\
	\footnotesize$[3',3]$ & \footnotesize$\frac{2304}{7} N^3+\frac{288}{7} N^2-\frac{837}{7} N+\frac{465}{14}$ & \footnotesize$[3,2,1]$ &\footnotesize $\frac{2304}{7} N^3-\frac{6624}{7} N^2+\frac{5499}{7} N+\frac{4404}{35}$ \\
	\footnotesize$[3^2]$ & \footnotesize$\frac{2304}{7} N^3-\frac{5472}{7} N^2+\frac{3483}{7} N+\frac{3263}{105}$ & \footnotesize$[4',1^2]$ & \footnotesize$\frac{2304}{7} N^3+\frac{288}{7} N^2-\frac{837}{7} N+\frac{51179}{210}$ \\
	\footnotesize$[4',2']$ & \footnotesize$\frac{2304}{7} N^3+\frac{3744}{7} N^2+\frac{1179}{7} N+\frac{33}{10}$ & \footnotesize$[4,1^2]$ & \footnotesize$\frac{2304}{7} N^3-\frac{5472}{7} N^2+\frac{3483}{7} N+\frac{31103}{105}$ \\
\footnotesize	$[4,2']$ & \footnotesize$\frac{2304}{7} N^3-\frac{864}{7} N^2-\frac{741}{7} N+\frac{116}{3}$ & \footnotesize$[2'^2,1^2]$ &\footnotesize $\frac{2304}{7} N^3-\frac{864}{7} N^2-\frac{741}{7} N+\frac{6808}{35}$ \\
	\footnotesize$[2'^3]$ & \footnotesize$\frac{2304}{7} N^3+\frac{1440}{7} N^2-\frac{549}{7} N+\frac{2021}{105}$ & \footnotesize$[2',2,1^2]$ &\footnotesize $\frac{2304}{7} N^3-\frac{4320}{7} N^2+\frac{1851}{7} N+\frac{21533}{42}$ \\
	\footnotesize$[4',2]$ & \footnotesize$\frac{2304}{7} N^3+\frac{1440}{7} N^2-\frac{549}{7} N+\frac{535}{21}$ &\footnotesize $[2^2,1^2]$ &\footnotesize $\frac{2304}{7} N^3-1440 N^2+\frac{13851}{7} N-\frac{9719}{105}$ \\
	\footnotesize$[4,2]$ & \footnotesize$\frac{2304}{7} N^3-\frac{4320}{7} N^2+\frac{1851}{7} N+\frac{917}{10}$ &\footnotesize $[3',1^3]$ &\footnotesize $\frac{2304}{7} N^3-\frac{3168}{7} N^2+\frac{603}{7} N+\frac{25437}{35}$ \\
	\footnotesize$[2'^2,2]$ &\footnotesize $\frac{2304}{7} N^3+\frac{288}{7} N^2-\frac{837}{7} N+\frac{6539}{210}$ & \footnotesize$[3,1^3]$ &\footnotesize $\frac{2304}{7} N^3-\frac{8928}{7} N^2+\frac{10683}{7} N+\frac{85933}{210}$ \\
	\footnotesize$[2',2^2]$ & \footnotesize$\frac{2304}{7} N^3-\frac{3168}{7} N^2+\frac{603}{7} N+\frac{6946}{35}$ & \footnotesize$[2',1^4]$ & \footnotesize$\frac{2304}{7} N^3-\frac{7776}{7} N^2+\frac{7899}{7} N+\frac{100267}{105}$ \\
	\footnotesize$[2^3]$ & \footnotesize$\frac{2304}{7} N^3-\frac{8928}{7} N^2+\frac{10683}{7} N-\frac{47987}{210}$ & \footnotesize$[2,1^4]$ &\footnotesize $\frac{2304}{7} N^3-\frac{13536}{7} N^2+\frac{25659}{7} N-\frac{36003}{70}$ \\
	\footnotesize$[5,1]$ &\footnotesize $\frac{2304}{7} N^3-\frac{3168}{7} N^2+\frac{603}{7} N+\frac{667}{7}$ & \footnotesize$[1^6]$ & \footnotesize$\frac{2304}{7} N^3-\frac{19296}{7} N^2+\frac{53019}{7} N-\frac{94922}{21}$ \\
	\footnotesize $[3',2',1]$ & \footnotesize $\frac{2304}{7} N^3+\frac{2592}{7} N^2+\frac{123}{7} N+\frac{1628}{105}$ \\ \bottomrule
\caption{Explicit values of the 6d $a$ anomaly for all A-type $k=6$ orbi-instantons.}
\label{tab:a6d}
\end{longtable}%
\renewcommand{\arraystretch}{1}%
Finally, we note that the (incomplete) hierarchy induced by $\Delta a>0$ must be equivalent to the procedure used in \cite{Heckman:2015ola} to derive some allowed flows between orbi-instantons,  namely to subtracting the UV and IR 6d anomaly polynomials.\footnote{Computing $a$ can only tell us if a given Kac label sits higher than another in the hierarchy, but not if and when a flow bifurcates. Therefore it is not enough to obtain the complete RG flow hierarchy.}

\subsection{\texorpdfstring{Embedding the $\su{6}$ nilpotent orbits Hasse diagram}{Embedding the su(6) nilpotent orbits Hasse diagram}}
\label{sub:su6}

Having constructed an explicit hierarchy of RG flows,  we can now analyze its most salient features, and propose \emph{decorations} to visually convey physical properties of the flows. This is where the usefulness of the Type IIA pictures stands out.

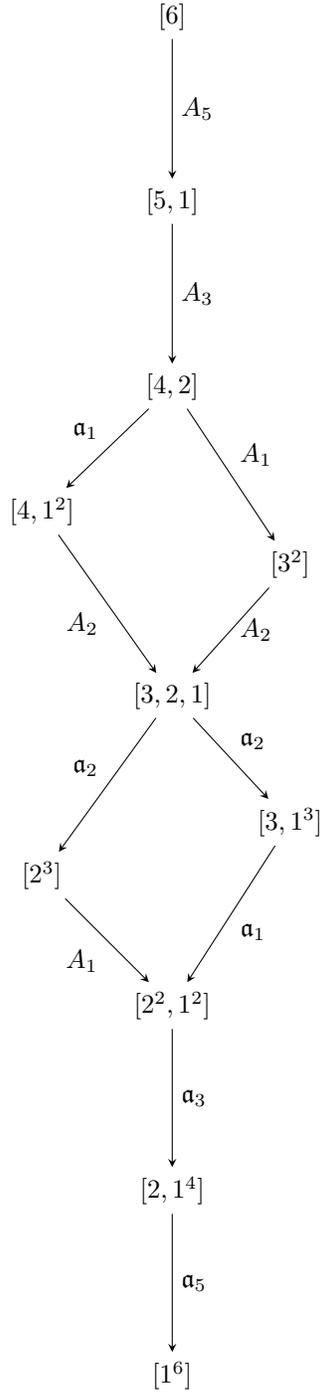
\begin{figure}
	\centering
	\footnotesize
	\begin{tikzpicture}[node distance=70pt]
	  \tikzstyle{arrow} = [->,>=stealth]
	  \node (n1) {$[6]$};
	  \node (n2) [below of=n1] {$[5,1]$};
	  \node (n3) [below of=n2] {$[4,2]$};
	  \node (n4) [below left=30pt and 20pt of n3] {$[4,1^2]$};
	  \node (n5) [below right=50pt and 20pt of n3] {$[3^2]$};
	  \node (n6) [below=100pt of n3] {$[3,2,1]$};
	  \node (n7) [below= 80pt of n5] {$[3,1^3]$};
	  \node (n8) [below=120pt of n4] {$[2^3]$};
 	  \node (n9) [below=100pt of n6] {$[2^2,1^2]$};
	  \node (n10) [below of=n9] {$[2,1^4]$};
	  	  \node (n11) [below of=n10] {$[1^6]$};
  
	  \draw [arrow] (n1)--(n2) node [midway, left] {} node[midway, right] {$A_5$}; 

	  \draw [arrow] (n2)--(n3) node [midway, left] {} node[midway, right] {$A_3$}; 

	  	  \draw [arrow] (n3)--(n4) node [midway, left] {} node[midway, above left] {$\mathfrak{a}_1$}; 
	  	  	  	  \draw [arrow] (n3)--(n5) node [midway, left] {} node[midway, above right] {$A_1$}; 

	  \draw [arrow] (n4)--(n6) node [midway, above right=0pt and 1pt] {} node[midway, below left] {$A_2$};
	  	  \draw [arrow] (n5)--(n6) node [midway, above right=0pt and 1pt] {} node[midway, right] {$A_2$};

	  \draw [arrow] (n6)--(n7) node [midway, right] {} node[midway, above right] {$\mathfrak{a}_2$};  
	  	  \draw [arrow] (n6)--(n8) node [midway, right] {} node[midway, above left] {$\mathfrak{a}_2$};  
	  \draw [arrow] (n7)--(n9) node [midway, left] {} node[midway, below right] {$\mathfrak{a}_1$}; 
	  \draw [arrow] (n8)--(n9) node [midway, left] {} node[midway, below left] {$A_1$}; 

	  \draw [arrow] (n9)--(n10) node [midway, left] {} node[midway, right] {$\mathfrak{a}_3$}; 
	  	  \draw [arrow] (n10)--(n11) node [midway, left] {} node[midway, right] {$\mathfrak{a}_5$}; 
	\end{tikzpicture}
	\caption{The $\su{6}$ nilpotent orbits Hasse diagram (in terms of integer partitions of $6$) with its KP transitions \cite{KP1}.  We draw the diagram in a staggered form (whenever a flow bifurcates) to visually convey which IR SCFT (i.e. Kac label) has larger 6d $a$ anomaly.  See table \ref{tab:a6d} for the actual values of $a$ at fixed $N$.}
	\label{fig:su6hasse}
  \end{figure}

As we have already noticed in section \ref{subsub:IIAk6} we can embed the Hasse diagram of $\su{6}$ nilpotent orbits into the bottom part of the homomorphisms Hasse diagram for $k=6$.  Kac labels which map to integer partitions of $6$ in the obvious way (i.e. to nilpotent orbits) are colored in red.  The Hasse diagram of the nilpotent orbits is indicated by red dashed edges in figure \ref{fig:flowk6}, which also carry the information about which KP transition generates the RG flow at each step.  Notice that in our case the Hasse diagram is order-reversed, i.e. we have  implicitly applied the so-called Lusztig--Spaltenstein map \cite[Thm. 6.3.2]{collingwood1993nilpotent} to the original one in \cite[Sec. 3]{KP1} (because of our convention on transpose tableaux from table \ref{tab:k6Kacquiv}), which we reproduce in figure \ref{fig:su6hasse} for the convenience of the reader.

The identification between homomorphisms and nilpotent orbits of $\su{6}$ has a clear physical interpretation.  Each Kac label labeling nodes which are connected by dashed edges preserves, according to the algorithm explained below \eqref{eq:longk}, a rank-8 maximal subalgebra $\f$ of $E_8$ which contains a strong-coupling and a weak-coupling summand. These summands are written down in table \ref{tab:k6Kacquiv}. The weak-coupling summand is literally the centralizer of the nilpotent orbit associated with an integer partition of $k=6$ giving the boundary condition of $k$ D6's ending on the D8's, following the same logic as in \cite{Heckman:2016ssk}.  In the fourth column of that table we exhibit the ending pattern of the D6's on the D8's explicitly.  To each ending pattern there is an associated Young tableau (as explained e.g. in \cite[Sec. 3.1]{Heckman:2016ssk}).  From that we can readily perform the Higgsing/KP transition by sliding off to infinity along $x^{789}$ one or more trapped segments of D6, landing us onto the next Young tableau.  We can finally perform a few Hanany--Witten moves \cite[Sec. 3.2]{Hanany:1997sa} to go back to the configuration in the third column, i.e. a D8 with $n$ D6's ending on it should cross the $n$-th segment of D6's (with $n=0$ the zeroth segment which crosses the O8, which may contain zero D6-branes).

Therefore this portion of the hierarchy of RG flows (in which the Hasse diagram of $\su{6}$ nilpotent orbits embeds) realizes the possible ways of ``peeling off'' D8-branes from the ``maximal'' substack of $6$ for Kac label $[6]$ (see the first row in table \ref{tab:k6Kacquiv}), and placing them further inside the quiver compatibly with gauge anomaly cancellation, i.e.  D6 charge conservation \cite[Sec.  2 \& 3]{Hanany:1997sa}.  On the other hand, the D8's that sit on the O8 are responsible for the strong-coupling summand of the flavor symmetry, i.e. the one adjacent to the undecorated $-1$ curve (i.e. empty gauge algebra) corresponding to zeroth segment with $0$ D6-branes. These \emph{do not} generate the Hasse diagram of the nilpotent orbits of the $\mathfrak{su}$ or $\mathfrak{so}$ algebra. We will see an explicit example in section \ref{sub:6O8*}.

\subsection{\texorpdfstring{Physical meaning of $\uu{1}$ summands}{Physical meaning of u(1) summands}}
\label{sub:u1}

A $\uu{1}$ summand in the leftover flavor symmetry $\f$ is present if and only if there is more than one part in the Kac label, i.e. if we kill more than one node in the affine $E_8$ Dynkin. That is, $\f$ will be a non-semisimple regular subalgebra of $E_8$. 

It is interesting to understand the physical origin of these summands.  Looking at the IIA pictures of Kac labels which do contain a $\uu{1}$ it is tempting to think that this is always associated either with a single D8-brane crossing some D6 segment, or with the center-of-mass motion of more than one D8's (the center of mass obviously coinciding with the D8 in the case of a single brane), or even with the rotation symmetry of a single hypermultiplet in a complex representation. However we have to keep in mind that the D8 stack center-of-mass degree of freedom is removed from the 6d dynamics as per formula \eqref{eq:suflavor},\footnote{In the examples of table \ref{tab:k6Kacquiv} the center of mass is located in the first segment for all Kac labels participating in the $\su{6}$ nilpotent orbits Hasse diagram.  To see this, we temporarily depart from our standard notation introduced below \eqref{eq:fullymasslessE8}, and label the intervals from $i=1$. Call $f_i$ the number of D8's in the $i$-th interval; then the weighted sum $\sum_i i f_i  / 8 = 14/8$ (e.g. $[4.2]$ has $(4\cdot 1 + 2\cdot 2 + 2\cdot 3)/8 = 14/8$) is invariant throughout those Kac labels.  (The related quantity $n_{2,i} = -\sum_{j=1}^{i-1} j f_j$ \cite[Eq. (2.16)]{Cremonesi:2015bld} is the $F_2$ flux integer of the dual AdS$_7$ vacua of \cite{Apruzzi:2013yva, Apruzzi:2017nck}, i.e. $2\pi n_2 = \int_{S^2} F_2 - BF_0$, with $2\pi F_{0,i} = n_{0,i} = r_i - r_{i-1}$ \cite[Eq. (2.15)]{Cremonesi:2015bld} the Romans mass in the $i$-th interval with $r_i$ D6-branes.)} and this statement has a counterpart in the F-theory construction of the same theories: the $\uu{1}$ symmetry summand is actually ``delocalized'' in the geometry, i.e. is \emph{not} associated with a single component of the F-theory discriminant (see the discussion in \cite[Sec. 7]{Heckman:2016ssk} and \cite{Apruzzi:2020eqi}).

Moreover we have to distinguish ``candidate'' $\uu{1}$ symmetries which are visible on the tensor branch from genuine ones which persist at its origin (i.e. in the SCFT).  In \cite{Apruzzi:2020eqi} a general analysis was performed to determine which $\uu{1}$ summands persist in the SCFT as many candidate $\uu{1}$'s actually suffer from an Adler--Bell--Jackiw (ABJ) anomaly,  and thus do not constitute genuine global symmetries at the CFT point.\footnote{That is, for a hypermultiplet of charge $q$ under a flavor $\uu{1}$ summand and in a representation $\mathcal{R}$ of the gauge algebra there is a contribution of the form 
\begin{equation}
I_\text{ABJ} \supset \frac{q}{6} F_{\uu{1}} \Tr_{\mathcal{R}} F_\mathfrak{g}^3 \nonumber
\end{equation}
to the eight-form anomaly polynomial of the tensor branch quiver,  coming from a square diagram in 6d with one flavor current and three gauge currents. Here $F_{\uu{1}}$ and $F_\mathfrak{g}$ are the flavor $\uu{1}$  and gauge field strengths respectively.} 
There it was found that for orbi-instantons with generic $N,k$ and Kac label $[1^k]$ there is only one ABJ-free combination of candidate $\uu{1}$ symmetries (see around \cite[Eq. (4.17)]{Apruzzi:2020eqi}), under which only matter in the plateau of the quiver (i.e. when the gauge algebra rank stabilizes to $k$) is charged.  Indeed the global symmetry (neglecting the global structure of the flavor group) of A-type orbi-instantons was determined to be $E_8 \oplus \su{k}\oplus \uu{1}$ for $k>2$ and $E_8 \oplus \su{2}\oplus \su{2}$ for $k=2$.\footnote{The ABJ-free combination is absent for generic $k$ and $N=1$, i.e.  $[E_8]\  \overset{\mathfrak{su}(k)}{1}\ [\SU(k)]$,  and for $k=2$ and generic $N$,  i.e.  $[E_8]\ {1}\ \overset{\mathfrak{su}(1)}{2}\ \overset{\su{2}}{\underset{[N_\text{f}=1]}{2}}\ \overset{\su{2}}{2}\cdots \overset{\su{2}}{2}\ [\SU(2)]$. In the former case the right flavor symmetry algebra enhances as $\su{k}\oplus \uu{1} \to \su{k+1}$, whereas in the latter as $\su{2}\oplus \uu{1} \to \su{2}\oplus \su{2}$ \cite{Apruzzi:2020eqi}.} 

In M-theory this $\uu{1}$ arises as a subalgebra of the rotation symmetry of probe M5-branes inside the M9 (in absence of the orbifold), i.e. along $x^{7\ldots10}$ in table \ref{tab:M5}: $\so{4}=\su{2}_\text{L} \oplus \su{2}_\text{R}$. The latter factor is the algebra of the R-symmetry of the 6d SCFT, and adding the $\zz_k$ orbifold preserves a $\uu{1} \subset \su{2}_\text{L}$ for generic $k$ and the full $\su{2}_\text{L}$ for $k=2$ (whereas it preserves nothing in type D or E).  In F-theory this ABJ-free $\uu{1}$ is actually delocalized in the geometry \cite[Sec. 6]{Apruzzi:2020eqi}, and this is reflected in the fact that the anomaly-free combination is a \emph{sum} of the $\uu{1}$ carried by the single D8 (or hypermultiplet in another complex representation) and the $\uu{1}$ subalgebras inside the $\uu{N_\text{f}}$ symmetries of $N_\text{f}$ bifundamentals in complex representations (for $\su{r_i}$ gauge algebras with $r_i \geq 3$). Finally, since the $[1^k]$ theory is the progenitor of all other orbi-instantons by ``fission and fusion'' \cite{Heckman:2018pqx}, the above statements carry over to all other Kac labels. Therefore, when a label contains more than one part and thus its $\f$ contains (at least) one $\uu{1}$ factor, the latter combines with the one preserved by the orbifold to generate the ABJ-free combination.

\subsection{\texorpdfstring{Partially embedding the $\so{16}$ nilpotent orbits Hasse diagram}{Partially embedding the so(16) nilpotent orbits Hasse diagram}}
\label{sub:so16}

Certain Kac labels preserve a weak-coupling $\mathfrak{so}(2n)$ factor, with the possibilities $2n=16,\ldots,8$ being realized for $k=6$. These can only come from hypermultiplets in the (pseudo-real) fundamental representation of a $\mathfrak{usp}$ gauge algebra engineered by the zeroth stack of D6-branes crossing the O8 (in the absence of a stuck NS5-brane).  In the IIA setup these hypermultiplets are carried by D8's crossing the nonempty zeroth D6 segment.  E.g. consider Kac label $[2'^3]$ with electric quiver given in \eqref{eq:k62'3} (also reproduced below) and IIA setup on the right of figure \ref{fig:IIAk64'2'2'3}.  Peeling off iteratively D8's and placing them further inside the quiver we can reach a few other configurations with a smaller-rank $\mathfrak{so}$ flavor symmetry, but we are never able to generate the full Hasse diagram of $\mathfrak{so}(16)$ nilpotent orbits.  

Consider for instance the two flows 
\begin{equation}\label{eq:flowsso}
[2'^3] \xrightarrow{\mathfrak{d}_8} [2'^2,2] \xrightarrow{\mathfrak{d}_6} [4,2']\ , \quad [2'^3] \xrightarrow{\mathfrak{d}_8} [2'^2,2] \xrightarrow{\mathfrak{a}_1} [2'^2,1^2].
\end{equation}
The difference in the respective $a$ anomalies along the flow can be read off from \fref{tab:a6d}:
\begin{align}\label{eq:deltaa6d1}
\begin{split}
	Δa_{[2'^3] → [2'^2,2]} &= \frac{1152}{7} N^2+\frac{288}{7} N-\frac{2497}{210}\ , \\
	Δa_{[2'^2,2] → [4,2']}  &= \frac{1152}{7} N^2-\frac{96}{7} N-\frac{527}{70}\ , \\
	Δa_{[2'^2,2] → [2'^2,1^2]}  &= \frac{1152}{7} N^2-\frac{96}{7} N -\frac{34309}{210}\ . 
\end{split}
\end{align}
Clearly, these are positive for $N > k =6$. The electric quivers corresponding to these theories are:
\begin{subequations}\label{eq:someSOexam}
\begin{align}
&[2'^3] \leftrightarrow \f=\so{16}: & &\overset{\mathfrak{usp}(6)}{\underset{[N_\text{f}=8]}{1}}\ \overset{\mathfrak{su}(6)}{2}\ \overset{\mathfrak{su}(6)}{2}\ \overset{\mathfrak{su}(6)}{2} \cdots  \overset{\mathfrak{su}(6)}{2}\ [\SU(6)] \ , \\
&[2'^2,2] \leftrightarrow \mathfrak{f}=\mathfrak{so}(12)\oplus \mathfrak{su}(2) \oplus \mathfrak{u}(1): & &\overset{\mathfrak{usp}(4)}{\underset{[N_\text{f}=6]}{1}}\ \overset{\mathfrak{su}(6)}{\underset{[N_\text{f}=2]}{2}}\ \overset{\mathfrak{su}(6)}{2} \cdots  \overset{\mathfrak{su}(6)}{2}\ [\SU(6)]\ , \\
&[4,2'] \leftrightarrow \f = \so{8} \oplus \su{4} \oplus \uu{1}: & & \overset{\mathfrak{usp}(2)}{\underset{[N_\text{f}=4]}{1}}\ \overset{\mathfrak{su}(6)}{\underset{[N_\text{f}=4]}{2}}\ \overset{\mathfrak{su}(6)}{2}\ \overset{\mathfrak{su}(6)}{2} \cdots  \overset{\mathfrak{su}(6)}{2}\ [\SU(6)] \ , \\
&[2'^2,1^2]: \f = \so{14} \oplus \uu{1} : & & \overset{\mathfrak{usp}(4)}{\underset{[N_\text{f}=7]}{1}}\ \overset{\mathfrak{su}(5)}{2}\ \overset{\mathfrak{su}(6)}{\underset{[N_\text{f}=1]}{2}}\ \overset{\mathfrak{su}(6)}{2} \cdots  \overset{\mathfrak{su}(6)}{2}\ [\SU(6)] \ 
\end{align}
\end{subequations}
In the first flow we give vev to hypermultiplets charged under $\mathfrak{so}(16)$ and then to those charged under $\mathfrak{so}(12)$, breaking those symmetry factors to a subalgebra.  Likewise for the second. The IIA realization of the flows is in figures \ref{fig:IIAflow2'^32'^2242'} and \ref{fig:IIAflow2'^32'^222'^21^2}.
\begin{figure}
\centering
\raisebox{3pt}
{\begin{tikzpicture}[scale=1,baseline,every node/.style={inner sep=0,outer sep=0}]
		\tikzstyle{D6} = [solid,black];
		\tikzstyle{O8} = [densely dashed,black,thick];
		\tikzstyle{D8} = [solid,black,thick];
		\draw[O8] (-0.15,-.75)--(-0.15,.75) node[black,midway] {} node[black,midway] {};
		\draw[D8] (0,-.75)--(0,.75) node[black,midway, yshift=+30pt]{\footnotesize 8};
		
		\draw[D6] (-0.15,.45)--(.8,.45) node[] {};
		\draw[D6] (-0.15,.3)--(.8,.3) node[] {};
		\draw[D6] (-0.15,0.15)--(.8,0.15) node[] {};
		\draw[D6] (-0.15,0)--(.8,0) node[] {};
		\draw[D6] (-0.15,-0.15)--(.8,-0.15) node[] {};
		\draw[D6] (-0.15,-0.3)--(.8,-0.3) node[] {};
		
		\node[circle,minimum size =7pt,draw,fill=black] (r1) at (1.25,0){};
		\node[circle,minimum size =7pt,draw,fill=black] (r2) at (2.25,0) {};
		\node at (3,0) (r3) {};
		
		\draw[D6] (.8,.45)--(r1) (.8,0.3)--(r1) (.8,0.15)--(r1) (.8,0)--(r1) (.8,-0.15)--(r1) (.8,-0.3)--(r1);
		\draw[D6] (r1)--(r2) node [midway,yshift=5pt] {\footnotesize 6};
		
\end{tikzpicture}}
$\xrightarrow{\displaystyle \mathfrak{d}_8}$\hspace{.75cm}
\raisebox{3pt}
{\begin{tikzpicture}[scale=1,baseline,every node/.style={inner sep=0,outer sep=0}]
		\tikzstyle{D6} = [solid,black];
		\tikzstyle{O8} = [densely dashed,black,thick];
		\tikzstyle{D8} = [solid,black,thick];
		\draw[O8] (-0.15,-.75)--(-0.15,.75) node[black,midway] {} node[black,midway] {};
		\draw[D8] (0,-.75)--(0,.75) node[black,midway, yshift=30pt]{\footnotesize 6};
		\draw[D8] (0.4,-.75)--(0.4,.75) node[black,midway, yshift=30pt]{\footnotesize 1};
		\draw[D8] (0.6,-.75)--(0.6,.75) node[black,midway, yshift=30pt]{\footnotesize 1};
		
		\draw[D6] (0.4,.45)--(1,.45) node[] {};
		\draw[D6] (0.6,.3)--(1,.3) node[] {};
		\draw[D6] (-0.15,0.15)--(1,0.15) node[] {};
		\draw[D6] (-0.15,0)--(1,0) node[] {};
		\draw[D6] (-0.15,-0.15)--(1,-0.15) node[] {};
		\draw[D6] (-0.15,-0.3)--(1,-0.3) node[] {};
		
		\node[circle,minimum size =7pt,draw,fill=black] (r1) at (1.25,0){};
		\node[circle,minimum size =7pt,draw,fill=black] (r2) at (2.25,0) {};
		\node at (3,0) (r3) {};
		
		\draw[D6] (1,.45)--(r1) (1,0.3)--(r1) (1,0.15)--(r1) (1,0)--(r1) (1,-0.15)--(r1) (1,-0.3)--(r1);
		\draw[D6] (r1)--(r2) node [midway,yshift=5pt] {\footnotesize 6};
\end{tikzpicture}}
$\xrightarrow{\displaystyle \mathfrak{d}_6}$\hspace{.75cm}
\raisebox{3pt}
{\begin{tikzpicture}[scale=1,baseline,every node/.style={inner sep=0,outer sep=0}]
		\tikzstyle{D6} = [solid,black];
		\tikzstyle{O8} = [densely dashed,black,thick];
		\tikzstyle{D8} = [solid,black,thick];
		\draw[O8] (-0.15,-.75)--(-0.15,.75) node[black,midway] {} node[black,midway] {};
		\draw[D8] (0,-.75)--(0,.75) node[black,midway, yshift=30pt]{\footnotesize 4};
         \draw[D8] (0.2,-.75)--(0.2,.75) node[black,midway, yshift=30pt]{\footnotesize 1};
		\draw[D8] (0.4,-.75)--(0.4,.75) node[black,midway, yshift=30pt]{\footnotesize 1};
		\draw[D8] (0.6,-.75)--(0.6,.75) node[black,midway, yshift=30pt]{\footnotesize 1};
		\draw[D8] (0.8,-.75)--(0.8,.75) node[black,midway, yshift=30pt]{\footnotesize 1};
		
		\draw[D6] (0.6-0.4,.45)--(1,.45) node[] {};
		\draw[D6] (0.6-0.2,.3)--(1,.3) node[] {};
		\draw[D6] (0.6,0.15)--(1,0.15) node[] {};
		\draw[D6] (0.8,0)--(1,0) node[] {};
		\draw[D6] (-0.15,-0.15)--(1,-0.15) node[] {};
		\draw[D6] (-0.15,-0.3)--(1,-0.3) node[] {};
		
		\node[circle,minimum size =7pt,draw,fill=black] (r1) at (1.25,0){};
		\node[circle,minimum size =7pt,draw,fill=black] (r2) at (2.25,0) {};
		\node at (3,0) (r3) {};
		
		\draw[D6] (1,.45)--(r1) (1,0.3)--(r1) (1,0.15)--(r1) (1,0)--(r1) (1,-0.15)--(r1) (1,-0.3)--(r1);
		\draw[D6] (r1)--(r2) node [midway,yshift=5pt] {\footnotesize 6};
\end{tikzpicture}}
	  \caption{Type IIA brane configurations engineering the flows $[2'^3] \xrightarrow{\mathfrak{d}_8} [2'^2,2] \xrightarrow{\mathfrak{d}_6} [4,2']$ for $k=6$. The $\mathfrak{so}$ weak-coupling summand gets broken as $\so{16} \to \so{12} \to\so{8}$ with extra $\mathfrak{su}$ or $\mathfrak{u}$ summands.}
	  \label{fig:IIAflow2'^32'^2242'}
	\end{figure}
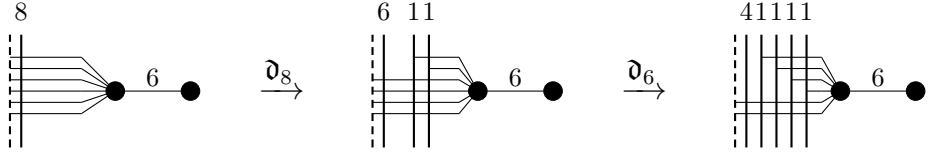
\begin{figure}
\centering
\raisebox{3pt}
{\begin{tikzpicture}[scale=1,baseline,every node/.style={inner sep=0,outer sep=0}]
		\tikzstyle{D6} = [solid,black];
		\tikzstyle{O8} = [densely dashed,black,thick];
		\tikzstyle{D8} = [solid,black,thick];
		\draw[O8] (-0.15,-.75)--(-0.15,.75) node[black,midway] {} node[black,midway] {};
		\draw[D8] (0,-.75)--(0,.75) node[black,midway, yshift=30pt]{\footnotesize 8};
		
		\draw[D6] (-0.15,.45)--(.8,.45) node[] {};
		\draw[D6] (-0.15,.3)--(.8,.3) node[] {};
		\draw[D6] (-0.15,0.15)--(.8,0.15) node[] {};
		\draw[D6] (-0.15,0)--(.8,0) node[] {};
		\draw[D6] (-0.15,-0.15)--(.8,-0.15) node[] {};
		\draw[D6] (-0.15,-0.3)--(.8,-0.3) node[] {};
		
		\node[circle,minimum size =7pt,draw,fill=black] (r1) at (1.25,0){};
		\node[circle,minimum size =7pt,draw,fill=black] (r2) at (2.25,0) {};
		\node at (3,0) (r3) {};
		
		\draw[D6] (.8,.45)--(r1) (.8,0.3)--(r1) (.8,0.15)--(r1) (.8,0)--(r1) (.8,-0.15)--(r1) (.8,-0.3)--(r1);
		\draw[D6] (r1)--(r2) node [midway,yshift=5pt] {\footnotesize 6};
		
\end{tikzpicture}}
$\xrightarrow{\displaystyle \mathfrak{d}_8}$\hspace{.75cm}
\raisebox{3pt}
{\begin{tikzpicture}[scale=1,baseline,every node/.style={inner sep=0,outer sep=0}]
		\tikzstyle{D6} = [solid,black];
		\tikzstyle{O8} = [densely dashed,black,thick];
		\tikzstyle{D8} = [solid,black,thick];
		\draw[O8] (-0.15,-.75)--(-0.15,.75) node[black,midway] {} node[black,midway] {};
		\draw[D8] (0,-.75)--(0,.75) node[black,midway, yshift=30pt]{\footnotesize 6};
		\draw[D8] (0.4,-.75)--(0.4,.75) node[black,midway, yshift=30pt]{\footnotesize 1};
		\draw[D8] (0.6,-.75)--(0.6,.75) node[black,midway, yshift=30pt]{\footnotesize 1};
		
		\draw[D6] (0.4,.45)--(1,.45) node[] {};
		\draw[D6] (0.6,.3)--(1,.3) node[] {};
		\draw[D6] (-0.15,0.15)--(1,0.15) node[] {};
		\draw[D6] (-0.15,0)--(1,0) node[] {};
		\draw[D6] (-0.15,-0.15)--(1,-0.15) node[] {};
		\draw[D6] (-0.15,-0.3)--(1,-0.3) node[] {};
		
		\node[circle,minimum size =7pt,draw,fill=black] (r1) at (1.25,0){};
		\node[circle,minimum size =7pt,draw,fill=black] (r2) at (2.25,0) {};
		\node at (3,0) (r3) {};
		
		\draw[D6] (1,.45)--(r1) (1,0.3)--(r1) (1,0.15)--(r1) (1,0)--(r1) (1,-0.15)--(r1) (1,-0.3)--(r1);
		\draw[D6] (r1)--(r2) node [midway,yshift=5pt] {\footnotesize 6};
\end{tikzpicture}}
$\xrightarrow{\displaystyle \mathfrak{a}_1}$\hspace{.75cm}
\raisebox{3pt}
{\begin{tikzpicture}[scale=1,baseline,every node/.style={inner sep=0,outer sep=0}]
		\tikzstyle{D6} = [solid,black];
		\tikzstyle{O8} = [densely dashed,black,thick];
		\tikzstyle{D8} = [solid,black,thick];
		\draw[O8] (-0.15,-.75)--(-0.15,.75) node[black,midway] {} node[black,midway] {};
		\draw[D8] (0,-.75)--(0,.75) node[black,midway, yshift=30pt]{\footnotesize 7};
		\draw[D8] (0.6,-.75)--(0.6,.75) node[black,midway, yshift=30pt]{\footnotesize 1};
		
		\draw[D6] (0.6,.45)--(1,.45) node[] {};
		\draw[D6] (0.6,.3)--(1,.3) node[] {};
		\draw[D6] (-0.15,0.15)--(1,0.15) node[] {};
		\draw[D6] (-0.15,0)--(1,0) node[] {};
		\draw[D6] (-0.15,-0.15)--(1,-0.15) node[] {};
		\draw[D6] (-0.15,-0.3)--(1,-0.3) node[] {};
		
		\node[circle,minimum size =7pt,draw,fill=black] (r1) at (1.25,0){};
		\node[circle,minimum size =7pt,draw,fill=black] (r2) at (2.25,0) {};
		\node at (3,0) (r3) {};
		
		\draw[D6] (1,.45)--(r1) (1,0.3)--(r1) (1,0.15)--(r1) (1,0)--(r1) (1,-0.15)--(r1) (1,-0.3)--(r1);
		\draw[D6] (r1)--(r2) node [midway,yshift=5pt] {\footnotesize 6};
\end{tikzpicture}}
	  \caption{Type IIA brane configurations engineering the flows $[2'^3] \xrightarrow{\mathfrak{d}_8} [2'^2,2] \xrightarrow{\mathfrak{a}_1} [2'^2,1^2]$ for $k=6$. The $\mathfrak{so}$ weak-coupling summand gets broken as $\so{16} \to \so{12} \to\so{14}$ with extra $\mathfrak{su}$ or $\mathfrak{u}$ summands.}
	  \label{fig:IIAflow2'^32'^222'^21^2}
	\end{figure}
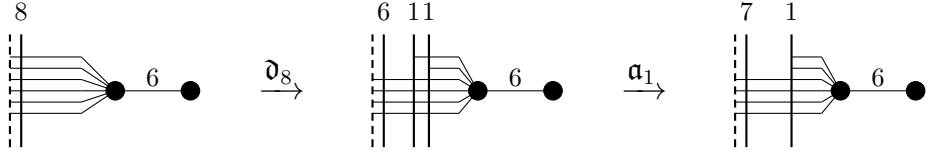
We immediately notice that, starting from $\so{16}$ and peeling off D8's, we can generate a few smaller-rank $\mathfrak{so}$'s, potentially together with other $\mathfrak{su}$ or $\mathfrak{u}$ summands: any smaller-rank $\mathfrak{so}$ summand is generated by sliding off to infinity a trapped segment of D6's that \emph{crosses} the O8 (see the first step in the flow of figure \ref{fig:IIAflow2'^32'^222'^21^2}), whereas the $\mathfrak{su}$ summands by those that \emph{do not} anymore,  i.e. after peeling off some D8's from the O-plane and sliding off to infinity trapped segments from those D8's to the first NS5 (see the second step in the flow of figure \ref{fig:IIAflow2'^32'^222'^21^2}). Moreover we know from the most general Kac label in \eqref{eq:longk} that the smallest-rank $\mathfrak{so}$ subalgebra we can ever hope to get is $\mathfrak{so}(8)$ (label $[4^{n_4},3^{n_3},2^{n_2},1^{n_1}, 2'^{n_{2'}}]$ for sufficiently high $k$, with $4,2'$ necessarily present). Looking at the Hasse diagram for $k=6$ in figure \ref{fig:flowk6} we recognize its presence with Kac label $[4,2']$, so we know that $\so{8}$ is one ``end'' to a partial Hasse diagram of $\so{16}$ nilpotent orbits.

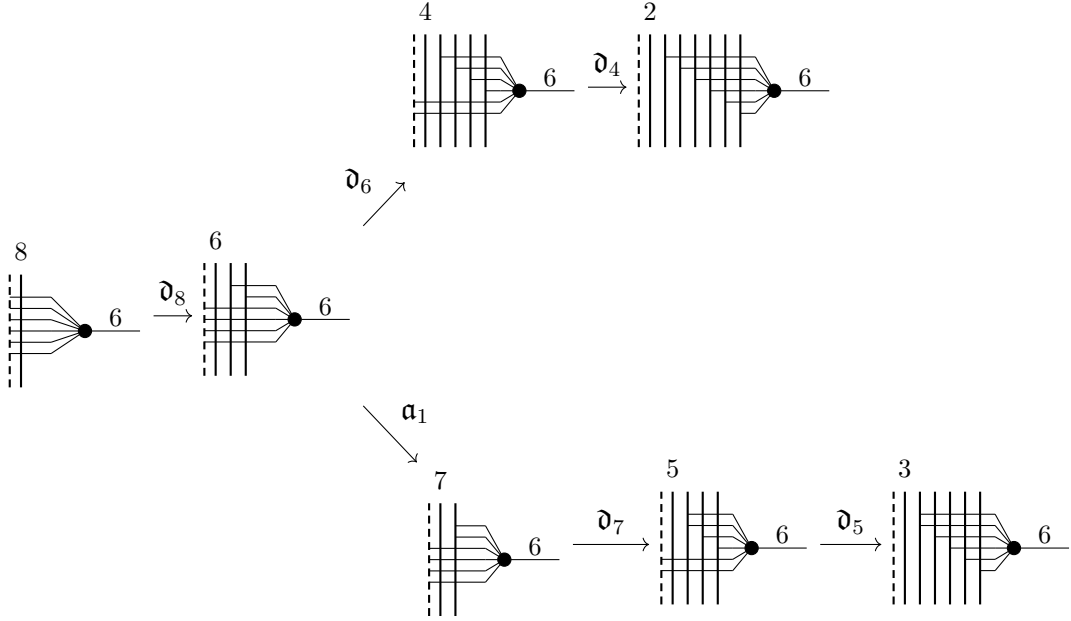
\begin{figure}[htbp!]
	\begin{tikzpicture}
		\matrix [row sep=15pt, column sep=15pt]
		{ 
			& & \node (n42) 
			{\raisebox{-30pt}
				{\begin{tikzpicture}[scale=1,baseline,every node/.style={inner sep=0,outer sep=0}]
						\tikzstyle{D6} = [solid,black];
						\tikzstyle{O8} = [densely dashed,black,thick];
						\tikzstyle{D8} = [solid,black,thick];
						\draw[O8] (-0.15,-.75)--(-0.15,.75) node[black,midway] {} node[black,midway] {};
						\draw[D8] (0,-.75)--(0,.75) node[black,midway, yshift=30pt]{\footnotesize 4};
						\draw[D8] (0.2,-.75)--(0.2,.75) node[black,midway, yshift=-30pt]{\footnotesize };
						\draw[D8] (0.4,-.75)--(0.4,.75) node[black,midway, yshift=-30pt]{\footnotesize };
						\draw[D8] (0.6,-.75)--(0.6,.75) node[black,midway, yshift=-30pt]{\footnotesize };
						\draw[D8] (0.8,-.75)--(0.8,.75) node[black,midway, yshift=-30pt]{\footnotesize };
						
						\draw[D6] (0.6-0.4,.45)--(1,.45) node[] {};
						\draw[D6] (0.6-0.2,.3)--(1,.3) node[] {};
						\draw[D6] (0.6,0.15)--(1,0.15) node[] {};
						\draw[D6] (0.8,0)--(1,0) node[] {};
						\draw[D6] (-0.15,-0.15)--(1,-0.15) node[] {};
						\draw[D6] (-0.15,-0.3)--(1,-0.3) node[] {};
						
						\node[circle,minimum size =5pt,draw,fill=black] (r1) at (1.25,0){};
						\node (r2) at (2,0) {};
						
						\draw[D6] (1,.45)--(r1) (1,0.3)--(r1) (1,0.15)--(r1) (1,0)--(r1) (1,-0.15)--(r1) (1,-0.3)--(r1);
						\draw[D6] (r1)--(r2) node [midway,yshift=5pt] {\footnotesize 6};
			\end{tikzpicture}}}; 
			& \node (n6)
			{
				\raisebox{-30pt}
				{\begin{tikzpicture}[scale=1,baseline,every node/.style={inner sep=0,outer sep=0}]
						\tikzstyle{D6} = [solid,black];
						\tikzstyle{O8} = [densely dashed,black,thick];
						\tikzstyle{D8} = [solid,black,thick];
						\draw[O8] (-0.55,-.75)--(-0.55,.75) node[black,midway] {} node[black,midway] {};
						\draw[D8] (-0.4,-.75)--(-0.4,.75) node[black,midway, yshift=30pt]{\footnotesize 2};
						\draw[D8] (-0.2,-.75)--(-0.2,.75) node[black,midway, yshift=-30pt]{\footnotesize };
						\draw[D8] (0,-.75)--(0,.75) node[black,midway, yshift=-30pt]{\footnotesize };
						\draw[D8] (0.2,-.75)--(0.2,.75) node[black,midway, yshift=-30pt]{\footnotesize };
						\draw[D8] (0.4,-.75)--(0.4,.75) node[black,midway, yshift=-30pt]{\footnotesize };
						\draw[D8] (0.6,-.75)--(0.6,.75) node[black,midway, yshift=-30pt]{\footnotesize };
						\draw[D8] (0.8,-.75)--(0.8,.75) node[black,midway, yshift=-30pt]{\footnotesize };
						
						\draw[D6] (-0.2,.45)--(1,.45) node[] {};
						\draw[D6] (0,.3)--(1,.3) node[] {};
						\draw[D6] (0.2,.15)--(1,.15) node[] {};
						\draw[D6] (0.4,.0)--(1,.0) node[] {};
						\draw[D6] (0.6,-0.15)--(1,-0.15) node[] {};
						\draw[D6] (0.8,-0.3)--(1,-0.3) node[] {};
						
						\node[circle,minimum size =5pt,draw,fill=black] (r1) at (1.25,0){};
						\node (r2) at (2,0) {};
						
						\draw[D6] (1,.45)--(r1) (1,0.3)--(r1) (1,0.15)--(r1) (1,0)--(r1) (1,-0.15)--(r1) (1,-0.3)--(r1);
						\draw[D6] (r1)--(r2) node [midway,yshift=5pt] {\footnotesize 6};
				\end{tikzpicture}}
			}; &\\
			\node (n23) {\raisebox{-30pt}
				{\begin{tikzpicture}[scale=1,baseline,every node/.style={inner sep=0,outer sep=0}]
						\tikzstyle{D6} = [solid,black];
						\tikzstyle{O8} = [densely dashed,black,thick];
						\tikzstyle{D8} = [solid,black,thick];
						\draw[O8] (0.25,-.75)--(0.25,.75) node[black,midway] {} node[black,midway] {};
						\draw[D8] (0.4,-.75)--(0.4,.75) node[black,midway, yshift=30pt]{\footnotesize 8};
						
						\draw[D6] (0.25,.45)--(.8,.45) node[] {};
						\draw[D6] (0.25,.3)--(.8,.3) node[] {};
						\draw[D6] (0.25,0.15)--(.8,0.15) node[] {};
						\draw[D6] (0.25,0)--(.8,0) node[] {};
						\draw[D6] (0.25,-0.15)--(.8,-0.15) node[] {};
						\draw[D6] (0.25,-0.3)--(.8,-0.3) node[] {};
						
						\node[circle,minimum size =5pt,draw,fill=black] (r1) at (1.25,0){};
						\node (r2) at (2,0) {};
						
						\draw[D6] (.8,.45)--(r1) (.8,0.3)--(r1) (.8,0.15)--(r1) (.8,0)--(r1) (.8,-0.15)--(r1) (.8,-0.3)--(r1);
						\draw[D6] (r1)--(r2) node [midway,yshift=5pt] {\footnotesize 6};
						
			\end{tikzpicture}}};
			
			& \node(n222) {\raisebox{-30pt}
				{\begin{tikzpicture}[scale=1,baseline,every node/.style={inner sep=0,outer sep=0}]
						\tikzstyle{D6} = [solid,black];
						\tikzstyle{O8} = [densely dashed,black,thick];
						\tikzstyle{D8} = [solid,black,thick];
						\draw[O8] (0.05,-.75)--(0.05,.75) node[black,midway] {} node[black,midway] {};
						\draw[D8] (0.2,-.75)--(0.2,.75) node[black,midway, yshift=30pt]{\footnotesize 6};
						\draw[D8] (0.4,-.75)--(0.4,.75) node[black,midway, yshift=-30pt]{\footnotesize };
						\draw[D8] (0.6,-.75)--(0.6,.75) node[black,midway, yshift=-30pt]{\footnotesize };
						
						\draw[D6] (0.4,.45)--(1,.45) node[] {};
						\draw[D6] (0.6,.3)--(1,.3) node[] {};
						\draw[D6] (0.05,0.15)--(1,0.15) node[] {};
						\draw[D6] (0.05,0)--(1,0) node[] {};
						\draw[D6] (0.05,-0.15)--(1,-0.15) node[] {};
						\draw[D6] (0.05,-0.3)--(1,-0.3) node[] {};
						
						\node[circle,minimum size =5pt,draw,fill=black] (r1) at (1.25,0){};
						\node (r2) at (2,0) {};
						
						\draw[D6] (1,.45)--(r1) (1,0.3)--(r1) (1,0.15)--(r1) (1,0)--(r1) (1,-0.15)--(r1) (1,-0.3)--(r1);
						\draw[D6] (r1)--(r2) node [midway,yshift=5pt] {\footnotesize 6};
			\end{tikzpicture}}}; & & &\\
			& &\node (n2211) {\raisebox{-30pt}
				{\begin{tikzpicture}[scale=1,baseline,every node/.style={inner sep=0,outer sep=0}]
						\tikzstyle{D6} = [solid,black];
						\tikzstyle{O8} = [densely dashed,black,thick];
						\tikzstyle{D8} = [solid,black,thick];
						\draw[O8] (0.25,-.75)--(0.25,.75) node[black,midway] {} node[black,midway] {};
						\draw[D8] (0.4,-.75)--(0.4,.75) node[black,midway, yshift=30pt]{\footnotesize 7};
						\draw[D8] (0.6,-.75)--(0.6,.75) node[black,midway, yshift=30pt]{\footnotesize };
						
						\draw[D6] (0.6,.45)--(1,.45) node[] {};
						\draw[D6] (0.6,.3)--(1,.3) node[] {};
						\draw[D6] (0.25,0.15)--(1,0.15) node[] {};
						\draw[D6] (0.25,0)--(1,0) node[] {};
						\draw[D6] (0.25,-0.15)--(1,-0.15) node[] {};
						\draw[D6] (0.25,-0.3)--(1,-0.3) node[] {};
						
						\node[circle,minimum size =5pt,draw,fill=black] (r1) at (1.25,0){};
						\node (r2) at (2,0) {};
						
						\draw[D6] (1,.45)--(r1) (1,0.3)--(r1) (1,0.15)--(r1) (1,0)--(r1) (1,-0.15)--(r1) (1,-0.3)--(r1);
						\draw[D6] (r1)--(r2) node [midway,yshift=5pt] {\footnotesize 6};
			\end{tikzpicture}}}; 
			& \node (n321) 
			{\raisebox{-30pt}
				{\begin{tikzpicture}[scale=1,baseline,every node/.style={inner sep=0,outer sep=0}]
						\tikzstyle{D6} = [solid,black];
						\tikzstyle{O8} = [densely dashed,black,thick];
						\tikzstyle{D8} = [solid,black,thick];
						\draw[O8] (0.05,-.75)--(0.05,.75) node[black,midway] {} node[black,midway] {};
						\draw[D8] (0.2,-.75)--(0.2,.75) node[black,midway, yshift=30pt]{\footnotesize 5};
						\draw[D8] (0.4,-.75)--(0.4,.75) node[black,midway, yshift=-30pt]{\footnotesize };
						\draw[D8] (0.6,-.75)--(0.6,.75) node[black,midway, yshift=-30pt]{\footnotesize };
						\draw[D8] (0.8,-.75)--(0.8,.75) node[black,midway, yshift=-30pt]{\footnotesize };
						
						\draw[D6] (0.4,.45)--(1,.45) node[] {};
						\draw[D6] (0.4,.3)--(1,.3) node[] {};
						\draw[D6] (0.6,0.15)--(1,0.15) node[] {};
						\draw[D6] (0.8,0)--(1,0) node[] {};
						\draw[D6] (0.05,-0.15)--(1,-0.15) node[] {};
						\draw[D6] (0.05,-0.3)--(1,-0.3) node[] {};
						
						\node[circle,minimum size =5pt,draw,fill=black] (r1) at (1.25,0){};
						\node (r2) at (2,0) {};
						
						\draw[D6] (1,.45)--(r1) (1,0.3)--(r1) (1,0.15)--(r1) (1,0)--(r1) (1,-0.15)--(r1) (1,-0.3)--(r1);
						\draw[D6] (r1)--(r2) node [midway,yshift=5pt] {\footnotesize 6};
			\end{tikzpicture}}}
			; & \node(n51) 
			{\raisebox{-30pt}
				{\begin{tikzpicture}[scale=1,baseline,every node/.style={inner sep=0,outer sep=0}]
						\tikzstyle{D6} = [solid,black];
						\tikzstyle{O8} = [densely dashed,black,thick];
						\tikzstyle{D8} = [solid,black,thick];
						\draw[O8] (-0.35,-.75)--(-0.35,.75) node[black,midway] {} node[black,midway] {};
						\draw[D8] (-0.2,-.75)--(-0.2,.75) node[black,midway, yshift=30pt]{\footnotesize 3};
						\draw[D8] (0,-.75)--(0,.75) node[black,midway, yshift=-30pt]{\footnotesize };
						\draw[D8] (0.2,-.75)--(0.2,.75) node[black,midway, yshift=-30pt]{\footnotesize };
						\draw[D8] (0.4,-.75)--(0.4,.75) node[black,midway, yshift=-30pt]{\footnotesize };
						\draw[D8] (0.6,-.75)--(0.6,.75) node[black,midway, yshift=-30pt]{\footnotesize };
						\draw[D8] (0.8,-.75)--(0.8,.75) node[black,midway, yshift=-30pt]{\footnotesize };
						
					     \draw[D6] (0,.45)--(1,.45) node[] {};
						\draw[D6] (0,.3)--(1,.3) node[] {};
						\draw[D6] (0.2,.15)--(1,.15) node[] {};
						\draw[D6] (0.4,.0)--(1,.0) node[] {};
						\draw[D6] (0.6,-0.15)--(1,-0.15) node[] {};
						\draw[D6] (0.8,-0.3)--(1,-0.3) node[] {};
						
						\node[circle,minimum size =5pt,draw,fill=black] (r1) at (1.25,0){};
						\node (r2) at (2,0) {};
						
						\draw[D6] (1,.45)--(r1) (1,0.3)--(r1) (1,0.15)--(r1) (1,0)--(r1) (1,-0.15)--(r1) (1,-0.3)--(r1);
						\draw[D6] (r1)--(r2) node [midway,yshift=5pt] {\footnotesize 6};
			\end{tikzpicture}}}; \\
		};
		
		\draw [->] (n42) -- (n6) node [midway, above] {$\mathfrak{d}_4$};
		\draw [->] (n2211) -- (n321) node [midway, above] {$\mathfrak{d}_7$};
		\draw [->] (n321) -- (n51) node [midway, above] {$\mathfrak{d}_5$};
		\draw [->] (n23) -- (n222) node [midway, above] {$\mathfrak{d}_8$};
		\draw [->] (n222) -- (n42) node [midway, above left] {$\mathfrak{d}_6$};
		\draw [->] (n222) -- (n2211) node [midway, above right] {$\mathfrak{a}_1$};
		
	\end{tikzpicture}
	\caption{Partial Hasse diagram of RG flows starting from Kac label $[2'^3]$ for $k=6$. In both termini of the bifurcation we end up with no $\mathfrak{so}$ summand, even though we started with the full $\so{16}$ at the other terminus.}
	\label{fig:so-web}
\end{figure}

On the other hand, looking at the flows in figures \ref{fig:IIAflow2'^32'^2242'} and \ref{fig:IIAflow2'^32'^222'^21^2}, one might expect the sequence of Higgsings to end when there are no D6's left in the zeroth segment, i.e. the $\mathfrak{usp}$ gauge algebra is empty.  If there are leftover D8's close to the O8 (i.e. crossing this empty segment), they must be responsible for a strong-coupling $\mathfrak{so}$ flavor summand. This expectation is indeed borne out. Looking back at the flows in \eqref{eq:flowsso}, we can push them further, and obtain:
\begin{subequations}
\begin{align}
&\underset{\usp{6}}{[2'^3]} \xrightarrow{\mathfrak{d}_8} \underset{\usp{4}}{[2'^2,2]} \xrightarrow{\mathfrak{d}_6} \underset{\usp{2}}{[4,2']} \xrightarrow{\mathfrak{d}_4} \underset{\usp{0}}{[6]}\ , \\
&\underset{\usp{6}}{[2'^3]} \xrightarrow{\mathfrak{d}_8} \underset{\usp{4}}{[2'^2,2]} \xrightarrow{\mathfrak{a}_1} \underset{\usp{4}}{[2'^2,1^2]} \xrightarrow{\mathfrak{d}_7} \underset{\usp{2}}{[3,2',1]}  \xrightarrow{\mathfrak{d}_5} \underset{\usp{0}}{[5,1]}\ ,
\end{align}
\end{subequations}
where we have indicated the gauge algebra on the $-1$ curve (i.e. zeroth segment of D6-branes crossing the O8) at each step. 
The difference in the $a$ anomaly for the last steps along the flow can be read off from \fref{tab:a6d}, and is given by:
\begin{align}
\begin{split}
	Δa_{[4,2'] → [6]} &= \frac{1152}{7} N^2-\frac{480}{7} N-\frac{19}{6}\ , \\
	Δa_{[2'^2,1^2] → [3,2',1]} &= \frac{1152}{7} N^2-\frac{480}{7} N+\frac{1423}{30}\ ,	\\
	Δa_{[3,2',1] → [5,1]} &= \frac{1152}{7} N^2-\frac{864}{7} N+\frac{10877}{210}\ .
\end{split}
\end{align}
These are again clearly positive for $N> k =6$. The electric quivers at the last step are given by \eqref{eq:651elec}:
\begin{subequations}
\begin{align}
&[6] \leftrightarrow \mathfrak{f}=\mathfrak{su}(6)\oplus \mathfrak{su}(3)\oplus\mathfrak{su}(2): & &[\SU(3)\times \SU(2)] \ {1}\ \overset{\mathfrak{su}(6)}{\underset{[N_\text{f}=6]}{2}}\ \overset{\mathfrak{su}(6)}{2} \cdots  \overset{\mathfrak{su}(6)}{2}\ [\SU(6)]\ ,  \\
&[5,1]\leftrightarrow \mathfrak{f}=\mathfrak{su}(5)\oplus \mathfrak{su}(4)\oplus\mathfrak{u}(1): & &[\SU(5)] \ {1}\ \overset{\mathfrak{su}(5)}{\underset{[N_\text{f}=4]}{2}}\ \overset{\mathfrak{su}(6)}{\underset{[N_\text{f}=1]}{2}}\ \overset{\mathfrak{su}(6)}{2} \cdots  \overset{\mathfrak{su}(6)}{2}\ [\SU(6)]\ .
\end{align}
\end{subequations}
We see that we have completely lost the weak-coupling $\mathfrak{so}$ summand. The Type IIA picture for both of these KP transitions is shown in \fref{fig:so-web}.
There are other $\mathfrak{d}_i$ transitions such that we end up with empty zeroth gauge algebra,
\begin{subequations}
\begin{align}
&\underset{\usp{2}}{[2',1^4]} \xrightarrow{\mathfrak{d}_7} \underset{\usp{0}}{[3,1^3]}\ ,  \quad \underset{\usp{2}}{[2,2',1^2]} \xrightarrow{\mathfrak{d}_6} \underset{\usp{0}}{[4,1^2]} \ ,
\end{align}
\end{subequations}
whose differences in $a$ anomalies along the flow are
\begin{align}
\begin{split}
	Δa_{[2',1^4] → [3,1^3]} &= \frac{1152 N^2}{7}-\frac{2784 N}{7}+\frac{114601}{210}	\ , \\
	Δa_{[2,2',1^2] → [4,1^2]} &= \frac{1152 N^2}{7}-\frac{1632 N}{7}+\frac{15153}{70}	\ ,
\end{split}
\end{align}
both of which are obviously positive for $N>k=6$.
The electric quivers are
\begin{subequations}
\begin{align}
&[3,1^3] \leftrightarrow \f=E_6 \oplus \su{2} \oplus \uu{1} : &&  [E_6] \  1 \ \overset{\mathfrak{su}(3)}{\underset{[N_\text{f}=2]}{2}}\ \overset{\mathfrak{su}(4)}{2}\ \overset{\mathfrak{su}(5)}{2}\ \overset{\mathfrak{su}(6)}{\underset{[N_\text{f}=1]}{2}} \cdots  \overset{\mathfrak{su}(6)}{2}\ [\SU(6)] \ ,  \\
&[4,1^2] \leftrightarrow \f=\so{10} \oplus \su{3} \oplus \uu{1} : && [\SO(10)] \  \overset{}{1}\ \overset{\mathfrak{su}(4)}{\underset{[N_\text{f}=3]}{2}}\ \overset{\mathfrak{su}(5)}{2}\ \overset{\mathfrak{su}(6)}{\underset{[N_\text{f}=1]}{2}} \cdots  \overset{\mathfrak{su}(6)}{2}\ [\SU(6)] \ .
\end{align}
\end{subequations}
Here we notice the second effect mentioned above: sometimes we can have an empty $\mathfrak{usp}$ gauge algebra (in position zero) and still an $\mathfrak{so}$ flavor summand, which is necessarily strongly coupled. Since we cannot perform other Higgsings in the zeroth segment, the $\mathfrak{d}_i$ transitions must end here, and to continue flowing we are forced to use $\mathfrak{a}_i$ transitions.  Another example of this is
\begin{equation}
[4,2] \leftrightarrow \f=\so{10} \oplus \su{2} \oplus \su{2} \oplus \uu{1} : [\SO(10)] \  \overset{}{1}\ \overset{\mathfrak{su}(4)}{\underset{[N_\text{f}=2]}{2}}\ \overset{\mathfrak{su}(6)}{\underset{[N_\text{f}=2]}{2}} \cdots  \overset{\mathfrak{su}(6)}{2}\ [\SU(6)] \ .
\end{equation}
Both Kac labels with strong-coupling $\mathfrak{so}$ summands are reached via $\mathfrak{d}$-type KP transitions. Then, since the product of this transition is an empty zeroth segment, we can no more use trapped D6's (in that segment) to Higgs the gauge algebra, so this is an endpoint for $\mathfrak{d}$-type KP transitions, and the flows lying below it in the Hasse diagram necessarily come from $\mathfrak{a}$-type KP transitions.  (Incidentally, $[4,2]$ offers an example where the $\uu{1}$ summand in $\f$ is clearly delocalized in the F-theory geometry, not being associated with any matter representation. Therefore it must be the ABJ-free combination under which matter in the plateau is charged. The same holds for $[4,2']$ and $[2'^2,2]$ in \eqref{eq:someSOexam}.)

The above observations \emph{i)} explain why it is \emph{not} possible to embed the full Hasse diagram of nilpotent orbits of $\so{16}$ inside the homomoprhisms Hasse diagram for $k=6$ (or higher $k$, since $\so{16}$ is the biggest $\mathfrak{so}$ summand we can ever generate from \eqref{eq:longk} and $k=6$ is already large enough to accommodate it via one or more of its Kac labels); and \emph{ii)} explain in terms of D6 Higgsings in IIA the $\mathfrak{d}$-type KP transitions that do appear in the homomorphisms Hasse diagram. In light of the above two points, we have decided not to highlight the few $\so{16}$ orbits that do appear in the homomorphisms Hasse.

\subsection{\texorpdfstring{The O8$^*$-plane}{The O8*-plane}}
\label{sub:6O8*}

\begin{figure}[ht!]
\centering
\begin{tikzpicture}[scale=1,baseline]
				\draw[fill=black] (0,0) circle (0.075cm) node[xshift=-0.4cm] {$\tfrac{1}{2} {\fontsize{0.25pt}{0.5pt}\selectfont \yng(1,1,1)}$};
				\draw[fill=black] (1,0) circle (0.075cm);
				\draw[fill=black] (2,0) circle (0.075cm);
				\draw[fill=black] (3,0) circle (0.075cm);
				
				\draw[solid,black,thick] (0,0)--(1,0) node[black,midway,xshift=0.2cm,yshift=0.2cm] {\footnotesize $6$};
				\draw[solid,black,thick] (1,0)--(2,0) node[black,midway,yshift=0.2cm] {\footnotesize $6$};
				\draw[loosely dotted,black,thick] (2,0)--(3,0) node[black,midway, ] {};
				\draw[solid,black,thick] (3,0)--(4,0) node[black,midway,yshift=0.2cm] {\footnotesize $6$};
				
				\draw[dashed,black,very thick] (0,-.5)--(0,.5) node[black,midway,yshift=+.75cm] {\footnotesize O8$^-$};
				\draw[solid,black,very thick] (0.5,-.5)--(0.5,.5) node[black,midway, xshift =0cm, yshift=+.75cm] {\footnotesize $8$};
\end{tikzpicture}
$\xrightarrow{\text{non-perturbative resolution}}$
\begin{tikzpicture}[scale=1,baseline]
				\draw[fill=black] (0,0) circle (0.075cm) node[xshift=-0.4cm] {$\tfrac{1}{2} {\fontsize{0.25pt}{0.5pt}\selectfont \yng(1,1,1)}$};
				\draw[fill=black] (1,0) circle (0.075cm);
				\draw[fill=black] (2,0) circle (0.075cm);
				\draw[fill=black] (3,0) circle (0.075cm);
				
				\draw[solid,black,thick] (0,0)--(1,0) node[black,midway,xshift=0.2cm,yshift=0.2cm] {\footnotesize $6$};
				\draw[solid,black,thick] (1,0)--(2,0) node[black,midway,yshift=0.2cm] {\footnotesize $6$};
				\draw[loosely dotted,black,thick] (2,0)--(3,0) node[black,midway, ] {};
				\draw[solid,black,thick] (3,0)--(4,0) node[black,midway,yshift=0.2cm] {\footnotesize $6$};
				
				\draw[dotted,black,very thick] (0,-.5)--(0,.5) node[black,midway,yshift=+.75cm] {\footnotesize O8$^*$};
				\draw[solid,black,very thick] (0.5,-.5)--(0.5,.5) node[black,midway, xshift =0cm, yshift=+.75cm] {\footnotesize $9$};
\end{tikzpicture}
\caption{Type IIA engineering of the orbi-instanton for $k=6$ and Kac label $[3'^2]$. \textbf{Left:} naive expectation. \textbf{Right:} non-perturbative configuration in Type I' which involves the O8$^*$ \cite{Gorbatov:2001pw}.}
\label{fig:IIAk63'2}
\end{figure}
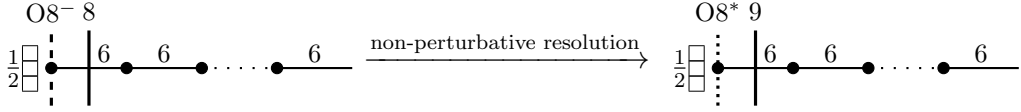

Consider Kac label $[3'^2]$: it preserves $\f=\su{9}$. This may sound surprising at first, as we cannot generate such a summand from the weak-coupling symmetry on a stack of D8's: even if all 8 of them sit together in one segment, they would generate at most $\su{8}$. One may then hope that its origin is inherently strongly coupled, so that the IIA origin is unclear. However this cannot be the case, as the electric quiver has a nonempty zeroth gauge algebra:
\begin{equation}
[3'^2] \leftrightarrow \mathfrak{f}=\mathfrak{su}(9): \quad \overset{\mathfrak{su}(6)}{\underset{[N_{\fontsize{0.25pt}{.5pt}\selectfont \frac{1}{2}\yng(1,1,1)}=1, N_\text{f}=9]}{1}}\ \overset{\mathfrak{su}(6)}{2}\ \overset{\mathfrak{su}(6)}{2} \cdots  \overset{\mathfrak{su}(6)}{2}\ [\SU(6)]\ .
\end{equation}
Because of the $\mathfrak{su}$ algebra on the zeroth segment, we know there must be a half-NS5 stuck on the O8, so that the IIA setup is the one in figure \ref{fig:IIAk63'2}, which is surprisingly similar to the one on the right of figure \ref{fig:IIAk64'2'2'3}, were it not for the presence of a half-hypermultiplet in the \emph{three-index} antisymmetric representation of $\su{6}$  due to which the gauge anomaly cancellation condition gets modified nontrivially with respect to familiar $\mathfrak{su}$ cases without such a representation.  (Notice that the $\fontsize{0.25pt}{.5pt}\selectfont \yng(1,1,1)$ of $\su{6}$ is pseudo-real and as such the internal symmetry of one half-hypermultiplet in this representation is trivial, $\so{2 \cdot \frac{1}{2}} = \so{1} = \emptyset$.)
For the zeroth segment that condition (which is written down e.g. in \cite[Eq. (3.10)]{Apruzzi:2017nck}) reads:
\begin{equation}
2 \cdot 6 = \frac{1}{2}(-6) + 6 +M\ ,
\end{equation}
where $-6$ is the quartic Casimir of the $\fontsize{0.25pt}{.5pt}\selectfont \yng(1,1,1)$ of $\su{6}$ \cite[Sec. 3]{okubo-patera}, and $M$ is the number of fundamental flavors (i.e. D8-branes) in that segment needed to cancel the gauge anomaly, equivalently to satisfy D6 charge conservation. Solving this simple equation gives us $M=9$, that is we must have 9 D8-branes there, not 8. This is only possible if the O-plane is of the type O8$^*$ rather than O8$^-$, given the former has precisely $-9$ D8 charge. 

Thus we discover that harmless-looking Kac labels can in fact hide surprising facts in IIA, or better the non-perturbative completion of Type I' which contains the O8$^*$ \cite{Seiberg:1996bd}.  One may now wonder whether it is possible for $k=6$ to generate the full $\su{9}$ nilpotent orbits Hasse diagram by successively peeling off D8's from the 9 stack, similarly to what we saw in section \ref{sub:su6} for $\su{6}$. However the $\su{9}$ summand cannot be Higgsed further via $\mathfrak{a}_i$ or $A_i$ KP transitions (as would be required in the $\su{9}$ nilpotent orbits Hasse diagram \cite{KP1}). Instead, the next Higgs branch RG flow is implemented by a $\mathfrak{d}_8$ KP transition to Kac label $[2'^2,2]$, whose electric quiver is
\begin{equation}
[2'^2,2] \leftrightarrow \f = \so{12}\oplus \su{2} \oplus \uu{1}: \quad  \overset{\mathfrak{usp}(4)}{\underset{[N_\text{f}=6]}{1}}\ \overset{\mathfrak{su}(6)}{\underset{[N_\text{f}=2]}{2}}\ \overset{\mathfrak{su}(6)}{2} \cdots  \overset{\mathfrak{su}(6)}{2}\ [\SU(6)]\ .
\end{equation}
This is not the end of the story though. In fact $[3'^2]$ obviously generalizes to $[3'^{k/3}]$ for any $k$ which is a multiple of 3: only those $k$ can ever preserve an $\su{9}$. We will see in section \ref{sub:genO8*} that depending on the actual value of $k/3$ we can have configurations with or without the O8$^*$. For a configuration with the O8$^*$ (i.e. with $\su{9}$), for $k$ a multiple of 3 and $k\geq 9$, we will see that the homomorphisms Hasse diagram can and indeed \emph{does} accommodate the full $\su{9}$ nilpotent orbits Hasse diagram, as expected. As a preview, consider Kac label $[3'^3]$ for $k=9$; its electric quiver is
\begin{equation}
[3'^3] \leftrightarrow \mathfrak{f}=\mathfrak{su}(9): \quad 1\ \overset{\su{9}}{\underset{[N_\text{f}=9]}{2}}\ \overset{\su{9}}{2} \cdots  \overset{\su{9}}{2}\ [\SU(9)]\ .
\end{equation}
The $\su{9}$ has weak-coupling origin, and as such can be further Higgsed via $\mathfrak{a}_i$ or $A_i$ KP transitions which correspond, in the type IIA engineering, to peeling off branes from this stack of 9 D8-branes. This way we can generate the full Hasse diagram of nilpotent orbits of $\su{9}$.

\subsection{Combining small instanton transitions and Higgs branch RG flows}
\label{sub:6N}

Finally, let us comment on the existence of ``mixed'' flows and how to detect them. By this we mean that we keep the type of the orbi-instanton fixed (A-type throughout this paper), the order $k$ of the M-theory orbifold, but we vary both the boundary condition (i.e. we flow among Kac labels) \emph{and} $N$, i.e. we allow for small instanton transitions reducing the number of tensor multiplets in the theory.

This is the approach followed in \cite{Frey:2018vpw} and \cite{Giacomelli:2022drw}; both references analyzed a few examples for low $k$ and proposed a hierarchy of Higgs branch RG flows between Kac labels which also involves some (or all) $N\to N-1$ transitions (until $N=1$). In appendix \ref{app:RG} we will see explicitly how the proposed hierarchy of \cite[Sec. 4]{Frey:2018vpw} for $k=4$ fits into our formalism.

Here we content ourselves with explaining how to transition from the homomorphisms Hasse diagram for fixed $k$ at number $N$ of tensor multiplets, to that at $N-1$ (so that this generates the full list of diagrams all the way to $N=1$). To do this, we subtract from each node of the diagram at $N$ all nodes of the same diagram at $N-1$,  and find all the nodes (i.e. Kac labels) in the former that allow a quiver subtraction with \emph{all} nodes of the latter. This defines the Kac labels from where (by doing a small instanton transition $N\to N-1$) we can land on the top label(s) of the next diagram. Activating Higgs branch RG flows one can then generate the full $N-1$ Hasse diagram, and so on.

For each $k$ we color such nodes in green. In the case of $k=6$ the lowest node of the $N$ diagram from which we can jump to the $N-1$ diagram is $[6]$ as shown in \fref{fig:flowk6}. A feature of the RG flows that first appears at $k=6$ is that there are two possible starting points for the flow between orbi-instantons, and it is not possible to flow from one to the other. Naturally, to get the full RG flow at $N-1$ we must be able to flow to \emph{both} of these ``top nodes'', and our green highlighted nodes are precisely these.  Among the flow hierarchies that we have explicitly computed, we see the same behavior persists for all $k \geq 6$.

\subsection{Legend of decorations}
\label{sub:6legend}

To sum up, the decorations appearing in the hierarchy of RG flows for $k=6$ at fixed $N$ in figure \ref{fig:flowk6} are as follows.
\begin{itemize}
\item Each node represents an orbi-instanton, labeled by its Kac label.
\item All edges represent allowed Higgs branch RG flows (i.e. $\Delta a>0$) and are labeled by the product of quiver subtraction between the 3d magnetic quivers associated with the 6d electric ones (giving the tensor branch description of the orbi-instantons). These subtractions can in turn be understood as KP transitions.
\item Red nodes denote Kac labels which correspond to the nilpotent orbits (i.e. integer partitions) of $\su{6}$.
\item Red dashed edges correspond to KP transitions between such nilpotent orbits, i.e. to the $\su{6}$ Hasse diagram.
\item Green nodes denote Kac labels (i.e. SCFTs) from which, by performing a small instanton transition $N\to N-1$,  we can jump to the homomorphisms Hasse diagram of orbi-instantons for the same $k$ but with $N-1$ tensor multiplets (rather than $N$ as we started from).
\end{itemize}

\section{\texorpdfstring{Subtleties for high $k$}{Subtleties for high k}}
\label{sec:highk}

In this section we assume that $k$ be large, i.e. we want to describe phenomena which only arise for sufficiently high $k$ and are (partially) absent in the $k=6$ case study of section \ref{sec:k=6}. We expect all interesting phenomena to take place already in the window $6 \leq k \lesssim 20$, so that one would not learn anything new by taking $k \gg 1$. We have checked that this is indeed the case with an algorithmic scan, and we have collected the $\text{Hom}(\zz_k,E_8)$ Hasse diagrams for $k=2,\ldots,20$ in appendix \ref{app:RG}. %
\begin{figure}[ht!]
	\centering
	\vspace{1cm}
	\def\svgwidth{0.8\columnwidth}
	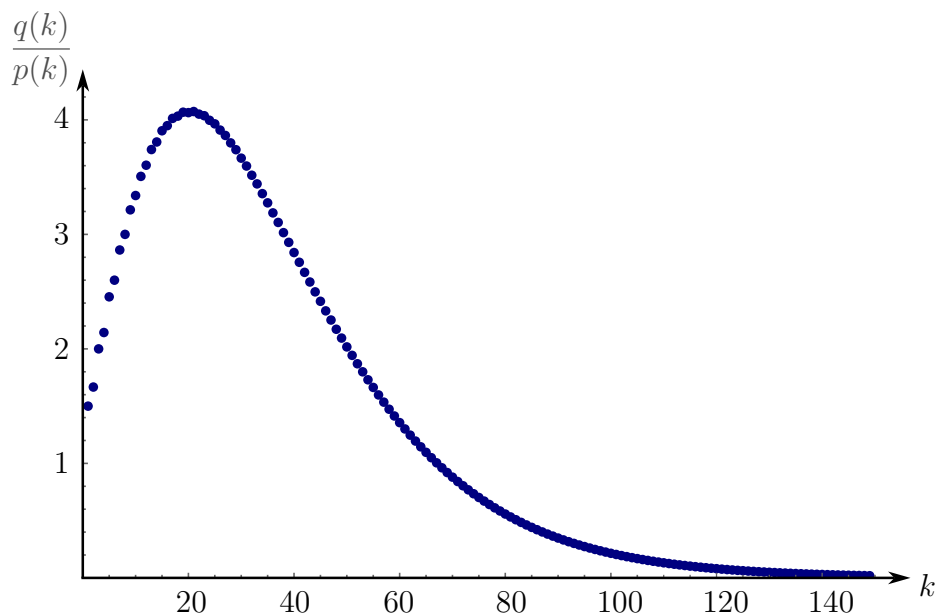
	\caption{Ratio of the number of Kac labels $q(k)$ and the number of integer partitions $p(k)$ for $k=2,…,150$.}
	\label{fig:ratio}
\end{figure}%

A very heuristic argument for the above statement is the following. One can compute the ratio between the number $q(k)$ of Kac labels of $k$ and the number $p(k)$ of integer partitions of $k$, whose generating functions read respectively
\begin{subequations}
\begin{align}
\mathcal{Z}_\text{Kac}(k) &= \sum_{k=0}^\infty q(k) x^k = \frac{1}{(1-x) \left(1-x^2\right)² \left(1-x^3\right)² \left(1-x^4\right)² (1-x^5)(1-x^6)}\ ,\\
\mathcal{Z}_\text{partitions}(k) &=  \sum_{k=0}^\infty p(k) x^k = \prod_{n=1}^\infty\frac{1}{1-x^n} = \frac{1}{(x;x)_\infty}\ ,
\end{align}
\end{subequations}
with $(q;q)_\infty$ the $q$-Pochhammer symbol.  Plotting this ratio for $k=2,\ldots,150$ in \fref{fig:ratio}, we see that it peaks around $k=20$, reaches $1$ around $k=70$ and quickly goes to zero afterwards.  Since the number of Kac labels $q(k)$ for a given $k$ increases rapidly up to $k \sim 20$ compared to that of integer partitions $p(k)$ of $k$, one could heuristically expect most new features to appear in this interval.

\subsection{Embedding $\mathfrak{su}$ Hasse diagrams}
\label{sub:smallhasse}

The biggest $\mathfrak{su}$ summand preserved by the general Kac label \eqref{eq:longk} is $\su{9}$ when $k$ is a multiple of 3, and $\su{8}$ when it is not.  In light of the results of section \ref{sub:su6},  it is then natural to ask whether we can always embed the full $\su{9}$ or $\su{8}$ nilpotent orbits Hasse diagram inside the homomorphisms Hasse diagram for any sufficiently high $k$. 
From $k\geq 7$ onward the one-to-one correspondence between a subset of Kac labels and integer partitions of $k$ ceases to exist, because the former only use integers up to $6$ whereas the latter up to $k$. One may then think that this introduces a difficulty in identifying the Hasse diagram of $\su{8}$ or $\su{9}$ nilpotent orbits inside the full homomorphisms Hasse diagram. However one can resort once again to the Type IIA intuition whereby the nilpotent orbits are one-to-one with the ways in which we can peel off D8-branes from a ``maximally populated'' stack. Let us work out the cases $k=7,8,9$ explicitly before moving on to a more general discussion for $k > 9$.

\subsubsection{\texorpdfstring{Case $k=7$}{Case k=7}}

The maximal stack of D8's is realized when $7$ D8's sit together. The associated Kac label and electric quiver are
\begin{equation}\label{eq:k7start}
	[4',3'] \leftrightarrow \mathfrak{f}=\su{7}\oplus \su{2}  \oplus \mathfrak{u}(1) : \quad [\SU(2)]\  1\ \overset{\su{7}}{\underset{[N_\text{f}=7]}{2}} \overset{\su{7}}{2} \cdots  \overset{\su{7}}{2}\ [\SU(7)]\ .
\end{equation}
Rearranging these $7$ D8's generates quivers whose gauge algebra ranks can then be partial-ordered to generate a Hasse diagram, which terminates at 
\begin{equation}\label{eq:k7end}
	[1^7] \leftrightarrow \mathfrak{f}=E_8 : \quad [E_8]\ 1\  \overset{\su{1}}{2}\ \overset{\su{2}}{2}\ \overset{\su{3}}{2}  \cdots  \overset{\su{7}}{\underset{[N_\text{f}=1]}{2}}\ \overset{\su{7}}{2}  [\SU(7)]\ .
\end{equation}
The differences of the ranks of the gauge algebras $\left\{r_{i+1}-r_i\right\}_i$ for each theory are one-to-one with partitions of the integer $7$, thus generating the Hasse diagram of $\su{7}$ nilpotent orbits. For example,  Kac label in \fref{eq:k7start} corresponds to the partition $[7]$, and \fref{eq:k7end} corresponds to $[1^7]$. This forms a part of the full hierarchy of RG flows of the $k=7$ case, colored in red (nodes and dashed edges). See figure \ref{fig:k78}.

\subsubsection{\texorpdfstring{Case $k=8$}{Case k=8}}
For $k=8$, a stack of $8$ D8's can be realized in the electric quiver of the following Kac label:
\begin{equation}\label{eq:k8start}
	[4'^2] \leftrightarrow \mathfrak{f}=\su{8} \oplus \su{2} : \quad [\SU(2)]\ 1\ \overset{\su{8}}{\underset{[N_\text{f}=8]}{2}} \overset{\su{8}}{2} \cdots  \overset{\su{8}}{2}\ [\SU(8)]\ .
\end{equation}
The differences in ranks is again in one-to-one correspondence with the partitions of $8$ and can be partial-ordered to obtain the full Hasse diagram of nilpotent orbits of $\su{8}$. As before, this forms a part of the full RG flow and is highlighted in red in figure \ref{fig:k78}.

\subsubsection{\texorpdfstring{Case $k=9$}{Case k=9}}
Kac label $[3'^3]$ realizes a stack of $9$ D8's; Kac label and the electric quiver are:
\begin{equation}\label{eq:k9start}
	[3'^3] \leftrightarrow \mathfrak{f}=\su{9} : \quad 1\ \overset{\su{9}}{\underset{[N_\text{f}=9]}{2}}\  \overset{\su{9}}{2} \cdots  \overset{\su{9}}{2}\ [\SU(9)]\ .
\end{equation}
Similarly to $k=7,8$, the differences in ranks correspond to the partitions of $9$ and generate the Hasse diagram of nilpotent orbits of $\su{9}$. This is highlighted in red in figure \ref{fig:k9}.

\subsubsection{\texorpdfstring{Generic $k$}{Generic k}}
\label{sec:generick_nilpotent_hasse}

Unlike for $k \le 9$, it is no longer possible to have a stack of $k$ D8's for $k \ge 10$. So one must carefully identify the maximal stack of $m$ D8's to find the embedding of the Hasse diagram of nilpotent orbits of $\su{m}$ in the full RG flows hierarchy. To this end, we propose the following algorithm.
\begin{enumerate}
\item If $k = 0 \mod 3$, the Kac label $[3'^{k/3}]$ preserves $\su{9}$. This corresponds to three different electric quivers depending on three discrete choices $l=0,1,2$, where $k/3 \equiv l \mod 3$:
	\begin{subequations}\label{eq:su9starting}
		\begin{align}
			&k/3 = 0 \mod 3: && \overset{}{1}\ \overset{\su{9}}{2}\  \overset{\su{18}}{2}\ \cdots  \overset{\su{k-9}}{2}\ \overset{\su{k}}{\underset{[N_\text{f}=9]}{2}} \cdots \ \overset{\su{k}}{2}\ [\SU(k)]\ , \label{eq:kmult33'k/3a}\\
			&k/3 = 1 \mod 3: && \overset{\su{3}}{\underset{[N_{\fontsize{0.5pt}{1pt}\selectfont \yng(1,1)}=1]}{1}}\ \overset{\su{12}}{2}\  \overset{\su{21}}{2}\ \cdots  \overset{\su{k-9}}{2}\ \overset{\su{k}}{\underset{[N_\text{f}=9]}{2}} \cdots \ \overset{\su{k}}{2}\ [\SU(k)]\ , \label{eq:kmult33'k/3b}\\
			&k/3 = 2 \mod 3: && \overset{\mathfrak{su}(6)}{\underset{[N_{\fontsize{0.25pt}{.5pt}\selectfont \frac{1}{2}\yng(1,1,1)}=1]}{1}}\ \overset{\mathfrak{su}(15)}{2}\  \overset{\su{24}}{2}\ \cdots  \overset{\mathfrak{su}(k-9)}{2}\ \overset{\mathfrak{su}(k)}{\underset{[N_\text{f}=9]}{2}}\ \cdots\ \overset{\mathfrak{su}(k)}{2}\ [\SU(k)] \ .
			\label{eq:kmult33'k/3c}
		\end{align}
	\end{subequations}
	All of them contain a maximal stack of $m=9$ D8's (i.e. there is an O8$^*$ in the Type IIA setup). 	By subsequently peeling off  D8-branes from the stack of 9 we can generate the $\su{9}$ Hasse diagram, akin to what was done for $k=6$ in table \ref{tab:k6Kacquiv}.
	
\item If $k \neq 0 \mod 3$, then $\su{8}$ is preserved by a stack of $m=8$ D8's on a $-2$ curve. The electric quivers and the corresponding Kac labels can be further identified as follows:
	\begin{enumerate}
	\item[\emph{a)}] if $k = 0 \mod 2$,  the Kac label that preserves $\su{8}$ is  $[4'^x,2'^{k/2-2x}]$, where $x=2,3,\ldots,\lfloor k/4 \rfloor$:		
	
	\begin{enumerate}
	\item[\emph{i)}] if $x= 0 \mod 2$, the electric quiver reads
		\begin{equation}\label{eq:generick-2ai}
		\overset{\usp{k-4x}}{1}\ \overset{\su{k-4x+8}}{2}\ \cdots \overset{\su{k}}{\underset{[N_\text{f}=8]}{2}} \cdots \ \overset{\su{k}}{2}\ [\SU(k)]\ .
		\end{equation}
		In this case, there is a second SCFT corresponding to $[3'^x,2'^{k/2-3x/2}]$ which has the same electric quiver, and differs only by the $6$d theta angle as will be discussed in section \ref{sub:theta}.

	\item[\emph{ii)}] If $x \neq 0 \mod 2$,  the electric quiver reads instead
		\begin{equation}\label{eq:generick-2aii}
		\overset{\su{k-4x+4}}{\underset{[N_{\fontsize{0.5pt}{1pt}\selectfont \yng(1,1)}=1]}{1}}\ \overset{\su{k-4x+12}}{2}\ \cdots \overset{\su{k}}{\underset{[N_\text{f}=8]}{2}} \cdots \ \overset{\su{k}}{2}\ [\SU(k)]\ .
		\end{equation}
	\end{enumerate}

	\item[\emph{b)}] If $k \neq 0 \mod 2$,  the Kac label is $[3'^3,2'^{(k-9)/2}]$, and the electric quiver reads
		\begin{align}\label{eq:generick-2b}
	 \overset{\su{k-8}}{1}\ \overset{\su{k}}{\underset{[N_\text{f}=8]}{2}}\ \overset{\su{k}}{2} \cdots \ \overset{\su{k}}{2}\ [\SU(k)]\ .
		\end{align}
	\end{enumerate}

\item Additionally, if $k = 3 \text{ or } 6 \mod 8$, there is an alternate way to realize an $\su{8}$ flavor algebra with a Kac label of the form $[3'^x, 2'^y]$. 
	\begin{enumerate}
	\item[\emph{a)}] If $k = 3 \mod 8$, $4x - k = 1 \mod 3$ and the 6d quiver is 
	\begin{equation}\label{eq:generick-3a}
		\overset{\su{3}}{\underset{[N_{\fontsize{0.5pt}{1pt}\selectfont \yng(1,1)}=1]}{1}}\ \overset{\su{11}}{2}\ \overset{\su{19}}{2}\ \cdots \overset{\su{k-8}}{2}\  \overset{\su{k}}{\underset{[N_\text{f}=8]}{2}}\ \overset{\su{k}}{2} \cdots \ \overset{\su{k}}{2}\ [\SU(k)]\ ;
	\end{equation}
	\item[\emph{b)}] if $k = 6 \mod 8$, $4x - k = 2 \mod 3$ and the 6d quiver is 
	\begin{equation}\label{eq:generick-3b}
		\overset{\mathfrak{su}(6)}{\underset{[N_{\fontsize{0.25pt}{.5pt}\selectfont \frac{1}{2}\yng(1,1,1)}=1]}{1}}\ \overset{\su{14}}{2}\ \overset{\su{22}}{2}\ \cdots \overset{\su{k-8}}{2}\  \overset{\su{k}}{\underset{[N_\text{f}=8]}{2}}\ \overset{\su{k}}{2} \cdots \ \overset{\su{k}}{2}\ [\SU(k)]\ .
	\end{equation}
	\end{enumerate}
These are similar to the theories in \fref{eq:su9starting}, but have $8$ D8's instead of $9$.
\end{enumerate}
It is clear from the above that there can be multiple starting points for embedding the $\su{8}$ nilpotent orbits Hasse diagram into the full RG flow hierarchy for large-enough $k$.  

As an example, the $\su{8}$ Hasse diagram for $k=10$ starts from two SCFTs that only differ by the $\theta$ angle on the tensor branch (as will be discussed at length in section \ref{sub:theta}). The electric quiver for both is the one in \eqref{eq:generick-2ai}:
\begin{equation}
\begin{rcases} [4'^2,2']_{\theta=0}\\  [3'^2,2'^2]_{\theta=\pi} \end{rcases}: \quad \overset{\usp{2}}{1}\ \overset{\su{10}}{\underset{[N_\text{f}=8]}{2}}\  \overset{\su{10}}{2} \cdots  \overset{\su{10}}{2}\ [\SU(10)]\ .
\end{equation}
The starting point $[3'^2,2'^2]$ is colored in blue in figure \ref{fig:k10}, whereas $[4'^2,2']$ in red (as the $\su{8}$ Hasse). For $k=14$, there are four theories that act starting point for $\su{8}$:
\begin{itemize}
	\item two theories differing by $\theta$ angle,  whose electric quiver is again the one in \eqref{eq:generick-2ai}:
	\begin{equation}\label{eq:k14theta1}
\begin{rcases} [4'^2,2'^3]_{\theta=0}\\  [3'^2,2'^4]_{\theta=\pi} \end{rcases}: \quad \overset{\usp{6}}{1}\ \overset{\su{14}}{\underset{[N_\text{f}=8]}{2}}\  \overset{\su{14}}{2} \cdots  \overset{\su{14}}{2}\ [\SU(14)]\ .
	\end{equation}
	This $\su{8}$ Hasse diagram is colored in magenta in figure \ref{fig:k14}.
	\item One theory with electric quiver given by \eqref{eq:generick-2aii}:
	\begin{equation}\label{eq:k14theta2}
		[4'^3,2']:\quad \overset{\su{6}}{\underset{[N_{\fontsize{0.5pt}{1pt}\selectfont \yng(1,1)}=1]}{1}}\ \overset{\su{14}}{\underset{[N_\text{f}=8]}{2}}\ \overset{\su{14}}{2}\ \cdots \overset{\su{14}}{2}\ [\SU(k)]\ .
	\end{equation}
		This $\su{8}$ Hasse diagram is colored in red in figure \ref{fig:k14}.
	\item One with quiver given by \eqref{eq:generick-3b}:
	\begin{equation}\label{eq:k14theta3}
		[3'^4,2']: \quad \overset{\mathfrak{su}(6)}{\underset{[N_{\fontsize{0.25pt}{.5pt}\selectfont \frac{1}{2}\yng(1,1,1)}=1]}{1}}\ \overset{\su{14}}{\underset{[N_\text{f}=8]}{2}}\ \overset{\su{14}}{2}\ \cdots\overset{\su{14}}{2}\ [\SU(k)]\ .
	\end{equation}
	This $\su{8}$ Hasse diagram is colored in blue in figure \ref{fig:k14}.
\end{itemize}
The various $\su{8}$ or $\su{9}$ nilpotent orbits Hasse diagrams are highlighted in multiple colors (with dashed edges) in all figures of appendix \ref{app:RG}.  For each figure we explain which Hasse appears (potentially more than once) and in which color, as done for figure \ref{fig:k14}.

As a final observation,  it is worth noting that for every $k \geq 10$ there always is another theory whose Type IIA engineering features an O$8^*$ and a stack of $8$ D8's plus a single D8-brane away from it. This corresponds to Kac label $[3'^x,1^y]$, and is similar to \eqref{eq:su9starting} except that the tail is modified from a plateau of $\su{k}$'s to a ramp of length $y$ and step size $1$ from $\su{3x}$ to $\su{k=3x+y}$, with $8$ D8's at the end of the beginning of the ramp and a single D8 at the beginning of the plateau.  More explicitly:
\begin{subequations}\label{eq:highknot33'1}
	\begin{align}
		&x = 0 \mod 3: &  {1}\ \overset{\su{9}}{2}\ \cdots  \overset{\su{k-9}}{2}\ \overset{\su{3x}}{\underset{[N_\text{f}=8]}{2}}\ \overset{\su{3x+1}}{2} \cdots \overset{\su{k}}{\underset{[N_\text{f}=1]}{2}} \ \overset{\su{k}}{2}\ \cdots \ \overset{\su{k}}{2}\ [\SU(k)]\ , \\
		&x = 1 \mod 3: &  \overset{\su{3}}{\underset{[N_{\fontsize{0.5pt}{1pt}\selectfont \yng(1,1)}=1]}{1}}\ \overset{\su{12}}{2}\ \cdots  \overset{\su{k-9}}{2}\ \overset{\su{3x}}{\underset{[N_\text{f}=8]}{2}}\ \overset{\su{3x+1}}{2} \cdots \overset{\su{k}}{\underset{[N_\text{f}=1]}{2}} \ \overset{\su{k}}{2}\ \cdots \ \overset{\su{k}}{2}\ [\SU(k)]\ , \\
		&x =2 \mod 3: & \overset{\mathfrak{su}(6)}{\underset{[N_{\fontsize{0.25pt}{.5pt}\selectfont \frac{1}{2}\yng(1,1,1)}=1]}{1}}\ \overset{\mathfrak{su}(15)}{2}\ \cdots  \overset{\mathfrak{su}(k-9)}{2}\ \overset{\su{3x}}{\underset{[N_\text{f}=8]}{2}}\ \overset{\su{3x+1}}{2} \cdots \overset{\su{k}}{\underset{[N_\text{f}=1]}{2}} \ \overset{\su{k}}{2}\ \cdots \ \overset{\su{k}}{2}\ [\SU(k)]\ .
\end{align}
\end{subequations}
There are indeed several theories for a given $k$ that proceed via splitting of this stack of $8$ D8's. However, this stack eventually combines with the single D8 and thus does not generate the Hasse diagram of $\su{8}$.

\subsection{\texorpdfstring{The role of the O8$^*$}{The role of the O8*}}
\label{sub:genO8*}

\begin{figure}
\centering
\begin{tikzpicture}[scale=1,baseline]
				\draw[fill=black] (0,0) circle (0.075cm);
				\draw[fill=black] (1,0) circle (0.075cm);
				\draw[fill=black] (2,0) circle (0.075cm);
				\draw[fill=black] (3,0) circle (0.075cm);
				\draw[fill=black] (4,0) circle (0.075cm);
				\draw[fill=black] (5,0) circle (0.075cm);
				\draw[fill=black] (6,0) circle (0.075cm);
				\draw[fill=black] (7,0) circle (0.075cm);
				
				\draw[solid,black,thick] (1,0)--(2,0) node[black,midway,yshift=0.2cm] {\footnotesize $9$};
				\draw[loosely dotted,black,thick] (2,0)--(3,0) node[black,midway,yshift=0.2cm] {};
				\draw[solid,black,thick] (3,0)--(4,0) node[black,midway,yshift=0.2cm] {\footnotesize $k-9$};
				\draw[solid,black,thick] (4,0)--(5,0) node[black,midway,xshift=0.2cm,yshift=0.2cm] {\footnotesize $k$};
				\draw[solid,black,thick] (5,0)--(6,0) node[black,midway,yshift=0.2cm] {\footnotesize $k$};
				\draw[loosely dotted,black,thick] (6,0)--(7,0) node[black,midway, ] {};
				\draw[solid,black,thick] (7,0)--(8,0) node[black,midway,yshift=0.2cm] {\footnotesize $k$};
				
				\draw[dotted,black,very thick] (0,-.5)--(0,.5) node[black,midway,yshift=+.75cm] {\footnotesize O8$^*$};
				\draw[solid,black,very thick] (4.5,-.5)--(4.5,.5) node[black,midway, xshift =0cm, yshift=+.75cm] {\footnotesize $9$};
\end{tikzpicture}
\hspace*{-.228cm}
\begin{tikzpicture}[scale=1,baseline]
				\draw[fill=black] (0,0) circle (0.075cm) node[xshift=-0.3cm] {${\fontsize{0.25pt}{0.5pt}\selectfont \yng(1,1)}$};
				\draw[fill=black] (1,0) circle (0.075cm);
				\draw[fill=black] (2,0) circle (0.075cm);
				\draw[fill=black] (3,0) circle (0.075cm);
				\draw[fill=black] (4,0) circle (0.075cm);
				\draw[fill=black] (5,0) circle (0.075cm);
				\draw[fill=black] (6,0) circle (0.075cm);
				\draw[fill=black] (7,0) circle (0.075cm);
				
				\draw[solid,black,thick] (0,0)--(1,0) node[black,midway,yshift=0.2cm] {\footnotesize $3$};
				\draw[solid,black,thick] (1,0)--(2,0) node[black,midway,yshift=0.2cm] {\footnotesize $12$};
				\draw[loosely dotted,black,thick] (2,0)--(3,0) node[black,midway,yshift=0.2cm] {};
				\draw[solid,black,thick] (3,0)--(4,0) node[black,midway,yshift=0.2cm] {\footnotesize $k-9$};
				\draw[solid,black,thick] (4,0)--(5,0) node[black,midway,xshift=0.2cm,yshift=0.2cm] {\footnotesize $k$};
				\draw[solid,black,thick] (5,0)--(6,0) node[black,midway,yshift=0.2cm] {\footnotesize $k$};
				\draw[loosely dotted,black,thick] (6,0)--(7,0) node[black,midway, ] {};
				\draw[solid,black,thick] (7,0)--(8,0) node[black,midway,yshift=0.2cm] {\footnotesize $k$};
				
				\draw[dotted,black,very thick] (0,-.5)--(0,.5) node[black,midway,yshift=+.75cm] {\footnotesize O8$^*$};
				\draw[solid,black,very thick] (4.5,-.5)--(4.5,.5) node[black,midway, xshift =0cm, yshift=+.75cm] {\footnotesize $9$};
\end{tikzpicture}
\hspace*{-.45cm}
\begin{tikzpicture}[scale=1,baseline]
				\draw[fill=black] (0,0) circle (0.075cm) node[xshift=-0.4cm] {$\tfrac{1}{2} {\fontsize{0.25pt}{0.5pt}\selectfont \yng(1,1,1)}$};
				\draw[fill=black] (1,0) circle (0.075cm);
				\draw[fill=black] (2,0) circle (0.075cm);
				\draw[fill=black] (3,0) circle (0.075cm);
				\draw[fill=black] (4,0) circle (0.075cm);
				\draw[fill=black] (5,0) circle (0.075cm);
				\draw[fill=black] (6,0) circle (0.075cm);
				\draw[fill=black] (7,0) circle (0.075cm);
				
				\draw[solid,black,thick] (0,0)--(1,0) node[black,midway,yshift=0.2cm] {\footnotesize $6$};
				\draw[solid,black,thick] (1,0)--(2,0) node[black,midway,yshift=0.2cm] {\footnotesize $15$};
				\draw[loosely dotted,black,thick] (2,0)--(3,0) node[black,midway,yshift=0.2cm] {};
				\draw[solid,black,thick] (3,0)--(4,0) node[black,midway,yshift=0.2cm] {\footnotesize $k-9$};
				\draw[solid,black,thick] (4,0)--(5,0) node[black,midway,xshift=0.2cm,yshift=0.2cm] {\footnotesize $k$};
				\draw[solid,black,thick] (5,0)--(6,0) node[black,midway,yshift=0.2cm] {\footnotesize $k$};
				\draw[loosely dotted,black,thick] (6,0)--(7,0) node[black,midway, ] {};
				\draw[solid,black,thick] (7,0)--(8,0) node[black,midway,yshift=0.2cm] {\footnotesize $k$};
				
				\draw[dotted,black,very thick] (0,-.5)--(0,.5) node[black,midway,yshift=+.75cm] {\footnotesize O8$^*$};
				\draw[solid,black,very thick] (4.5,-.5)--(4.5,.5) node[black,midway, xshift =0cm, yshift=+.75cm] {\footnotesize $9$};
\end{tikzpicture}
\caption{Type IIA configurations with O8$^*$ engineering the electric quivers in \eqref{eq:su9starting}.}
\label{fig:IIAsu8su9hasse}
\end{figure}
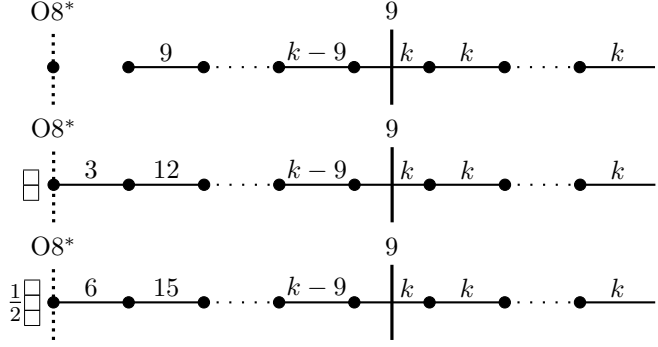

The Type IIA setups engineering the electric quivers in \eqref{eq:su9starting} are drawn in figure \ref{fig:IIAsu8su9hasse}, and they all involve the O8$^*$.\footnote{The three cases in \eqref{eq:su9starting} have already appeared in \cite[Fig. 1]{Hanany:2018vph}, and before that in \cite[Fig. 22]{Zafrir:2015rga} and \cite[Sec. 6]{Hayashi:2015zka}, in the context of lifting 5d SCFTs to 6d SCFTs, though none of these references stressed the role of the O8$^*$ to obtain a consistent Type IIA configuration as we do here.}  Notice that these are just the starting points of the $\su{9}$ nilpotent orbits Hasse diagram, so that all Kac labels along this diagram will also involve an O8$^*$.  Indeed, generically any Kac label \eqref{eq:kaccoord} satisfying
\begin{equation}
n_{3'} \geq n_{4'}\ , \quad n_{3'} - n_{4'} \geq 2 n_{2'}
\end{equation}
involves an O8$^*$, as first noted by \cite[Sec. 3.6]{Cabrera:2019izd}. 

\subsection{\texorpdfstring{Theories that differ by $\theta$ angle}{Theories that differ by theta angle}}
\label{sub:theta}

As noticed in \cite{Mekareeya:2017jgc}, there exist orbi-instantons that share the same 6d electric quiver, despite their M-theory origin suggests they are \emph{different}  SCFTs (their Kac labels are different).\footnote{To conclusively prove this statement one could study the spectra of BPS strings of the two SCFTs or construct their Higgs branch operator spectra, and show that they are different. The latter approach has been followed in a different yet related context in \cite{Distler:2022yse} (for $(D,D)$ conformal matter theories). }  It is proposed that the subtle difference between the electric quivers is captured by the so-called 6d $\theta$ angle of the gauge theory, where two possible choices of this parameter correspond to the two different SCFTs, which are thus distinguished even on their tensor branches.

For 6d $(1,0)$ gauge theories, the $\theta$ angles are the coefficients of the terms
\begin{equation}\label{eq:efflag}
\mathcal{L}_\text{eff} \supset \theta_i \Tr F_i^3
\end{equation}
in the effective Lagrangian on the tensor branch (for those $\Tr F_i^3$ which are nonvanishing), with 
\begin{equation}\label{eq:thetai}
\theta_i = \frac{x_{i+1}^{10} - x_{i}^{10}}{l_\text{s}^3}\ ,
\end{equation}
and $x_i^{10}$ the position along the M-theory circle in table \ref{tab:M5} of the $i$-th NS5-brane. The latter ``bounds'' from one side the $i$-th finite segment of D6-branes giving rise to the $i$-th non-Abelian vector multiplet with field strength $F_i$.  We also have a triplet of Fayet--Iliopoulos (FI) terms and a gauge coupling $1/g_i^2$ \cite{Seiberg:1996qx}, 
given respectively  by 
\begin{equation}\label{eq:FIcoup}
w_i^{7,8,9} = \frac{x_{i+1}^{7,8,9} - x_{i}^{7,8,9}}{l_\text{s}^3}\ , \quad \frac{1}{g_i^2} = \frac{x_{i+1}^{6} - x_{i}^{6}}{l_\text{s}^3\, g^\text{IIA}_\text{s}} = \phi_{i+1} - \phi_i\ ,
\end{equation}
with $\phi_i$ the scalar in the $i$-th tensor multiplet contributed by the $i$-th NS5/M5.  Both $\theta$ angles and FI terms are dynamical quantities in 6d, and play a crucial role in canceling the anomalies of $\uu{1}$ subalgebras of $\uu{r_i}$ gauge algebras (thus becoming the familiar $\su{r_i}$ appearing throughout this paper).\footnote{An anomalous transformation of the $\theta$ angle is presumed to give mass to the $\uu{1}$'s via a St\"uckelberg term \cite[Eq. 2.6]{Hanany:1997gh}, thereby rendering them massive and removing from the low-energy spectrum of the gauge theory the gauge anomaly they suffer from  \cite{Berkooz:1996iz,Park:2011wv}.} Notice also that the $\theta$ term in \eqref{eq:efflag} is nonzero only for $\mathfrak{u}$ or $\mathfrak{su}$,  since the cubic Casimir of the adjoint of $\mathfrak{usp}$ vanishes.\footnote{In fact, for all simple Lie algebras but $A_n$,  see e.g.\cite[Tab. 1]{vanRitbergen:1998pn}.}

\cite{Mekareeya:2017jgc} identified a whole family of A-type $k=2l+8$ orbi-instantons whose tensor branches only differ by the 6d $\theta$ angle, where the two SCFTs are given by Kac labels $[4'^2,2'^l]_{\theta=0}$ and $[3'^2,2'^{1+l}]_{\theta=\pi}$.  In both cases the 6d electric quiver reads
\begin{equation}\label{eq:thetaquiv}
\begin{rcases} [4'^2,2'^l]_{\theta=0}\\  [3'^2,2'^{1+l}]_{\theta=\pi} \end{rcases} \leftrightarrow \f = \su{8} \oplus \uu{1}: \ \overset{\mathfrak{usp}(2l)}{1}\ \overset{\mathfrak{su}(2l+8)}{\underset{[N_\text{f}=8]}{2}}\ \overset{\mathfrak{su}(2l+8)}{2} \cdots  \overset{\mathfrak{su}(2l+8)}{2}\ [\SU(2l+8)]\ .
\end{equation}
In field theory terms, this $\theta$ angle can be identified with a ``choice'' of gauge theory when the zeroth gauge algebra on the tensor branch is nonempty, since $\pi_5(\USp(2l))= \zz_2$ \cite{bott,mimura-toda}.\footnote{This can be quickly seen as follows.  We can realize a sphere as a homogeneous space given by the fibration
\begin{align}
&\SO(n-1)\hookrightarrow \SO(n) \to \frac{\SO(n)}{\SO(n-1)}\cong S^{n-1} \ , \nonumber \\
&\SU(n-1)\hookrightarrow \SU(n) \to \frac{\SU(n)}{\SU(n-1)}\cong S^{2n-1} \ , \nonumber \\
&\USp(n-1)\hookrightarrow \USp(n) \to \frac{\USp(n)}{\USp(n-1)}\cong S^{4n-1}\ . \nonumber
\end{align}
Passing to long exact sequences
\begin{equation}
\cdots \to \pi_i(\SO(n-1)) \to \pi_i (\SO(n)) \to \pi_i (S^{n-1}) \to \pi_{i-1}(\SO(n-1)) \to \cdots \nonumber
\end{equation}
and knowing that $\pi_m(S^n) = 0$ for $m<n$ and $\pi_2(\SO(2))=\pi_2(\SU(2))=0$, we get the wanted result (remembering the exceptional isomorphisms between Lie groups for small $n$).} The choice is given in practice by two different embeddings of $\su{2l+8}$ into the $\so{2(2l+8)}$ global symmetry of a $\mathfrak{usp}$ gauge algebra with $2l+8$ fundamentals. When $l=0$, i.e. when the zeroth $\mathfrak{usp}$ gauge algebra is empty, we cannot use this argument to distinguish between the two choices, but we nonetheless have two different embeddings of $\su{8}$ inside the $E_8$ of $[E_8]1$ at our disposal \cite[Sec. 3.3]{Mekareeya:2017jgc}.  Indeed notice that when $l=0$ the 6d SCFTs are distinguished by the commutant of the subalgebra of $E_8$ that gauges the E-string: $\f=\su{8}\oplus\su{2}$ for $[4'^2]_{\theta=0}$ and $\f=\su{8}\oplus \uu{1}$ for $[3'^2,2']_{\theta=\pi}$.  This is not the case for $l\neq 0$ since we have $\su{8}\oplus\uu{1}$ for both SCFTs. At the level of 6d dynamics, the SCFTs should be distinguished by the nonperturbative spectrum of instanton strings (D2-branes wrapping $x^{016}$ with vanishing tension), which is affected by the $\zz_2$ choice of $\theta$ \cite{Heckman:2017uxe,Gukov:2020btk}.

The constraint imposing that two different SCFTs specified by different Kac labels have the same electric quiver has been solved in full generality in  \cite[Eq. (3.161)]{Cabrera:2019izd}.  There it was also shown that two such 6d SCFTs, in spite of sharing the same electric quiver, differ by their 3d magnetic quivers at infinite coupling, which are thus capable of capturing the discrete choice of $\theta$. However the difference is very subtle in that it manifests itself as a $\zz_2$ graph automorphism exchanging the two tails at the right of the magnetic quivers \cite[Eq. (3.164)]{Cabrera:2019izd}, which we reproduce here below (before performing $N$ small instanton transitions):
\begin{align}
{\scriptstyle 1 - 2 - \cdots -(k-1) - k - (N+g_1) -(2N+g_2) -(3N+g_3) - (4N+g_4) - (5N+g_5) - \overset{\overset{\scriptstyle 3N+g_8}{\vert}}{(6N+g_6)}-(4N+g_7)-2N}
\end{align}
where \cite[Eq. (3.125)]{Cabrera:2019izd}
\begin{equation}\label{eq:theta-gi}
\begin{split} 
&g_j = \sum_{i=1}^{6-j} i n_{i+j} +2 n_{2'} +n_{4'} +\frac{6-j}{2}(n_{3'}+n_{4'})\ , \quad j=1,\ldots,6\ ; \\
& g_7 = n_{2'}\ ; \quad g_8 = n_{2'}+\frac{1}{2}(n_{4'}-n_{3'})\ .
\end{split}
\end{equation}
Performing $N$ small instanton transitions yields
\begin{equation}
	1 - 2 - \cdots -(k-1) -\overset{\verT{N}}{k} - g_1 -g_2-g_3-g_4-g_5- \overset{\vernocirc{g_8}}{g_6} - g_7 \ ,
\end{equation}
where $\overset{\curvearrowright}{N}$ indicates a $\U(N)$ gauge node with an adjoint.  In this presentation, going from $\theta=0$ to $\pi$ simply means swapping $g_7$ with $g_8$.

Here we identify an infinite class of solutions to that constraint and exhibit explicitly the 6d electric quivers distinguished solely by the choice of $\theta$ angle, according to the two above mechanisms (i.e. depending on whether the zeroth $\mathfrak{usp}$ algebra is empty or not). Thus, we can go beyond the example \eqref{eq:thetaquiv} constructed in \cite{Mekareeya:2017jgc}.  

Given a choice of three non-negative integers $\{m,n,l\}$ such that $m+n \equiv 0 \mod 2$ (and without loss of generality we may assume $m > n$),  the infinite class corresponds to Kac labels of the form 
\begin{equation}\label{eq:θangle}
[4'^m, 3'^n, 2'^l,\{p_i\}]_{θ=0} \ , \quad  [4'^n, 3'^m, 2'^{l+(m-n)/2},\{p_i\}]_{θ=π}\ .
\end{equation}
These two 6d SCFTs do not preserve the same $\f$ if $n=l=0$ or $l=0$.  The electric quiver associated with both labels is:
\begin{equation}\label{eq:theta-quiver}
\footnotesize 
	\overset{\usp{2l}}{1}\ \overset{\su{2l+8}}{2}\ \overset{\su{2l+16}}{2} \cdots \overset{\su{2l+4(m-n)}}{\underset{[N_\text{f}=1]}{2}} \cdots
	\overset{\su{k_0-7}}{2} \overset{\su{k_0}}{\underset{[N_\text{f}=7-p_1]}{2}}\ \overset{\su{k_0+p_1}}{\underset{[N_\text{f}=p_1-p_2]}{2}}\ \overset{\su{k_0+p_1+p_2}}{\underset{[N_\text{f}=p_2-p_3]}{2}} \cdots \overset{\su{k}}{\underset{[N_\text{f}=p_d]}{2}}\ [\SU(k)]\ ,
\end{equation}
where $k_0 \equiv 2l+4m+3n$,  all values of $l=0,1,2,…,\lfloor (k-8)/2 \rfloor$ are allowed, and the $\{p_i\}_{i=1}^d$ are the parts of a partition $[p_i]$ of $k-k_0$ in terms of the unprimed integers $1,\ldots,6$ only. 

There are three types of ramps (in the ranks of the gauge algebras) in the above electric quiver: \emph{i)} from $2l$ to $2l+4(m-n)$ in steps of length $8$, \emph{ii)} from $2l+4(m-n)$ to $k_0$ in steps of length $7$, and \emph{iii)} a composite ramp depending on the partition $[p_i]$.
Different partitions have different numbers $d$ of parts so that the length of the last ramp varies from partition to partition.
The flavors sit at the transition between the ramps, and can easily be fixed by requiring gauge anomaly cancellation of each of the $\su{r_i}$ algebras ($\usp{2l}$ is automatically anomaly-free), yielding the above electric quiver.
Clearly, this family can only exist for $k\geq 8$; for $m=2,n=0$, this collapses onto \eqref{eq:thetaquiv}.  

As an example,  take once again $k=14$, i.e. $l=0,\ldots,3$. Let us classify all pairs of orbi-instantons that have the same electric quiver based on $l$. We have:
\begin{itemize}
	\item $l=0$: $11$ quivers (i.e. 11 pairs of orbi-instantons sharing the same electric quiver) associated with Kac labels $[3'^2,2',\{p_i\}]_{\theta=\pi}$ and $[4'^2,\{p_i\}]_{\theta=0}$, where $[p_i]$ are the eleven integer partitions of $14-(2\cdot 0+8) = 6$, i.e.
	\begin{equation}\label{eq:k14usp0}
	[p_i] = [6],[5,1],[4,2],[4,1^2],[3^3],[3,2,1],[3,1^4],[2^3],[2^2,1^2],[2,1^2],[1^6]\ .
	\end{equation}
	\item $l=1$: $5$ quivers associated with Kac labels  $[3^2, 2'^2, \{p_i\}]_{\theta=\pi}$ and $[4'^2, 2',\{p_i\}]_{\theta=0}$, where $[p_i]$ are the four integer partitions of $14-(2\cdot 1+8) = 4$, i.e.
	\begin{equation}\label{eq:k14usp2}
	[p_i] = [4],[3,1],[2^2],[2,1^2],[1^4]\ .
	\end{equation}

	\item $l=2$: $2$ quivers associated with Kac labels  $[3^2, 2'^3, \{p_i\}]_{\theta=\pi}$ and $[4'^2, 2'^2,\{p_i\}]_{\theta=0}$, where $[p_i]$ are the two integer partitions of $14-(2\cdot 2+8) = 2$, i.e. $[p_i]=[2],[1^2]$.  Explicitly:
	\begin{subequations}\label{eq:k14usp4}
	\begin{align}
		&\begin{rcases} [4'^2,2'^2,2]_{\theta=0} \\ 
		[3'^2,2'^3,2]_{\theta=\pi} \end{rcases}: &&\overset{\mathfrak{usp}(4)}{1}\ \overset{\mathfrak{su}(12)}{\underset{[N_\text{f}=6]}{2}} \ \overset{\mathfrak{su}(14)}{\underset{[N_\text{f}=2]}{2}}\ \cdots  \overset{\mathfrak{su}(14)}{2}\ [\SU(14)]\ ,\\
		&\begin{rcases} [4'^2,2'^2,1²]_{\theta=0} \\
		[3'^2,2'^3,1²]_{\theta=\pi}
		\end{rcases}: &&\overset{\mathfrak{usp}(4)}{1}\ \overset{\mathfrak{su}(12)}{\underset{[N_\text{f}=7]}{2}}\ \overset{\mathfrak{su}(13)}{2} \ \overset{\mathfrak{su}(14)}{\underset{[N_\text{f}=1
		]}{2}}\ \cdots  \overset{\mathfrak{su}(14)}{2}\ [\SU(14)]\ .
	\end{align}
	\end{subequations}
	
	\item $l=3$: $1$ quiver associated with Kac labels  $[3^2, 2'^4]_{\theta=\pi}$ and $[4'^2, 2'^3]_{\theta=0}$.
\end{itemize}
We have found that the orbi-instantons for generic $k$ and fixed $l$ form a partial Hasse diagram of nilpotent orbits of $\su{k-8-2l}$,  excluding the partitions which use integers larger than $8$. Additionally, for a given $l$, the class of orbi-instantons for fixed $(m,n)$ which have $k-k_0 \leq 6$ form themselves a Hasse diagram of $\su{k-k_0}$ nilpotent orbits, where the partitions associated with the orbits are nothing but the $[p_i]$.\footnote{This is because, for $k-k_0 \leq 6$, partitions in terms of the integers $1,…,6$ only are the same as the total number of integer partitions.} We highlight in different colors (one for every choice of $l$) these Hasse diagrams embedded in the RG flow hierarchies of appendix \ref{app:RG}.

To illustrate how this works, consider the example of $k=19$. Using \eqref{eq:θangle}, we see that for $l=0$ the integers $(m,n)$ can either be $(2,0), (3,1)$ or $(4,0)$. For $(m,n)=(2,0)$ the $[p_i]$ must be all partitions of $19-2\cdot 2=15$ into integers $1,…,6$, which are $44$ in total. Next, $(m,n)=(3,1)$ implies $[p_i]$ are all integer partitions of $19-3\cdot4-1\cdot3=4$ using integers $1,…,6$, which are $5$ in total. Finally for $(m,n)=(4,0)$, the $[p_i]$ are all partitions of $19-4\cdot4=3$, which are $3$ in total. These $44+5+3=52$ orbi-instantons all correspond to electric quivers starting with an empty $\mathfrak{usp}$ gauge algebra on the $-1$ curve as can be inferred from \eqref{eq:theta-quiver}, since $l=0$.  Examining the difference of ranks of adjacent gauge algebras, we see that it falls into one of the following three patterns (corresponding to the three possibilities for $(m,n)$): $\{8,7,p_i\}$, $\{8,p_i\}$ or $\{8,8,p_i\}$, where $\{p_i\}$ are the parts of $[p_i]$. On disregarding the first integer $8$ which is always present,  the differences of adjacent ranks of these $52$ orbi-instantons precisely make up the integer partitions of $19-4\cdot2=11$ (where the $2$ arises because the smallest possibility for $(m,n)$ is $(2,0)$), except the partitions which use integers larger than $8$.   Interestingly enough,  the partial ordering of the electric quivers obtained via magnetic quiver subtraction corresponds precisely to the partial ordering of these integer partitions given by the difference in ranks of these orbi-instantons. (This is similar to the observation at the end of section \ref{subsub:IIAk6}.) These partitions are in turn one-to-one with the nilpotent orbits of $\su{11}$, i.e. we have a partial Hasse diagram of $\su{11}$ which does not include the four top nodes corresponding to partitions $[11], [10,1], [9,2], [9,1²]$, since these use integers larger than $8$.

Looking at the quiver in \eqref{eq:theta-quiver}, it is easy to see why integers larger than $8$ cannot appear in the difference of adjacent ranks. As another example for $k=19$, consider $l=3$. This gives only one possibility, $(m,n)=(2,0)$, which implies that $[p_i]$ are the $7$ integer partitions of $19-2\cdot4-3\cdot2=5$. Again, from \eqref{eq:theta-quiver} we see that the difference in adjacent ranks is of the form $\{8,p_i\}$ which, on disregarding the first $8$, corresponds to the partitions of the integer $5$. As expected, partial-ordering these orbi-instantons via magnetic quiver subtraction indeed gives the Hasse diagram of nilpotent orbits of $\su{5}$.

Let us conclude this section with a field theory observation.  It would be interesting to understand what is the remnant of the discrete $\theta$ angle in the SCFT (i.e. at the origin of the tensor branch) --especially in relation to $(-1)$-form ``symmetries''-- since the two possible choices of $\theta$ correspond to two distinct SCFTs. A possible starting point is an adaptation of the general analysis of \cite{Gukov:2020btk} to the $(1,0)$ orbi-instanton case.\footnote{We would like to thank N.~Mekareeya,  Y.~Tachikawa, and L.~Tizzano for discussion on this point.} We leave this question as an interesting avenue for future work.

\section{Conclusions}
\label{sec:conc}

In this paper we have determined the hierarchy of Higgs branch RG flows between orbi-instantons of type A at fixed $k,N$.  The flows are between orbi-instantons (i.e. SCFTs) defined by different boundary conditions in M-theory, i.e. allowed Kac labels for the given $k$.  The partial ordering on the hierarchy is given by the quiver subtraction operation, i.e. a flow between two Kac labels is allowed if the associated 3d magnetic quivers can be subtracted in a consistent way.  For each proposed flow we have checked that the $a$-theorem is verified, i.e. $\Delta a = a_\text{UV} - a_\text{IR}>0$.  We have explicitly produced hierarchies for $k=2,\ldots,20$, and highlighted a few key facts such as the possibility to fully embed an $\su{8}$ or $\su{9}$ nilpotent orbits Hasse diagram into the hierarchy,  to partially embed the $\so{16}$ Hasse, and to further embed other $\mathfrak{su}$ Hasse diagrams of flows between orbi-instantons which have different $\theta$ angle on the tensor branch, in spite of sharing the same electric quiver (i.e. perturbative spectrum of operators).

As a mathematical statement, these hierarchies should be understood as Hasse diagrams of $\text{Hom}(\zz_k,E_8)$, for which there are no known results in the literature. Similarly to what happens for nilpotent orbits of classical or exceptional Lie algebras, we have found that the edges in these Hasse diagrams are given by Kraft--Procesi transitions of type $A_i$ or $\mathfrak{a}_i, \mathfrak{d}_i$, although we are not always transitioning from nilpotent orbit to nilpotent orbit of a fixed Lie algebra (as is the case for \cite{KP0,KP1,KP2}).  Therefore we are unsure as to the meaning of these transitions (produced via quiver subtraction) in the homomorphisms Hasse. A possible hint is the following. It is known \cite[Thm. A]{malin-ostrik-vybornov} that the minimal degeneration singularities of the $E_8[[z]]$-orbits in the \emph{affine Grassmannian} of $E_8$ are all either Kleinian of type A or minimal of type given by a subdiagram of the $E_8$ Dynkin diagram.\footnote{We are grateful to P.~Levy for discussions on this and related points, and for suggesting a potential relation with the work of \cite{malin-ostrik-vybornov}.  When the group is non-simply-laced there exist other singularities called quasi-minimal \cite{malin-ostrik-vybornov}.} (By $E_8[[z]]$ we mean the group of matrices in $E_8$ with coefficients in the ring of formal power series in $z$,  usually denoted $\cc[[z]]$. For an introduction to affine Grassmannians accessible to physicists see e.g.  \cite[Sec. 2.3]{Bourget:2021siw}.) This is precisely what we found in our $\text{Hom}(\zz_k,E_8)$ Hasse. The connection between affine Grassmannians of Lie groups and moduli spaces of 3d $\mathcal{N}=4$ quivers has already been put forward in \cite{Bourget:2021siw}, so we expect the (slices of the) affine Grassmannian of $E_8$ to play a role in Higgs branches of 6d electric quivers obtained from O8-D8-D6-NS5 brane systems with 8 supercharges. It would be extremely interesting to explore this further.

An obvious generalization of our work is the construction of hierarchies of RG flows for D-type orbi-instantons, which involve M5-branes probing an $E_8$ wall and the $\cc^2/\Gamma_\text{D} = \cc^2/\mathbb{D}_{k}$ orbifold, with $\mathbb{D}_{k}$ the binary dihedral group of order $4k$ associated with the $D_{k+2}$ singularity.  (The corresponding Type IIA configurations will contain an O6-plane crossing the O$8^-$ \cite{Hanany:1997gh}, as well as the more exotic ON$^0$ \cite{Hanany:1999sj}.\footnote{The (Type IIA) ON$^0$ wraps the same directions as an NS5 (see table \ref{tab:M5}), being (T-dual to) the S-dual of an O5$^-$ (the ON$^-$) with a single NS5 on top.  Placed at the left end of an electric quiver without O6-planes or the O8 as follows
\begin{center}
\begin{tikzpicture}[scale=1,baseline]
\node[label={[label distance=-0.5cm]180:ON$^0$}] at (0,0) {};
\draw[fill=white] (0.5,0) circle (0.1cm);
\draw[fill=black] (1.5,0) circle (0.1cm);
\draw[fill=black] (2.5,0) circle (0.1cm);
\draw[fill=black] (3.5,0) circle (0.1cm);
\draw[fill=black] (4.5,0) circle (0.1cm);

\draw[solid,black,thick] (0.6,-.05)--(1.5,-.05) node[black,midway,yshift=-0.2cm] {\footnotesize $k$};
\draw[solid,black,thick] (0.6,.05)--(1.5,.05) node[black,midway,yshift=0.2cm] {\footnotesize $k$};
\draw[solid,black,thick] (1.5,0)--(2.5,0) node[black,midway,yshift=0.2cm] {\footnotesize $2k$};
\draw[loosely dotted,thick,black] (2.5,0)--(3.5,0) node[black,midway] {};
\draw[solid,black,thick] (3.5,0)--(4.5,0) node[black,midway,yshift=0.2cm] {\footnotesize $2k$};
\draw[solid,black,thick] (4.5,0)--(5.5,0) node[black,midway,yshift=0.2cm] {\footnotesize $2k$};
\end{tikzpicture}
,
\end{center}
it engineers an electric quiver with D-type Dynkin diagram shape,
\begin{center}
	\begin{tikzpicture}[scale=1]
	  \tikzstyle{arrow} = [->,>=stealth]
	  \node (n1) {$\overset{\su{2k}}{2} \, \overset{\su{2k}}{2}  \cdots  \overset{\su{2k}}{2}\, [\SU(2k)]\ ,$};
	  	  \node (n4) [below left=-15pt and -10pt of n1] {$\overset{\su{k}}{2}$};
	  \node (n5) [above left=-10pt and -10pt of n1] {$\overset{\su{k}}{2}$};
	\end{tikzpicture}
\end{center}
which is allowed by the general rules in \cite[App. D]{Heckman:2015bfa}. Adding O6 and O8, there are four gauge anomaly free combinations \cite[Sec. 6]{Hanany:1999sj}, of which the two relevant ones for our present purposes are O$8^-$-O$6^\mp$ with an ON$^0$ stuck at their intersection in the zeroth segment.  (Notice that any two of these orientifold projections always imply the third, as can be seen by taking three T-dualities on the system of \cite[Sec. 7.4]{Gaiotto:2008ak}. We are grateful to A.~Hanany and M.~Sperling for pointing this out to us.) The electric quiver engineered by the former (with all 8 D8's on top of the O8) can be found in \cite[Eq. (5.1)]{Hayashi:2015vhy},  whereas the latter in \cite[Eq. (5.16)]{Hayashi:2015vhy}.  Whenever O6 and O8 have the same charge the quiver is linear (i.e. has A-type Dynkin shape), whereas when they have opposite charge it has D-type shape \cite{Hanany:1999sj}.  The corresponding 3d magnetic quivers have been worked out in \cite[Sec. 4.2]{Sperling:2021fcf}.  (There also exists an $\widetilde{\text{O}6}^-$ which is equivalent to an O$6^-$ with a half-D6 stuck on it; it is allowed only for odd Romans mass \cite{Hyakutake:2000mr,Bergman:2001rp}. If the Romans mass is instead even --including the case where it is vanishing-- only the O$6^-$ is allowed, whence the above list of possibilities.)}) This poses two immediate difficulties however. First, the work of Kac \cite{kac1990infinite} cannot be directly generalized to homomorphisms in $\text{Hom}(\mathbb{D}_{k},E_8)$, since it is impossible to grade a Lie algebra by a non-Abelian group.\footnote{See footnote \ref{foot:grade}.} Hence we do not have at our disposal a set of ``natural'' Kac labels to define the SCFTs. One may proceed as follows.  To find the embeddings of $\mathbb{D}_{k}$ into the $E_8$ group we first embed its cyclic subgroup $\zz_{2k} \subset \mathbb{D}_k$, which is done via Kac labels. For each label we have an element $a$ of a maximal torus of $E_8$,  generating a subgroup of order $k$. We must then find elements $w$ of the Weyl group $W(E_8)$ which satisfy $w(a)=a^{-1}$: these are the elements of the form $-w'$ where $w'$ belongs to the centralizer of $a$ in $W(E_8)$.  We can pick one representative $b$ of each conjugacy class $w$, and ask how many of those have representatives in $E_8$ which are of order 2, namely find which of the homomorphisms in $\text{Hom}(\mathbb{D}_{k},E_8)$ from \cite{Frey:2018vpw} (called $\text{Hom}(\text{Dic}_{k},E_8)$ there) have kernels consisting of $b^2=a^n$ (see \cite[Sec. 7.3.3]{Heckman:2015bfa} for their F-theory realization).\footnote{We are grateful to P.~Levy for suggesting this argument to us.} Second, the full set of rules to obtain the 3d ortho-symplectic magnetic quivers of D-type orbi-instantons coming from O8-D8-O6-D6-NS5 brane systems has not been worked out yet (to the best of our knowledge), even though this difficulty may more easily be overcome by extending the rules proposed in \cite{Cabrera:2019dob,Bourget:2020xdz,Hanany:2022itc} to also include the O8 (see e.g. \cite[Sec. 4]{Sperling:2021fcf} for progress in this direction). Once this is done, we can again apply the quiver subtraction algorithm  to construct all possible RG flows, and check compatibility with the $a$-theorem. For E-type orbi-instantons, the 3d magnetic quiver technology is currently unavailable. (We expect there to be a finite number of hierarchies of RG flows since the binary tetrahedral, octahedral, and icosahedral groups $\Gamma_{E_{6,7,8}}$ have a finite, non-parametric order. See e.g. \cite[Sec. 7.4.1]{Heckman:2015bfa} for the F-theory realization of orbi-instantons in $\text{Hom}(\Gamma_{E_8},E_8)$.)

Finally, we wish to comment on two less obvious generalizations of the present work. One direction involves studying \cite{fazzi-giri-massive} the hierarchy of Higgs branch RG flows between so-called ``massive E-string theories'' \cite[Sec. 5.2]{Bah:2017wxp},\footnote{These theories have appeared even earlier in \cite{DelZotto:2014hpa,Zafrir:2015rga,Ohmori:2015tka,Hayashi:2015zka},  though they were not given a name there.} which are generalizations of \eqref{eq:fullymasslessE8} with nonzero Romans mass,  the total number of D8's being less than 8. These systems do not allow for an M-theory construction,  but do so both in massive Type IIA (hence the name) and F-theory, where the left flavor symmetry factor is $[E_{1+(8-n_0)}]$ for $n_0=1,\ldots,8$ ($8-n_0$ being the total number of D8's, which are close to the O8$^-$):
\begin{equation}\label{eq:massive}
[E_{1+(8-n_0)}]\ {1}\ \overset{\mathfrak{su}(n_0)}{2}\ \overset{\mathfrak{su}(2n_0)}{2} \cdots \overset{\mathfrak{su}((N-1)n_0)}{2}\  [\SU(Nn_0)]\ .
\end{equation}
As is clear, the difference with \eqref{eq:fullymasslessE8} stems from the absence of a plateau of $\mathfrak{su}$ algebras. Therefore we take $Nn_0$ to mean $k$.  Indeed notice that \eqref{eq:massive} comes from truncating the more general \emph{massless} quiver in \cite[Eq. (5.71)]{Mekareeya:2017jgc} with $8-n_0+n_0=8$ D8's, namely
\begin{equation}\label{eq:massEntail}
[E_{1+(8-{n}_0)}]\ 1\ \overset{\mathfrak{su}({n}_0)}{2}\ \overset{\mathfrak{su}(2{n}_0)}{2}\ \overset{\mathfrak{su}(3{n}_0)}{2} \cdots \ \overset{\mathfrak{su}((m-1){n}_0)}{2} \ \underset{[N_\text{f}={n}_0]}{\overset{\mathfrak{su}(m{n}_0)}{2}}\ \overset{\mathfrak{su}(m{n}_0)}{2} \cdots \overset{\mathfrak{su}(m{n}_0)}{2} \ [\SU(m{n}_0)]\ ,
\end{equation}
where obviously $k=mn_0$.\footnote{\eqref{eq:massEntail} is the electric quiver of $[n_0^m]$ for $n_0=1,\ldots,6$,  $[4'^m,3'^m]$ for $n_0=7$, and $[4'^{2m}]$ for $n_0=8$.} Putting $m=N$ and truncating at the leftmost $\overset{\mathfrak{su}(m{n}_0)}{2}$ in position $m$, we obtain \eqref{eq:massive} with $k=Nn_0$. However notice that this $k$ is \emph{not} the order of the $\cc^2/\zz_k$ orbifold but rather the number of semi-infinite D6's in the rightmost segment, since as we said there is no M-theory engineering of that theory.  Here
\begin{equation}\label{eq:seiberglist}
E_{1+(8-n_0)} = \{ E_8, E_7, E_6, \mathfrak{so}(10),\mathfrak{su}(5),\mathfrak{su}(3)\oplus \mathfrak{su}(2), \mathfrak{su}(2)\oplus \mathfrak{u}(1), \mathfrak{su}(2)\}_{n_0=1}^{8}
\end{equation}
is the list of \cite{Seiberg:1996bd} (excluding the $\tilde{E}_1 = \uu{1}$ and $E_0 = \emptyset$ cases of \cite{Morrison:1996xf}).  In analogy with \eqref{eq:E8partition}, one can define $E_{1+(8-{n}_0)}$ Kac labels for all $n_0$ (now associated with $\zz_{Nn_0}$-gradings of $E_{1+(8-n_0)}$) by once again exploiting the results of \cite{kac1990infinite}.  Although $E_8$ has a unique affine form,  $E_8^{(1)}$ in \eqref{eq:E81}, this is not true for some of the algebras in the above list, and one has to consider both \emph{untwisted and twisted} affine versions,\footnote{The twisted affine algebras also make an appearance in \cite{Giacomelli:2020gee} in the construction of 4d $\mathcal{N}=3$ S-folds from A-type orbi-instantons. It would be interesting to study whether a potential connection exists.} namely
\begin{equation}\label{eq:kacmoodylist}
E_8^{(1)}\ , \ E_7^{(1)}\ , \ E_{6}^{(2)}\ , \ E_6^{(1)}\ , \ D_5^{(2)}\ , \ D_5^{(1)}\ , \ A_4^{(2)}\ , \ A_4^{(1)}\ , \ A_2^{(2)}\ ,\ A_2^{(1)}\ ,  \ A_1^{(1)}\ ,
\end{equation}
and direct sums (also with $\uu{1}$) as per \eqref{eq:seiberglist}: the superscript $k'=1$ ($k'=2$) selects the untwisted (twisted) version \cite[Sec. 7.9]{fuchs2003symmetries}. (See table \ref{tab:kacmoody} for definitions.)
\begin{table}[ht!]
\centering
\renewcommand{\arraystretch}{2.5}
\begin{tabular}{cc}
$A_1^{(1)}:$ & $\node{1}{\alpha_0} \Leftrightarrow \node{1}{\alpha_1}$ \\
$A_2^{(1)}:$ & $\begin{array}{c} \raisebox{-5pt}{\rotatebox{45}{$\rule{10pt}{0.4pt}$}} \node{1}{\alpha_0}\raisebox{2.5pt} {\rotatebox{-45}{$\rule{10pt}{0.4pt}$}} \\[-7pt] \node{1}{\alpha_1}-\node{1}{\alpha_2} \end{array}$\\
$A_2^{(2)}:$ & $\node{2}{\alpha_0}\Llleftarrow\node{1}{\alpha_1}$ \\
$A_4^{(1)}:$ & $\begin{array}{c} \raisebox{-10.5pt}{\rotatebox{30}{$\rule{30pt}{0.4pt}$}} \node{1}{\alpha_0}\raisebox{4pt} {\rotatebox{-30}{$\rule{30pt}{0.4pt}$}} \\[-7pt] \node{1}{\alpha_1}-\node{1}{\alpha_2}-\node{1}{\alpha_3}-\node{1}{\alpha_4} \end{array}$ \\
$A_4^{(2)}:$ & $\node{2}{\alpha_0}\Leftarrow \node{2}{\alpha_1} \Leftarrow \node{1}{\alpha_2}$\\
$D_5^{(1)}:$ &  $\node{1}{\alpha_1}-\node{2\ver{1}{\alpha_0}}{\alpha_2}-\node{2\ver{1}{\alpha_{5}}}{\alpha_{3}}-\node{1}{\alpha_{4}}$  \\ 
$D_5^{(2)}:$ &  $ \node{1}{\alpha_0}\Leftarrow \node{1}{\alpha_1}-\node{1}{\alpha_2}-\node{1}{\alpha_4}\Rightarrow\node{1}{\alpha_5}$  \\ 
$E_6^{(1)}:$  & $\node{1}{\alpha_1}-\node{2}{\alpha_2}-\node{3\overset{\ver{1''}{\alpha_0}}{\ver{2''}{\alpha_6}}}{\alpha_3}-\node{2'}{\alpha_4}-\node{1'}{\alpha_5}$ \\
 $E_{6}^{(2)}:$ & $\node{1}{\alpha_0}-\node{2}{\alpha_1}-\node{3}{\alpha_2}\Leftarrow\node{2'}{\alpha_3}-\node{1'}{\alpha_4}$ \\
 $E_7^{(1)}:$ & $\node{1}{\alpha_0}-\node{2}{\alpha_1}-\node{3}{\alpha_2}-\node{4\ver{2''}{\alpha_7}}{\alpha_3}-\node{3'}{\alpha_4}-\node{2'}{\alpha_5}-\node{1'}{\alpha_6}$
\end{tabular}
\caption{The untwisted ($k'=1$) and twisted ($k'=2$) affine Dynkin diagrams of (summands of) the $E_{1+(8-n_0)}$ algebras listed in \eqref{eq:seiberglist}.}
\label{tab:kacmoody}
\end{table}
Accordingly, \eqref{eq:E8partition} gets modified to
\begin{equation}
Nn_0 = k'  \sum_{j=0}^{\rk\left(E_{1+(8-n_0)}\right)} a_j n_{j+1}\ ,
\end{equation}
with $k'=2$ being allowed only for even $Nn_0$, and $a_j$ the Coxeter labels appearing in table \ref{tab:kacmoody} (combining unprimed and primed ones by abuse of notation). We expect different massive E-string theories (for a given $E_{1+(8-n_0)}$) to be classified by Kac labels, and there to be a hierarchy of RG flows between them.

Second, the original motivation for this work was the study of the stability of the non-supersymmetric version of the Type IIA AdS$_7$ vacua dual to A-type orbi-instantons, constructed in full generality in \cite{Apruzzi:2013yva,Gaiotto:2014lca,Apruzzi:2017nck}.  (The Type IIA massless vacua come from a reduction to 10d of an orientifolded version of the 11d Freund--Rubin solution, i.e. AdS$_7\times S^4/(\zz_2 \times \Gamma_\text{ADE})$ \cite[Sec. 7]{DelZotto:2014hpa}.  For M5's probing just the $E_8$ wall,  i.e. for $k=1$ in type A, an AdS$_7 \times S^4/\zz_2$ vacuum was argued to exist already in \cite{Berkooz:1998bx}.) Indeed the O8 is the last missing ingredient in the analysis of \cite{Apruzzi:2019ecr}, which has found that AdS$_7$ vacua dual to D8-D6-NS5 or D8-O6-D6-NS5 brane systems are unstable (in their non-supersymmetric version) and non-scale-separated (in their supersymmetric version). We plan to come back to an extension of that analysis in presence of the O8$^-$ and O8$^*$ in \cite{companion}.

\section*{Acknowledgments}

We are indebted to G.~Bruno de Luca for collaboration in the early stages of this work, useful comments, and ongoing collaboration on related topics; to Simone Giacomelli and Noppadol Mekareeya for many useful discussions about magnetic quivers, and to the former for sharing with us some results from \cite{Giacomelli:2022drw} prior to publication and ongoing collaboration; to Paul Levy and Tom Rudelius for enlightening email correspondence; and to Alessandro Tomasiello for bringing to our attention reference \cite{Gorbatov:2001pw} and various useful comments.  The work of MF is supported in part by the European Union's Horizon 2020 research and innovation programme under the Marie Skłodowska-Curie grant agreement No. 754496 - FELLINI.  MF would like to thank SISSA, Trieste for hospitality during the completion of this work.  SG is supported in part by INFN and by MIUR-PRIN contract 2017CC72MK003. 

\appendix

\section{\texorpdfstring{RG flow Hasse diagrams for $k=2,\ldots,20$}{RG flow Hasse diagrams for k=2,...,20}}
\label{app:RG}

\begin{figure}
	\centering
	\begin{subfigure}[t]{0.1\textwidth}
		\centering
		\hspace*{-10pt}
		\includegraphics[height=0.5\textheight]{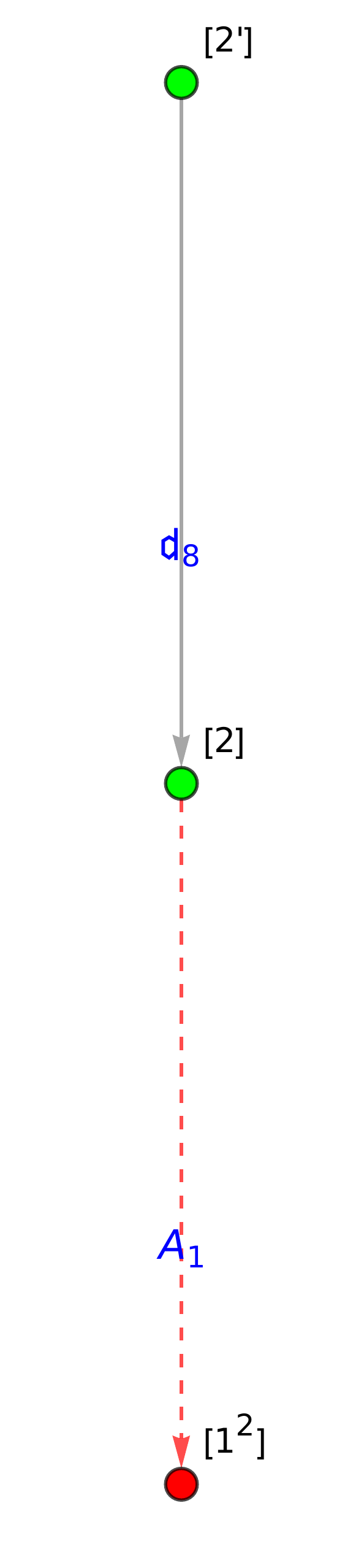}
		\caption{$k=2$}
	\end{subfigure}%
	~ 
	\begin{subfigure}[t]{0.1\textwidth}
		\centering
		\includegraphics[height=0.5\textheight]{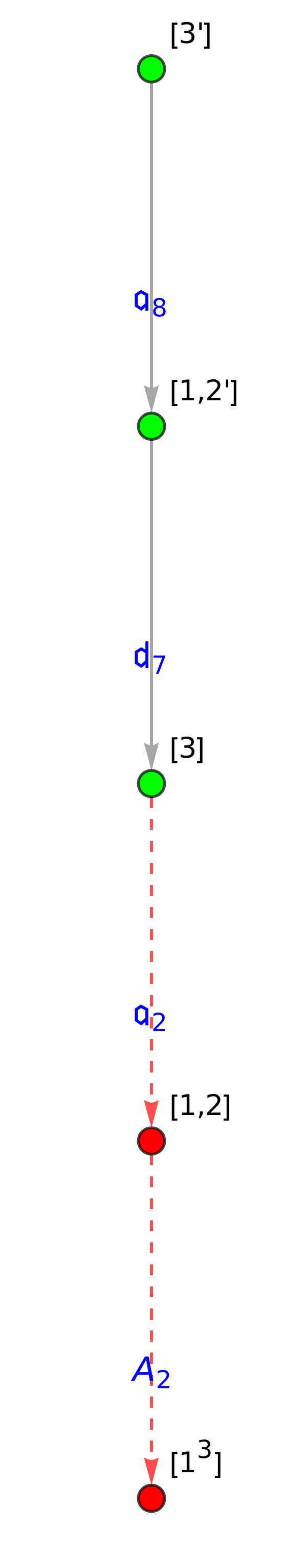}
		\caption{$k=3$}
	\end{subfigure}
	~ 
	\begin{subfigure}[t]{0.3\textwidth}
		\centering
		\includegraphics[height=0.6\textheight]{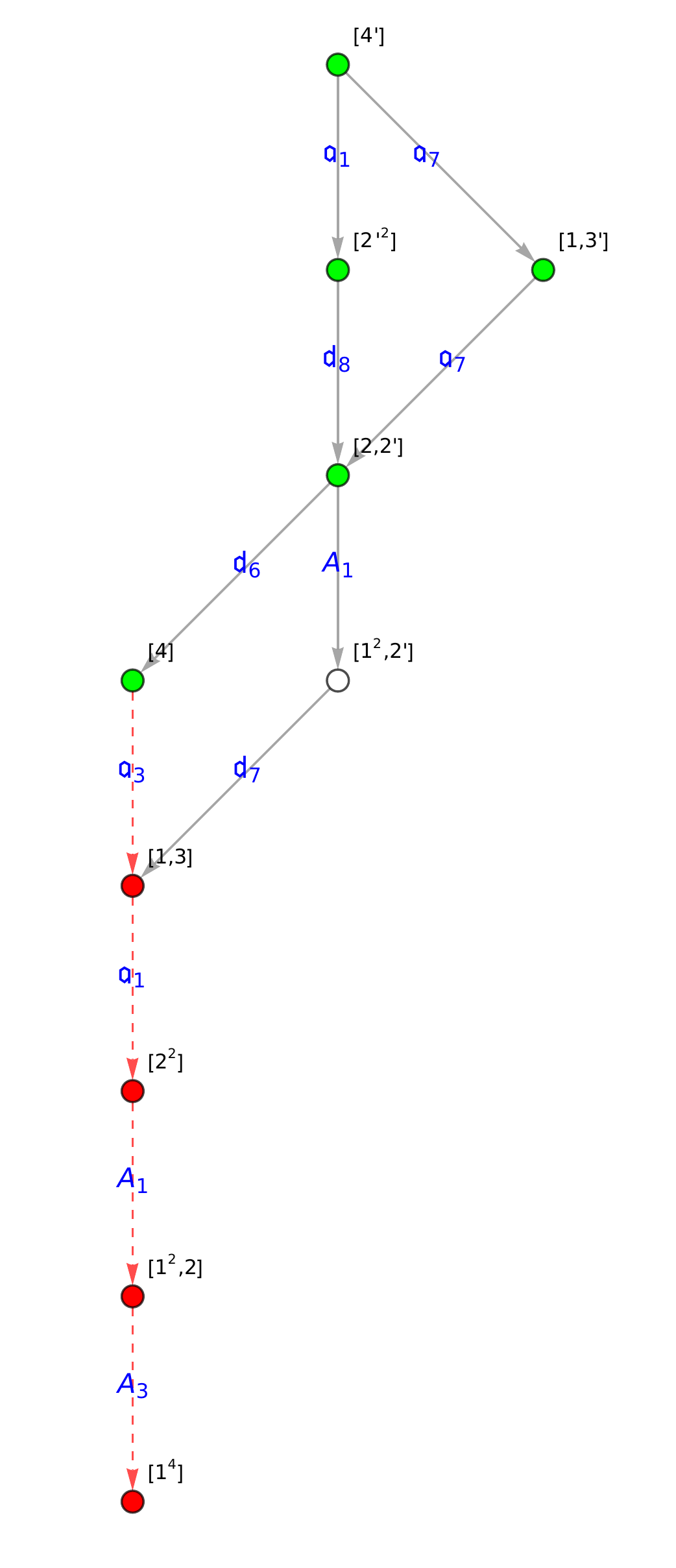}
		\caption{$k=4$}
	\end{subfigure}
	~ 
	\begin{subfigure}[t]{0.3\textwidth}
		\centering
		\includegraphics[height=0.8\textheight]{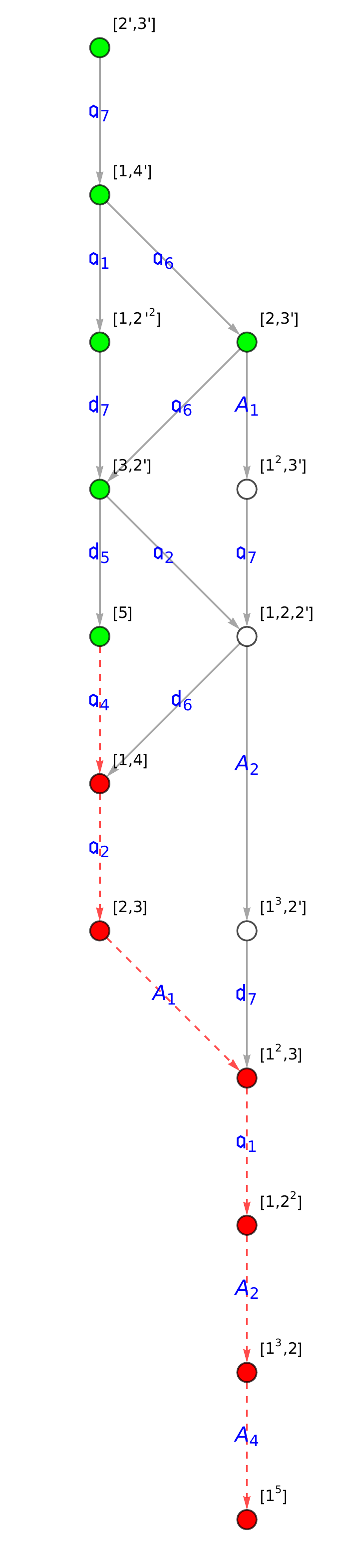}
		\caption{$k=5$}
	\end{subfigure}
	\caption{Hierarchy of RG flows between A-type orbi-instantons for $k=2,3,4,5$ at fixed $N$. Green nodes indicate theories that can flow into the top nodes at $N-1$, and therefore generate the full RG flow hierarchy at $N-1$. Red nodes and red dashed lines between them highlight the Hasse diagram of nilpotent orbits of the $\su{2},\su{3},\su{4},\su{5}$ flavor algebras that are realized on a stack of $2,3,4$ and $5$ D8-branes respectively.}
	\label{fig:k2345}
\end{figure}

In this appendix we collect the Hasse diagram of homomorphisms in $\text{Hom}(\zz_k,E_8)$ for $k=2,\ldots,20$.  The legend for the various colors and decorations in the RG flows is as follows. (The figures for $k\geq 7$ have been scaled down to fit in a single page; the interested reader can zoom in to inspect the details.) 
\begin{itemize}
	\item For $k=2,3,4,5$ (figure \ref{fig:k2345}) the list of colors and decorations is the same as that in section \ref{sub:6legend}. The red nodes and red dashed edges correspond to nilpotent orbits and KP transitions between them for $\su{2}$, $\su{3}$, $\su{4}$, $\su{5}$ respectively. 
	
In particular in figure \ref{fig:frey-rudelius} we have plotted the proposed hierarchy for $k=4$ from \cite[Fig. 1]{Frey:2018vpw}. Notice the difference with respect to our hierarchy: we have fixed $N$ whereas the latter reference describes a ``mixed'' flow between orbi-instantons with different Kac labels while also changing $N$ (i.e. performing small instanton transitions):\footnote{We would like to thank T.~Rudelius for email correspondence, and A.~Tomasiello for discussion on this point.} over each SCFT in their hierarchy we have superposed the number of tensor multiplets of the SCFT.  The flows are implemented by KP transitions of the minimal type written besides the edges in figure \ref{fig:frey-rudelius}; these transitions are found by quiver subtraction between the magnetic quivers (associated with the Kac labels) appearing in table \ref{tab:k=6freymagquiv}. Notice in particular the presence of $\mathfrak{e}_i$ KP transitions: these can only appear in the game whenever we perform a small $E_8$ instanton transition reducing $N$.
	
	\item For $k=7$ (figure \ref{fig:k78}), the colors and decorations correspond once again to those in section \ref{sub:6legend}.  The red nodes and red dashed edges correspond to nilpotent orbits and KP transitions between them for $\su{7}$, which starts from $[4',3']$ as noted in \fref{eq:k7start}
	
	\item For $k=8$ and $k=9$ (figure \ref{fig:k78} and \ref{fig:k9} respectively),  the nodes, edges, edge labels and the green nodes have the usual meaning as in section \ref{sub:6legend}. The red nodes and red dashed edges correspond to nilpotent orbits and KP transitions between them for $\su{8}$ and $\su{9}$ respectively. These flows start from $[4'^2]$ and $[3'^3]$ respectively as noted in equations \eqref{eq:k8start} and \eqref{eq:k9start}.  The yellow nodes indicate a pair of theories which have the same electric quiver but differ by the 6d $θ$ angle.
	
	\item For $k=10$ and $k=11$ (figure \ref{fig:k10} and figure \ref{fig:k11} respectively), the nodes, edges, edge labels and the green nodes have the usual meaning as in section \ref{sub:6legend}. Moreover:
	\begin{itemize}
		\item The blue nodes correspond to a pair of theories that also differ by their 6d $\theta$ angles, but have $\usp{2}$ as gauge algebra on the $-1$ curve.
		
		\item the red nodes and red dashed edges correspond to nilpotent orbits and KP transitions between them for $\su{8}$. For $k=10$, the starting point is $[4'^2,2']$, or equivalently $[3'^2, 2'^2]$ , as noted in \fref{eq:generick-2ai}. For $k=11$, the flow starts from $[3'^3, 2']$ as noted in \fref{eq:generick-2b}.
		
		\item The orange nodes and edges correspond to ``parallel flows'' between theories differing by the 6d $\theta$ angle, and have an empty gauge algebra on the $-1$ curve of their electric quivers. 
		
	\end{itemize}
	
	\item For $k=12$ (figure \ref{fig:k12}),  the nodes, edges, edge labels and the green nodes have the usual meaning as in section \ref{sub:6legend}. Moreover:
	\begin{itemize}
		\item the red nodes and red dashed edges correspond to nilpotent orbits and KP transitions between them for $\su{9}$. This starts from $[3'^4]$ as noted in \fref{eq:su9starting}.
		
		\item The orange nodes and edges correspond to parallel flows between theories differing by the 6d $\theta$ angle, and have an empty algebra group on the $-1$ curve of their electric quivers. 
		
		\item The blue nodes correspond to parallel flows between theories that differ by their 6d $θ$ angles, and have $\usp{2}$ as gauge algebra on the $-1$ curve. 
		
		\item The purple nodes correspond to a pair of theories that also differ by their 6d $θ$ angles, and have a $\usp{4}$ gauge algebra on the $-1$ curve.
	\end{itemize}
	
	\item For $k=13$ (figure \ref{fig:k13}),  the decorations are very similar to those to $k=12$, except that the red nodes and red dashed edges correspond to nilpotent orbits and KP transitions between them for $\su{8}$. The starting point of the flow is $[3'^3,2'^2]$ as noted in \fref{eq:generick-2b}.
	
	\item For $k=14$ (figure \ref{fig:k14}), the nodes, edges, edge labels and the green nodes have the usual meaning as in section \ref{sub:6legend}. Moreover:
	\begin{itemize} 		
		\item there are now four different ways of embedding the nilpotent orbits Hasse diagram of $\su{8}$ in this homomorphisms Hasse as discussed in equations \eqref{eq:k14theta1}-\eqref{eq:k14theta3}:
		\begin{itemize}
			\item the red nodes and red dashed edges correspond to $\su{8}$ Hasse diagrams starting from the electric quiver in \eqref{eq:k14theta2};
			
			\item the blue nodes and blue dashed edges correspond to $\su{8}$ Hasse diagrams starting from the electric quiver in \eqref{eq:k14theta3};
			
			\item the magenta nodes and magenta dashed edges correspond to $\su{8}$ Hasse diagrams starting from the electric quiver in \eqref{eq:k14theta1}. The two starting points for these flows differ by the 6d $θ$ angle and are highlighted by black nodes. The node $[4'^2, 2'^3]$ is black, magenta as well as green. For simplicity of presentation, we represent it simply as a green dot in \fref{fig:k14}.
		\end{itemize}
		
		\item There are four families of theories differing by the 6d $θ$ angle:
		
		\begin{itemize}
		\item the orange and light blue nodes with the corresponding edges are parallel flows between theories differing by their 6d $θ$ angles. These flows have an empty gauge algebra on the $-1$ curve and are discussed in \eqref{eq:k14usp0};
		
		\item the olive green nodes and edges correspond to parallel flows between theories that have a $\usp{2}$ gauge algebra on the $-1$ curve, and are discussed in \eqref{eq:k14usp2};
		
		\item the brown nodes and edges correspond to parallel flows between theories that have a $\usp{4}$ gauge algebra on the $-1$ curve, and are discussed in \eqref{eq:k14usp4};
		
		\item the black nodes correspond to the pair of theories that have a $\usp{6}$ gauge group on the $-1$ curve. As remarked above, The node $[4'^2, 2'^3]$ is black, magenta as well as green. For simplicity of presentation, we represent it simply as a green dot.
		
		\end{itemize}
	\end{itemize}
\item For $k=15$  (figure \ref{fig:k15}), the nodes, edges, edge labels and the green nodes have the usual meaning as in section \ref{sub:6legend}.
	\begin{itemize}
		\item In addition, the red nodes and dashed edges correspond to the $\su{9}$ Hasse diagram, which starts from $[3'^5]$ as noted in \fref{eq:su9starting}.
		\item The four families of theories differing by the 6d $θ$ angle are highlighted using the same colors as in the $k=14$ diagram.
	\end{itemize}

\item For $k=16$  (figure \ref{fig:k16}), the nodes, edges, edge labels and the green nodes have the usual meaning as in section \ref{sub:6legend}. Moreover:
\begin{itemize} 		
	\item there are three different starting points for the nilpotent orbits Hasse of $\su{8}$ as is evident from equations \eqref{eq:generick-2ai} and \eqref{eq:generick-2aii}. The ones that follow from \fref{eq:generick-2ai} are $[4'^2,2'^4]$ and $[4'^4,2']$, which additionally have an equivalent node that differ by the 6d $θ$ angle namely $[3'^2,2'^5]$ and $[3'^4,2'^2]$ respectively. \Fref{eq:generick-2aii} gives $[4'^3,2'^2]$. The nilpotent orbits Hasse diagram of $\su{8}$ originating from each of these nodes is highlighted with magenta (for $[4'^2, 2'^4]$ and $[3'^2,2'^5]$), red (for $[4'^3,2'^2]$), blue (for $[4'^4]$) and orange (for $[3'^4,2'^2]$) dashed edges respectively.
	
	\item The olive green, brown, black and yellow nodes correspond to theories and flows that differ by their 6d $θ$ angles. The electric quivers corresponding to these theories have $\usp{2}, \usp{4}, \usp{6}$ and $\usp{8}$ gauge groups on the $-1$ curve respectively.
\end{itemize}

\item For $k=17$  (figure \ref{fig:k17}), the nodes, edges, edge labels and the green nodes have the usual meaning as in section \ref{sub:6legend}. The rest of the decorations are the same as those for $k=15$. The nilpotent orbits Hasse diagram of $\su{8}$ starts from $[3'^3,2'^4]$ as expected from \fref{eq:generick-2b}.

\item For $k=18$  (figure \ref{fig:k18}), the nodes, edges, edge labels and the green nodes have the usual meaning as in section \ref{sub:6legend}.  In addition:
	\begin{itemize}
		\item the red nodes and dashed edges correspond to the $\su{9}$ Hasse diagram, which starts from $[3'^6]$ as noted in \fref{eq:su9starting}.
		\item The flows highlighted with cyan and orange correspond to parallel flows among theories that differ by the 6d $θ$ angle, and have an empty $-1$ curve in their electric quiver.
		\item Nodes and flows highlighted in olive green, brown, black, yellow and blue correspond to theories that differ by their 6d $θ$ angle and have the gauge group $\usp{2}, \usp{4}, \usp{6}, \usp{8}$ and $\usp{10}$ on the $-1$ curve of their electric quiver.
	\end{itemize}

\item For $k=19$  (figure \ref{fig:k19}), the nodes, edges, edge labels and the green nodes have the usual meaning as in section \ref{sub:6legend}.  In addition:
\begin{itemize}
	\item there are two different ways of embedding the nilpotent orbits Hasse diagram of $\su{8}$ in this homomorphisms Hasse. This follows from \fref{eq:generick-2b} -- which gives $[3'^3,2'^5]$, and \fref{eq:generick-3a} -- which gives $[3'^5,2'^2]$. These are highlighted with blue and red nodes and dashed edges respectively.
	\item The flows highlighted with cyan and orange correspond to parallel flows among theories that differ by the 6d $θ$ angle, and have an empty $-1$ curve in their electric quiver.
	\item Nodes and flows highlighted in olive green, brown, black, yellow and blue correspond to theories that differ by their 6d $θ$ angle and have the gauge group $\usp{2}, \usp{4}, \usp{6}, \usp{8}$ and $\usp{10}$ on the $-1$ curve of their electric quiver.
\end{itemize}

\item For $k=20$  (figure \ref{fig:k20}), the nodes, edges, edge labels and the green nodes have the usual meaning as in section \ref{sub:6legend}.  In addition:
\begin{itemize}
	\item there are four different ways of embedding the nilpotent orbits Hasse diagram of $\su{8}$ in this homomorphisms Hasse as
	 discussed in \fref{sec:generick_nilpotent_hasse}. These flows start from $[4'^2,2'^6]$ (and the equivalent $[3'^2,2'^7]$ that differs by the 6d $θ$ angle), $[4'^3,2'^4]$, $[4'^4,2'^2]$ (and the equivalent $[3'^4,2'^4]$ that differs by the 6d $θ$ angle) and $[4'^5]$. These follow from equations \eqref{eq:generick-2ai} and \eqref{eq:generick-2aii}. The flows are highlighted with dashed edges and nodes colored in red, blue, magenta and grey.	
	\item The flows highlighted with cyan and orange correspond to parallel flows among theories that differ by the 6d $θ$ angle, and have an empty $-1$ curve in their electric quiver.
	\item Nodes and flows highlighted in olive green, brown, black, yellow, blue and gray correspond to theories that differ by their 6d $θ$ angle and have the gauge group $\usp{2}, \usp{4}, \usp{6}, \usp{8}, \usp{10}$ and $\usp{12}$ on the $-1$ curve of their electric quiver.
\end{itemize}

\end{itemize}
\begin{figure}
	\centering
	\footnotesize
	\begin{tikzpicture}[node distance=70pt]
	  \tikzstyle{arrow} = [thick,->,>=stealth]
	  \node (n1) {$\overset{(N)}{[1^4] \leftrightarrow \f = E_8}$};
	  \node (n2) [below of=n1] {$\overset{(N-1)}{[2,1²]\leftrightarrow \f = E_7 \oplus \uu{1}}$};
	  \node (n3) [below of=n2] {$\overset{(N-2)}{[2^\prime, 1²]\leftrightarrow \f =\so{14}}$};
	  \node (n4) [below of=n3] {$\overset{(N-2)}{[3,1]\leftrightarrow \f =E_6 \oplus \su{2} \oplus \uu{1}}$};
	  \node (n5) [below left=30pt and 20pt of n4] {$\overset{(N-2)}{[2²]\leftrightarrow \f =E_7 \oplus \su{2}}$};
	  \node (n6) [below right=60pt and 20pt of n4] {$\overset{(N-3)}{[3',1]\leftrightarrow \f =\su{8}\oplus \uu{1}}$};
	  \node (n7) [below= 120pt of n4] {$\overset{(N-3)}{[2,2'] \leftrightarrow \f =\so{12}\oplus\su{2}\oplus\uu{1}}$};
	  \node (n8) [below of=n7] {$\overset{(N-3)}{[4]\leftrightarrow \f =\so{10}\oplus \su{4}}$};
	  \node (n9) [below of=n8] {$\overset{(N-4)}{[4']\leftrightarrow \f =\su{8}\oplus\su{2}}$};
	  \node (n10) [below of=n9] {$\overset{(N-4)}{[2'^2]\leftrightarrow \f =\so{16}}$};
  
	  \draw [arrow] (n1)--(n2) node [midway, left] {} node[midway, right] {$\mathfrak{e}_8$}; 

	  \draw [arrow] (n2)--(n3) node [midway, left] {} node[midway, right] {$\mathfrak{e}_7$}; 

	  	  \draw [arrow] (n3)--(n4) node [midway, left] {} node[midway, right] {$\mathfrak{d}_7$}; 

	  \draw [arrow] (n4)--(n5) node [midway, above left] {} node[midway, below right] {$\mathfrak{a}_1$};

	  \draw [arrow] (n4)--(n6) node [midway, above right=0pt and 1pt] {} node[midway, below left] {$\mathfrak{e}_6$};

	  \draw [arrow] (n5)--(n7) node [midway, left] {} node[midway, above right] {$\mathfrak{e}_7$}; 

	  \draw [arrow] (n6)--(n7) node [midway, right] {

	  } node[midway, above left] {$\mathfrak{a}_7$};  

	  \draw [arrow] (n7)--(n8) node [midway, left] {} node[midway, right] {$\mathfrak{d}_6$}; 

	  \draw [arrow] (n8)--(n9) node [midway, left] {} node[midway, right] {$\mathfrak{d}_5$}; 

	  \draw [arrow] (n9)--(n10) node [midway, left] {} node[midway, right] {$\mathfrak{a}_1$}; 
	\end{tikzpicture}
	\caption{The hierarchy of RG flows for A-type $k=4$ orbi-instantons from \cite[Fig. 1]{Frey:2018vpw} recast in the formalism of the present paper. The two staggered nodes are intended to convey that the higher IR SCFT (i.e. Kac label) has larger 6d $a$ anomaly.}
	\label{fig:frey-rudelius}
  \end{figure}
  
  \begin{table}
	\renewcommand{\arraystretch}{3}
	\centering
	\begin{tabular}{@{} c c c@{}}
	 \makecell{Kac \\label} & \makecell{Number of \\ tensor multiplets} & 3d magnetic quiver   \\ \toprule
	 $[1^4]$& $N$ & ${\scriptstyle 1-2-3-4-N - 2N - 3N - 4N - 5N - \overset{\overset{\scriptstyle 3N}{\vert}}{6N}- 4N - 2N}$ \\
	 $[2,1²]$& $N-1$ & ${\scriptstyle 1-2-3-4-N - 2(N-1) - 3(N-1) - 4(N-1) - 5(N-1) - \overset{\overset{\scriptstyle 3(N-1)}{\vert}}{6(N-1)}- 4(N-1) - 2(N-1)}$ \\
	 $[2^\prime, 1²]$ & $N-1$ & ${\scriptstyle 1-2-3-4-N - (2N-2) - (3N-4) - (4N-6) - (5N-8) - \overset{\overset{\scriptstyle 3N-5}{\vert}}{(6N-10)}- (4N-7) - (2N-4)}$ \\ 
	 $[3,1]$ & $N-2$ & ${\scriptstyle 1-2-3-4-N - (2N-3) - (3N-6)-(4N-8)-(5N-10)-\overset{\overset{\scriptstyle 3N-6}{\vert}}{(6N-12)}-(4N-8)-(2N-4)}$ \\
	 $[2²]$ & $N-2$ & ${\scriptstyle 1-2-3-4-N - 2(N-2) - 3(N-2) - 4(N-2) - 5(N-2) - \overset{\overset{\scriptstyle 3(N-2)}{\vert}}{6(N-2)}- 4(N-2) - 2(N-2)}$ \\
	 $[3^\prime, 1]$ & $N-3$ & ${\scriptstyle 1-2-3-4-N - (2N-3) - (3N-6) - (4N-9) - (5N-12) - \overset{\overset{\scriptstyle 3N-8}{\vert}}{(6N-15)}- (4N-10) - (2N-5)}$ \\
	 $[2,2^\prime]$ & $N-3$ & ${\scriptstyle 1-2-3-4-N - (2N-4) - (3N-7) - (4N-10) - (5N-13) - \overset{\overset{\scriptstyle 3N-8}{\vert}}{(6N-16)}- (4N-11) - (2N-6)}$ \\
	 $[4]$ & $N-3$ & ${\scriptstyle 1-2-3-4-N - 2(N-2) - (3N-8) - 4(N-3) - 5(N-3) - \overset{\overset{\scriptstyle 3(N-3)}{\vert}}{6(N-2)}- 4(N-2) - 2(N-2)}$ \\
	 $[4^\prime]$ & $N-4$ & ${\scriptstyle 1-2-3-4-N - (2N-4) - (3N-8) - (4N-12) - (5N-16) - \overset{\overset{\scriptstyle 3N-10}{\vert}}{(6N-20)}- (4N-14) - (2N-7)}$ \\
	 $[2'^2]$ & $N-4$ & ${\scriptstyle 1-2-3-4-N - (2N-4) - (3N-8) - (4N-12) - (5N-16) - \overset{\overset{\scriptstyle 3N-10}{\vert}}{(6N-20)}- (4N-14) - (2N-8)}$ \\
  
	 \bottomrule
	\end{tabular}
	\caption{Magnetic quivers corresponding to the $k=4$ Kac labels of figure \ref{fig:frey-rudelius}.}
	\renewcommand{\arraystretch}{1}
	\label{tab:k=6freymagquiv}
	\end{table}
\begin{figure}[htbp!]
	\centering 
	\begin{subfigure}[t]{0.4\textwidth}
	\centering
	\hspace*{-50pt}
	\includegraphics[height=0.8\textheight]{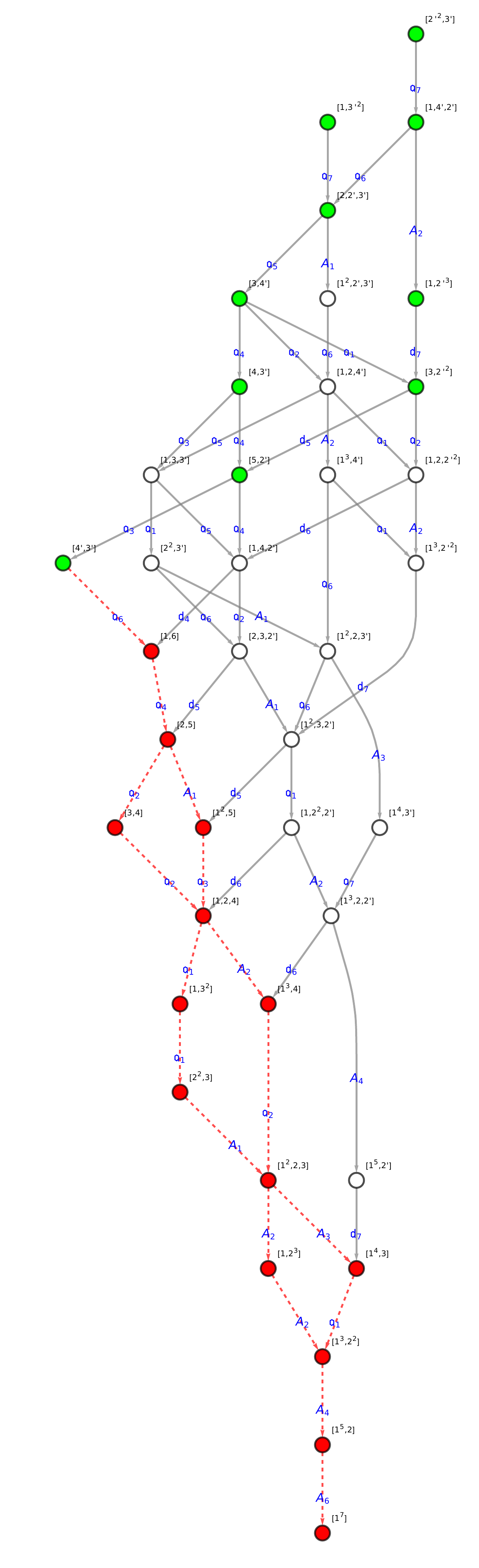}
	\caption{$k=7$}
\end{subfigure}
~
\begin{subfigure}[t]{0.4\textwidth}
	\centering
	\hspace*{5pt}
	\includegraphics[height=0.8\textheight]{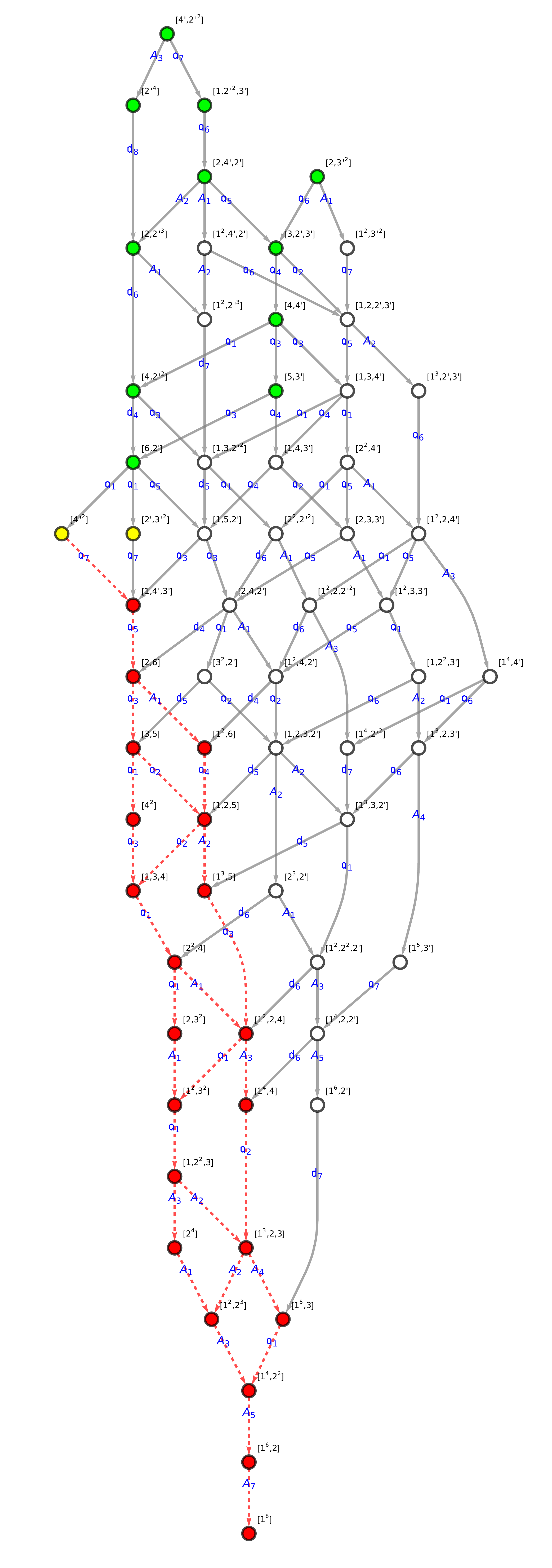}
	\caption{$k=8$}
\end{subfigure}%
	\caption{Hierarchy of RG flows between A-type orbi-instantons for $k=7$ and $k=8$ at fixed $N$. Green nodes indicate theories that can flow into all the top nodes at $N-1$, and therefore generate the full hierarchy at $N-1$. Red nodes and red dashed lines between them highlight the Hasse diagram of nilpotent orbits of the $\su{7}$ or $\su{8}$ flavor symmetry algebra that is realized on a stack of $7$ or $8$ D8-branes respectively.}
	\label{fig:k78}
\end{figure}

\begin{figure}
	\centering
	\includegraphics[height=0.95\textheight]{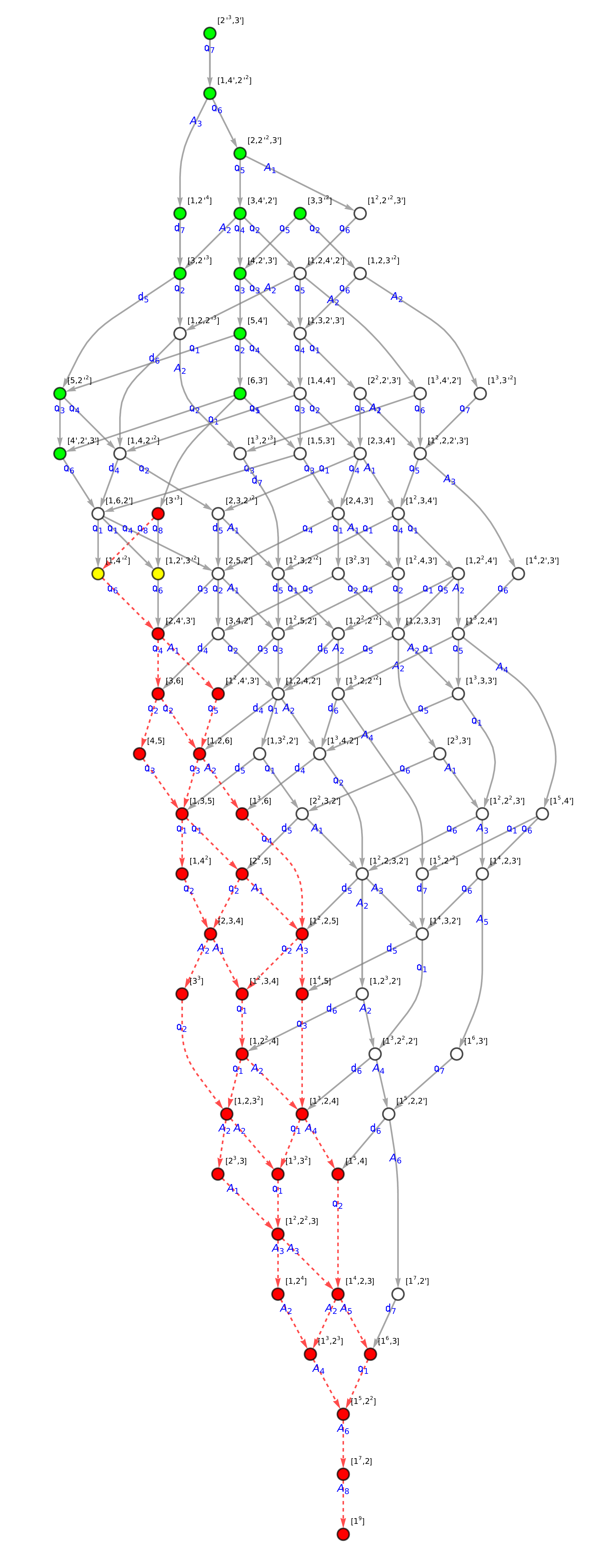}
	\caption{Hierarchy of RG flows between A-type orbi-instantons for $k=9$ and fixed $N$.}
	\label{fig:k9}
\end{figure}

\begin{figure}
	\centering
	\includegraphics[height=0.95\textheight]{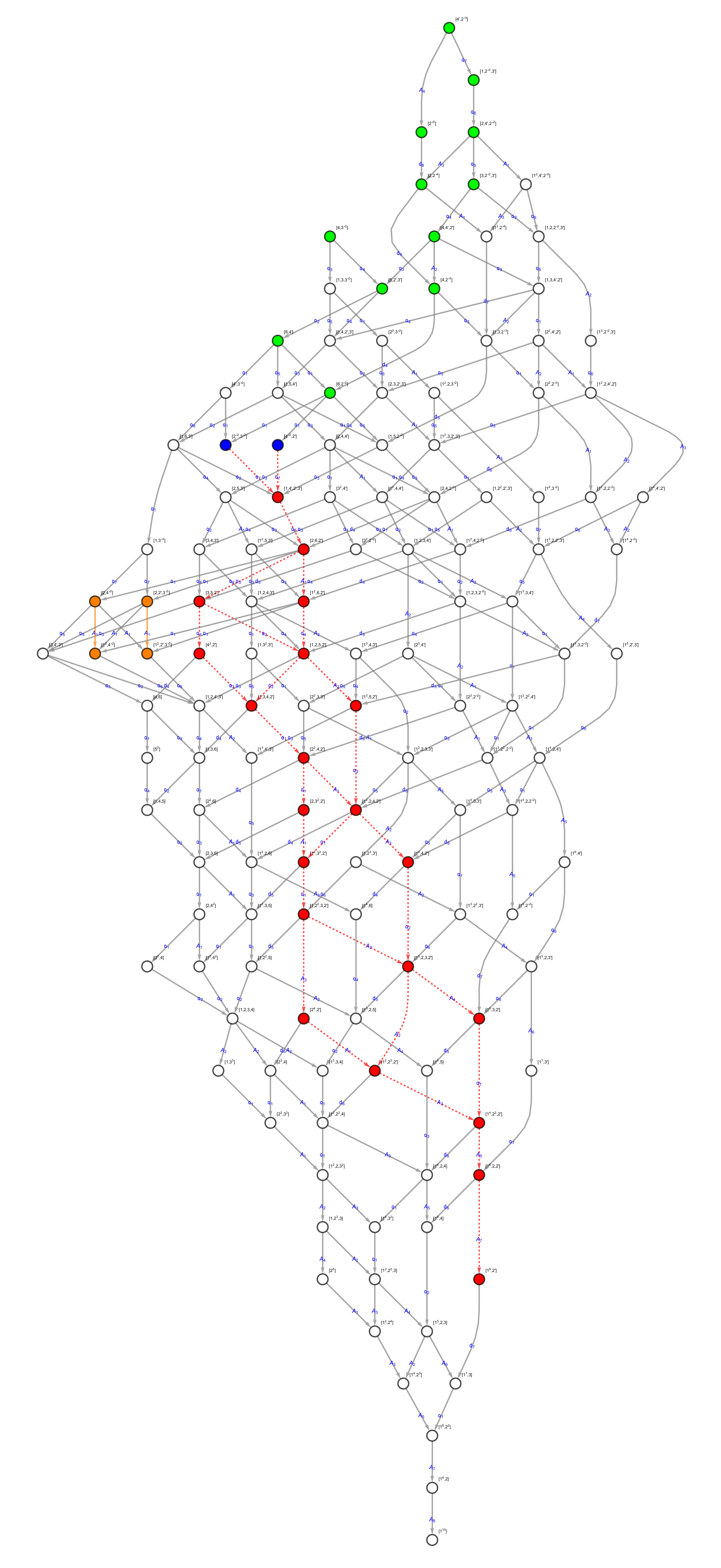}
	\caption{Hierarchy of RG flows between A-type  orbi-instantons for $k=10$ and fixed $N$.}
	\label{fig:k10}
\end{figure}

\begin{figure}
	\centering
	\includegraphics[height=0.95\textheight]{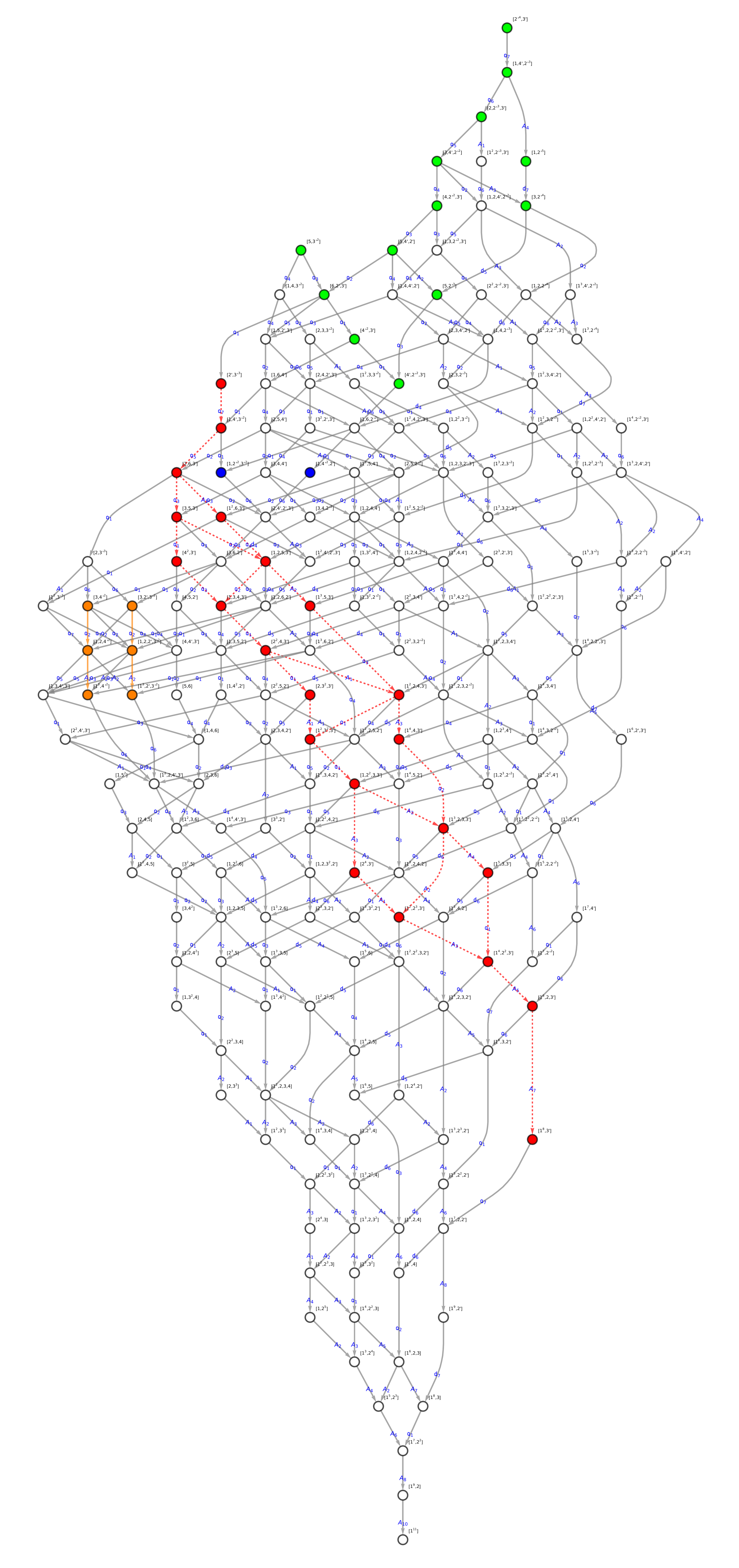}
	\caption{Hierarchy of RG flows between A-type  orbi-instantons for $k=11$ and fixed $N$.}
	\label{fig:k11}
\end{figure}

\begin{figure}
	\centering
	\includegraphics[height=0.95\textheight]{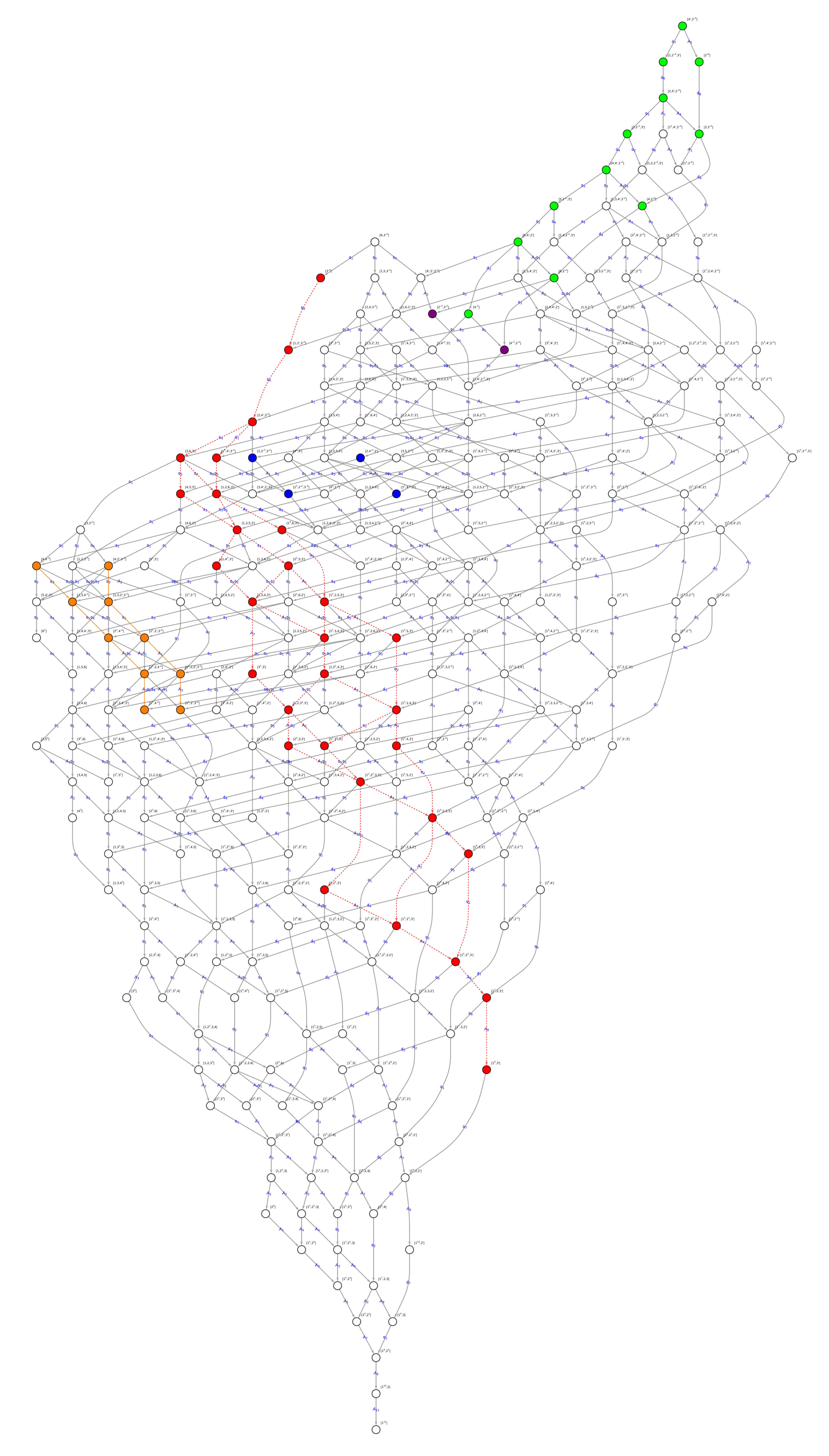}
	\caption{Hierarchy of RG flows between A-type  orbi-instantons for $k=12$ and fixed $N$.}
	\label{fig:k12}
\end{figure}

\begin{figure}
	\centering
	\includegraphics[height=0.95\textheight]{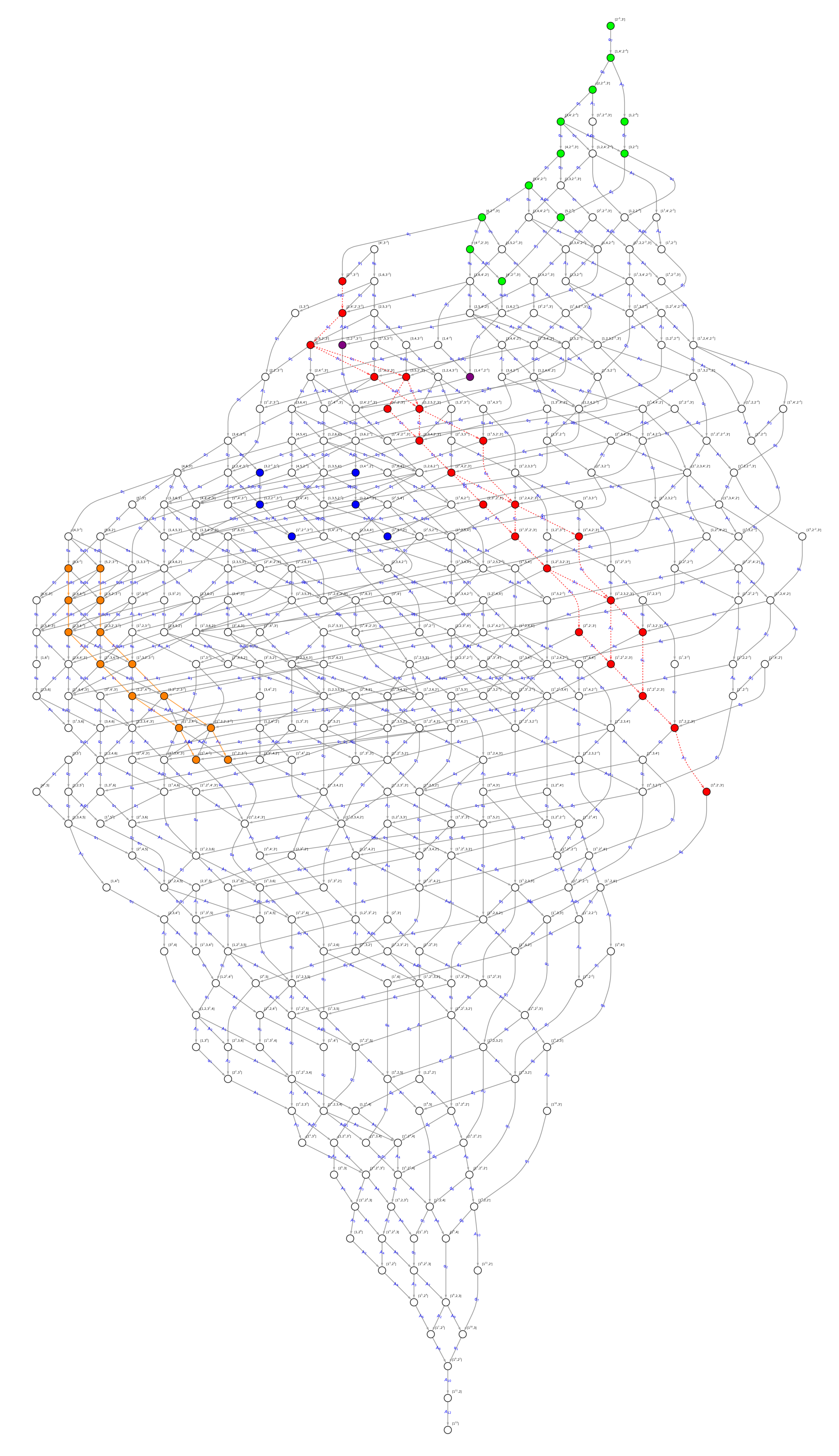}
	\caption{Hierarchy of RG flows between A-type  orbi-instantons for $k=13$ and fixed $N$.}
	\label{fig:k13}
\end{figure}

\begin{figure}
	\centering
	\includegraphics[height=0.95\textheight]{k14.pdf}
	\caption{Hierarchy of RG flows between A-type  orbi-instantons for $k=14$ and fixed $N$.}
	\label{fig:k14}
\end{figure}

\begin{figure}
	\centering
	\includegraphics[height=0.95\textheight]{k15.pdf}
	\caption{Hierarchy of RG flows between A-type  orbi-instantons for $k=15$ and fixed $N$.}
	\label{fig:k15}
\end{figure}

\begin{figure}
	\centering
	\includegraphics[width=\textwidth]{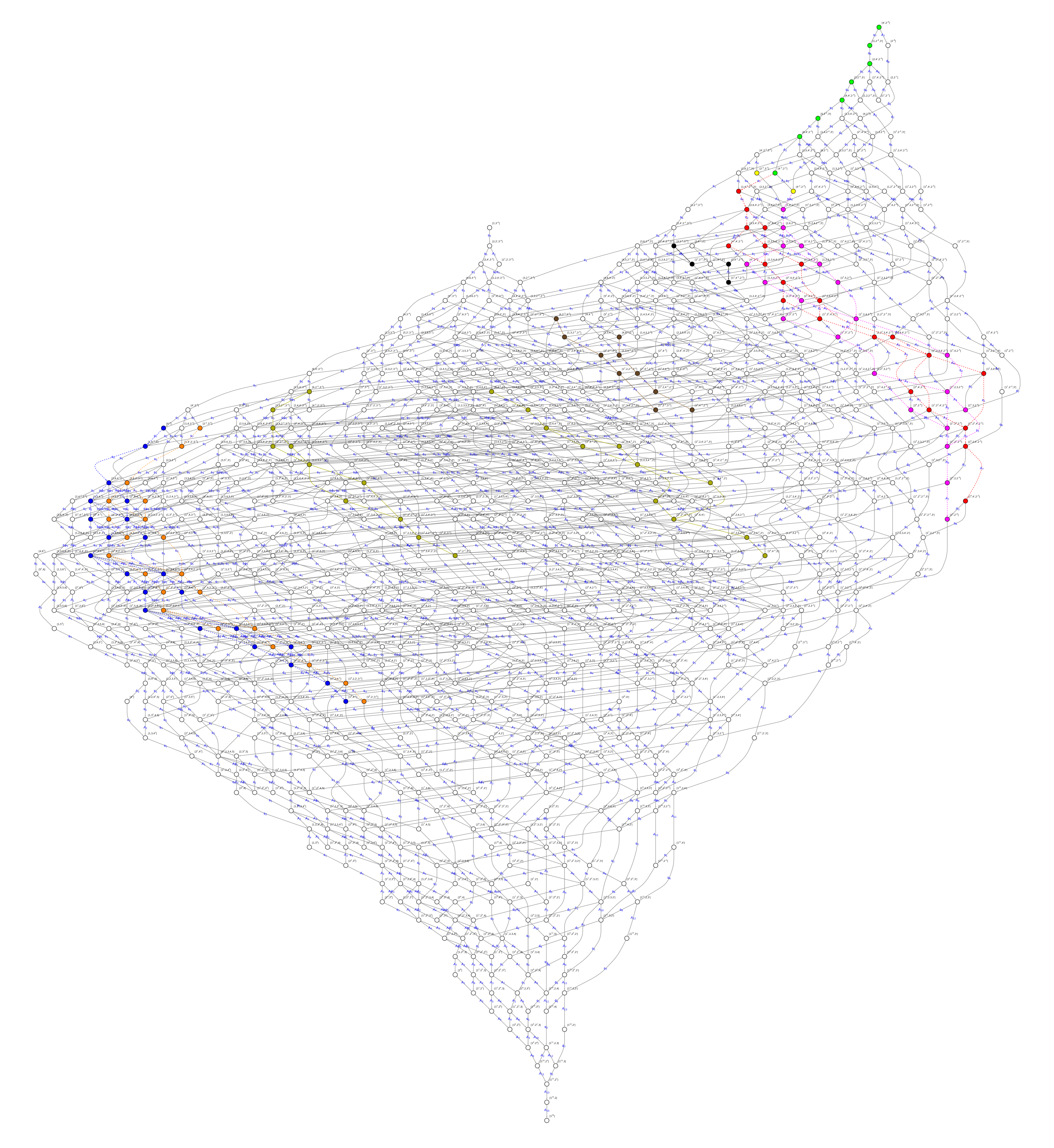}
	\caption{Hierarchy of RG flows between  A-type orbi-instantons for $k=16$ and fixed $N$.}
	\label{fig:k16}
\end{figure}

\begin{figure}
	\centering
	\includegraphics[width=\textwidth]{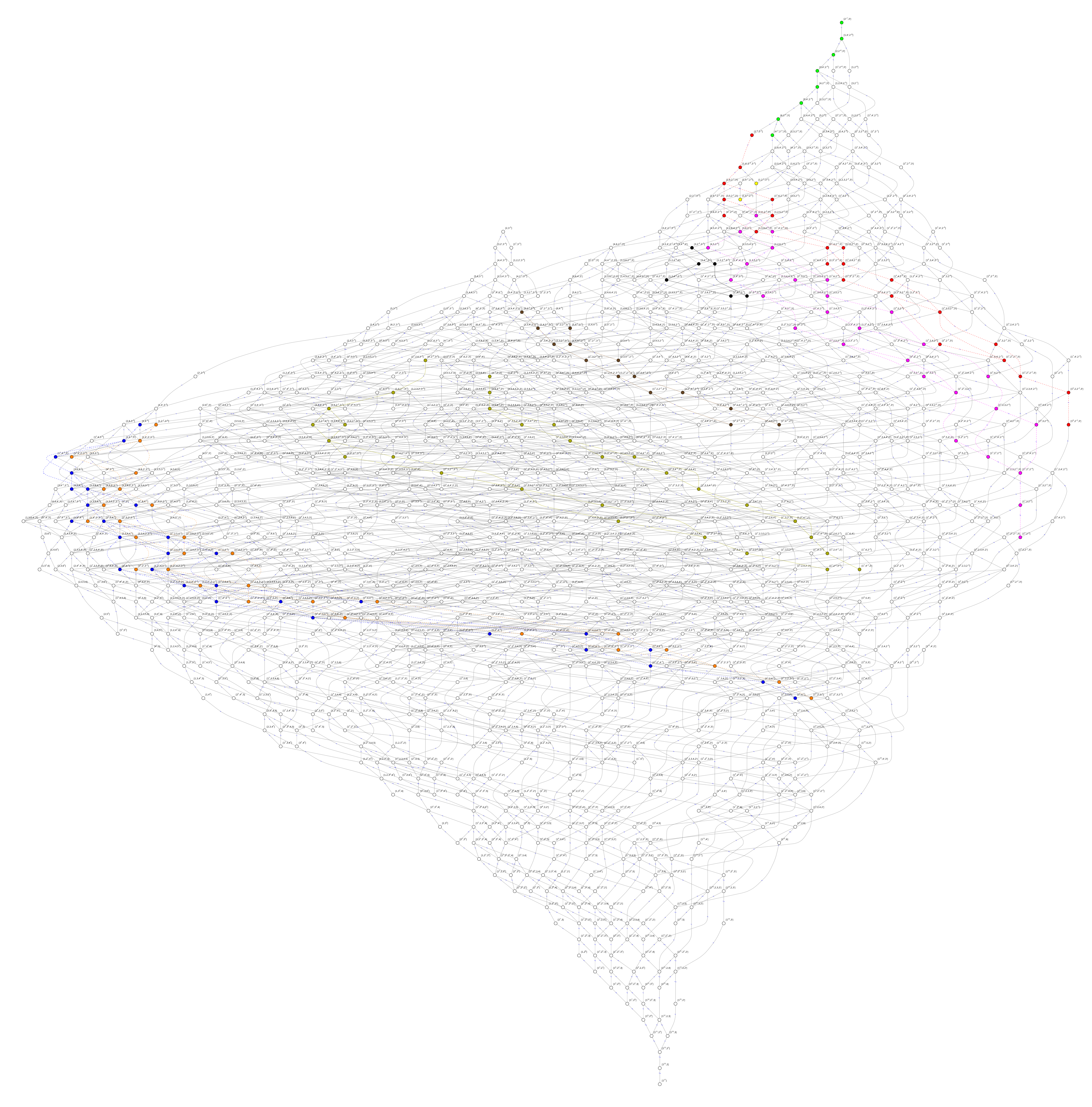}
	\caption{Hierarchy of RG flows between  A-type orbi-instantons for $k=17$ and fixed $N$.}
	\label{fig:k17}
\end{figure}

\begin{sidewaysfigure}
	\centering
	\includegraphics[width=\textwidth]{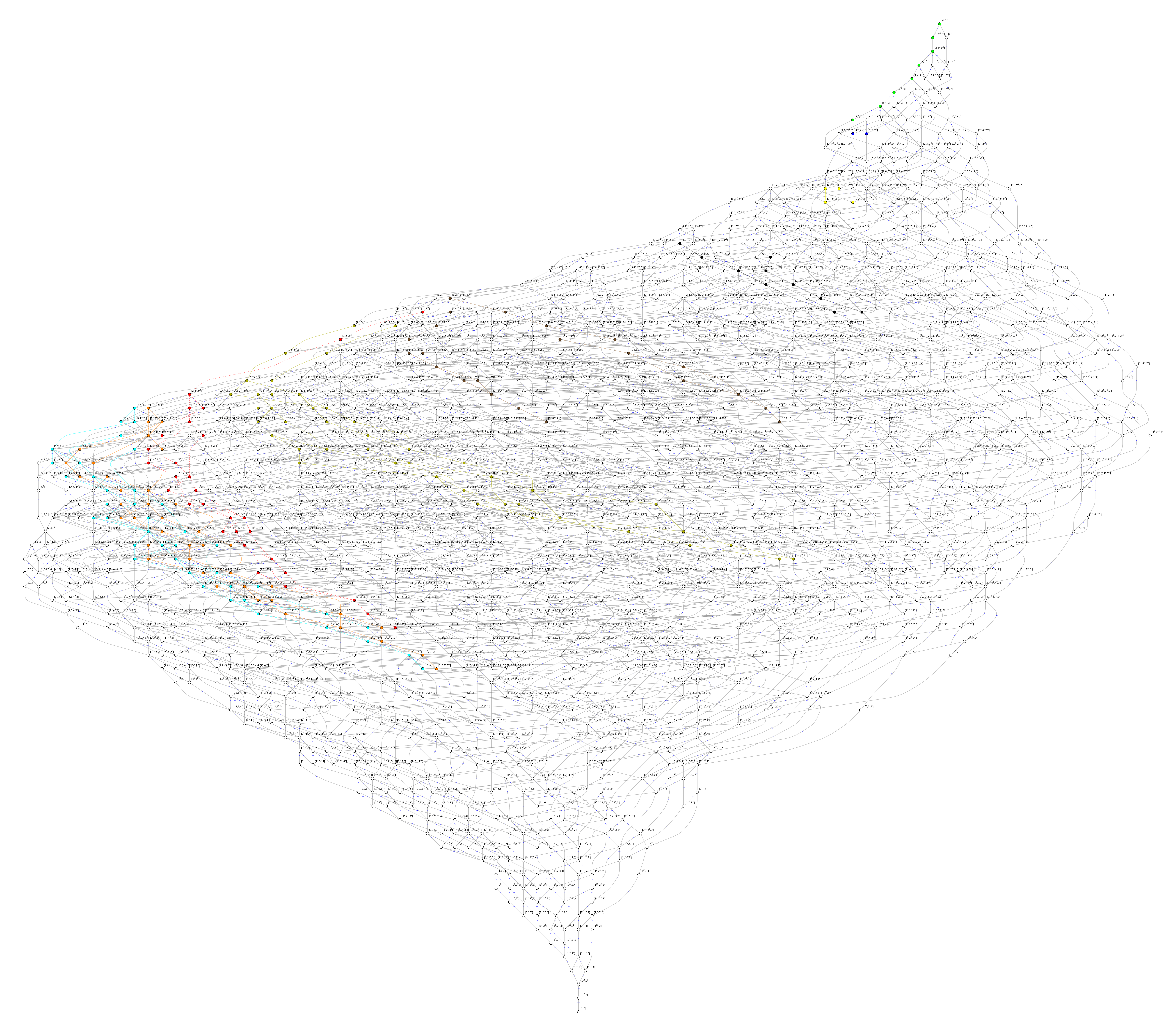}
	\caption{Hierarchy of RG flows between  A-type orbi-instantons for $k=18$ and fixed $N$.}
	\label{fig:k18}
\end{sidewaysfigure}

\begin{sidewaysfigure}
	\centering
	\includegraphics[width=\textwidth]{k19.pdf}
	\caption{Hierarchy of RG flows between  A-type orbi-instantons for $k=19$ and fixed $N$.}
	\label{fig:k19}
\end{sidewaysfigure}

\begin{sidewaysfigure}
	\centering
	\includegraphics[width=\textwidth]{k20.pdf}
	\caption{Hierarchy of RG flows between  A-type orbi-instantons for $k=20$ and fixed $N$.}
	\label{fig:k20}
\end{sidewaysfigure}

\clearpage

\bibliography{main}
\bibliographystyle{at}

\end{document}